\documentclass[a4paper,11pt]{article}
\pdfoutput=1 
\usepackage{jcappub}
\usepackage{graphicx}
\usepackage[normalem]{ulem}
\pdfoutput=1 
\usepackage[dvipsnames]{xcolor}
\usepackage{multirow}
\usepackage{makecell}
\usepackage{tikz}
\usetikzlibrary{shapes,arrows}

\title{Blind Observers of the Sky}

\makeatletter
\def\thickhline{%
  \noalign{\ifnum0=`}\fi\hrule \@height \thickarrayrulewidth \futurelet
   \reserved@a\@xthickhline}
\def\@xthickhline{\ifx\reserved@a\thickhline
               \vskip\doublerulesep
               \vskip-\thickarrayrulewidth
             \fi
      \ifnum0=`{\fi}}
\makeatother

\newlength{\thickarrayrulewidth}
\author[a,b]{Samuel Brieden,}
\author[a]{H\'ector Gil-Mar\'in,}
\author[a,c]{Licia Verde,}
\author[a,b,d]{Jos\'e Luis Bernal}

\newcommand{\hompc}{\,h\,{\rm Mpc}^{-1}}
\newcommand{\mpcoh}{\,h^{-1}\,{\rm Mpc}}

\newcommand{\underlying}{}
\newcommand{\blind}{^\prime}
\newcommand{\reference}{^\mathrm{ref}}
\newcommand{\shift}{^\mathrm{shift}}

\newcommand{\eff}{*}

\newcommand{\Nmocks}{100 }

\affiliation[1]{ICC, University of Barcelona, IEEC-UB, Mart\'i i Franqu\`es, 1, E-08028 Barcelona, Spain}
\affiliation[2]{Dept. de  F\'isica Qu\`antica i Astrof\'isica, Universitat de Barcelona, Mart\'i  i Franqu\`es 1, E-08028 Barcelona, Spain.}
\affiliation[3]{ICREA, Pg. Lluis Companys 23, Barcelona, E-08010, Spain.}
\affiliation[4]{Department of Physics and Astronomy, Johns Hopkins University, 3400 North Charles Street, Baltimore, Maryland 21218, USA}

\abstract{The concept of \textit{blind analysis}, a key procedure to  remove the human-based systematic error called \textit{confirmation bias}, has long been an integral part of data analysis in many research areas. In cosmology, \textit{blind analysis} is recently  making its entrance, as the field progresses into a fully fledged high-precision science. The credibility, reliability  and  robustness of results from  future sky-surveys will dramatically  increase if the effect of confirmation bias is kept under control by using an appropriate blinding procedure.
Here, we present a catalog-level blinding scheme for galaxy clustering data apt to be used in future spectroscopic galaxy surveys. We shift the individual galaxy positions along the line of sight based on 1) a geometric shift mimicking the Alcock-Paczynski effect and 2) a perturbative shift akin to redshift-space distortions. This procedure has several advantages. After combining the two steps above, it is almost impossible to accidentally unblind. The procedure induces a shift in cosmological parameters without changing the galaxies' angular positions, hence without interfering with the effects of angular systematics.  Since the method is applied at catalog level, there is no need to adopt several blinding schemes tuned to different summary statistics, likelihood choices or  types of analyses. By testing the method on mock catalogs and the BOSS DR12 catalog we demonstrate its performance in blinding galaxy clustering data for relevant cosmological parameters sensitive to the background expansion rate and the growth rate of structures. We publicly release our data products at \href{https://github.com/SamuelBrieden/BlindingCatalogs}{https://github.com/SamuelBrieden/BlindingCatalogs}}

\begin{document}

\maketitle

\section{Introduction}

Thanks to a coordinated observational and theoretical effort over the past two decades, cosmology has entered the precision era.  Precision cosmology rests on the extremely successful standard model of cosmology,  the flat $\Lambda$CDM model,  which provides a coherent  and precise description of a suite of observations ranging from the early to the late Universe. Two main components of the model,   dark energy with an equation of state  parameter $w=-1$ equivalent to a cosmological constant $\Lambda$, and cold dark matter (CDM), are poorly understood and may be considered as effective descriptions for the true model's ingredients, which are yet to be found. 

The goal to shed light on these components and their nature, as well as exploring other fundamental physics, is the main driver for forthcoming, observational efforts, especially for large-scale structure surveys as e.g. DESI \cite{aghamousa_desi_2016}, Euclid \cite{laureijs_euclid_2011}, LSST \cite{collaboration_science-driven_2017} and SKA~\cite{redbook}. 
Especially the LSS experiments listed above have the potential to induce a paradigm change in cosmology, since they will be sensitive to the cosmological parameters at nearly the same precision level as current CMB experiments as Planck \cite{Aghanim:2018eyx}.

%Cosmology in the future, still successful?
Unaccounted systematic errors may be interpreted as signs of new physics. This is a risk that should be mitigated at all costs. Conversely however, signs of new physics may appear as unexpected results, which the experimenter might attribute to systematic errors. It is therefore imperative to have an exquisite control on the systematic errors, which is ideally achieved by fulfilling the following conditions:
\begin{enumerate}
\item All potential sources of systematic errors need to be identified. 
\item Each systematic needs to be modeled appropriately.
\item Their effect on the posterior distribution needs to be explored and correctly included in the systematic error budget of the final result.
\end{enumerate}

While this is well known and a significant effort of the community is devoted to these  aims with important successes, a basic question remains to be answered:  at which point the search for further systematics may stop? This is the aspect more prone to  confirmation bias:  we, as scientists and human beings, unconsciously tend to accept results faster if they are in agreement with previous findings, than if they disagree. 

%What is blinding?
One promising way of avoiding experimenter's bias is to carry out blind data analyses, where the original data vector is transformed as to hide the true signal in a controlled way before running the analysis pipeline. If suitably implemented, this allows for studying the degeneracies of systematic effects with theoretical models while being blind to the underlying values of the model parameters.\footnote{In what follows we refer to the values of the model parameters preferred by the data as model's parameters {\it underlying} values. We prefer this nomenclature to {\it true} values which is often used, because, for example,  the model adopted could in principle be incorrect or  because of  sample variance the values preferred by the data might differ from the true ones. When dealing with mock data or when neglecting this possibility, there is a common trend to loosely identify {\it underlying} values with {\it true} values.} Once the analysis of the blinded data is finalised, i.e. the three conditions mentioned above are fulfilled, the analysis pipeline is frozen and either applied to the original data, or the results are unblinded.

% Where has blinding been used already?
The concept of blinding has already entered several fields of natural sciences. For the first time it was used in experimental particle physics in the 90's for the detection of the rare $K^0 \rightarrow \mu e$ decay \cite{PhysRevLett.70.1049}. They made use of the \textit{hidden signal box} technique, that excludes the parameter space from the analysis, where the signal is expected. The ``hidden box'' is opened after the free parameters related to the experiment are fixed. There are various blinding techniques used for different problems in particle physics, which are reviewed in \cite{klein_blind_2005}. An alternative  is adding fake signals. This method called ``salting" was used by the LIGO-Virgo collaboration to test whether the claimed sensitivity to detect a gravitational wave event was accurate \cite{ligo_scientific_collaboration_search_2010}.

%Blinding in cosmology
In cosmology we usually do not measure signals or events that can be separated from a background. Instead, we observe light from the sky, compute statistical quantities from its distribution and use these to  infer the underlying cosmological model.  Thus  different blinding techniques need to be developed for cosmology. In general, there are three different levels, at which cosmological data can be blinded:
\begin{itemize}
\item Level 1: at the catalog level.
\item Level 2: at the level of summary statistics.
\item Level 3: at the level of cosmological parameter inference.
\end{itemize} 
Level 3 blinding operates at the latest stage of the analysis. When fitting cosmological parameters to the model, an unknown offset is added to the parameters, so that the underlying values remain unknown until unblinding. This has the advantage that the step of unblinding can be done very fast. On the other hand, this means that it is also very easy to accidentally unblind. This approach  has been used in the analyses of distant supernovae \cite{conley_measurement_2006, zhang_blinded_2017}, as well as in the DES Y1 analysis \cite{des_collaboration_dark_2017}. Level 2 blinding operates at an earlier stage; it shifts the summary statistics by the difference vector between two different cosmologies. This blinding strategy is adopted for DES Y3 \cite{muir_blinding_2019}, since it is feasible for a multi-probe survey (e.g., weak lensing + galaxy clustering). Another approach for blinding astronomical data has been presented recently  \cite{sellentin_blinding_2019}, that operates at the level of the covariance matrix and is well suited for weak lensing analyses. Catalog-level or level 1 blinding operates at the earliest stage of the analysis which may be preferable in some cases. It has been used in the KiDS-450 weak lensing survey \cite{kuijken_gravitational_2015}, where the ellipticities of galaxies were systematically modified, as well as in the KiDS-450+Viking analysis \cite{hildebrandt_kids+viking-450_2019}, where the redshift distribution was shifted by a constant factor.

% Blinding guidelines
Independently of the level at which it  operates, it is desirable that a  blinding scheme meets the following criteria:
\begin{itemize}
\item it should modify the data in such a way that the effect on observables and best-fit parameters can be predicted easily.
\item it should make it difficult to unblind accidentally.
\item it should not interfere too much with delicate and key aspects of the analysis such as systematic weights.
\item it should not change significantly the shape of the posterior distribution.
\item  it should not be too numerically intensive.
\item ideally it should be easy to infer (to a sufficiently good approximation) unblind parameter constraints without the need to rerun the analysis pipeline again. This is  a``nice-to-have" but not a necessary condition. \footnote{The pipeline should be frozen before unblinding and can be re-run on the un-blinded data, but knowing approximately the expected results makes the process more transparent and easier to interpret. Another advantage is the possibility to finalize introduction sections and sketching results and conclusions sections  of the associated scientific papers even before the final runs on unblinded papers have finished.}
\end{itemize} 

% What is our work about?
Inspired by these criteria, in this work we introduce a novel method of catalog-level blinding targeted to spectroscopic galaxy surveys. The method might be interesting for clustering measurements from photometric surveys (as discussed in \cite{Blake:2004tr}, \cite{2012MNRAS.427.1891A}) as well, especially if the photometric errors 
are small enough. Most of the cosmological signal that a galaxy redshift survey carries is enclosed in two main physical effects: background metric evolution and growth of structures. The background evolution is usually tracked via the purely geometrical Alcock-Paczynski (AP) effect \cite{1979Natur.281..358A}, which is most visible on the baryon acoustic oscillation (BAO) feature. The most promising window into the growth of structures signal comes from the redshift-space distortions (RSD) effect (pioneered by \cite{kaiser_clustering_1987}  and also \cite{1972MNRAS.156P...1J}, \cite{1977ApJ...212L...3S}). 
The proposed blinding scheme affects both these signals.

\begin{figure}[t]
    \centering
    \includegraphics[width=\textwidth]{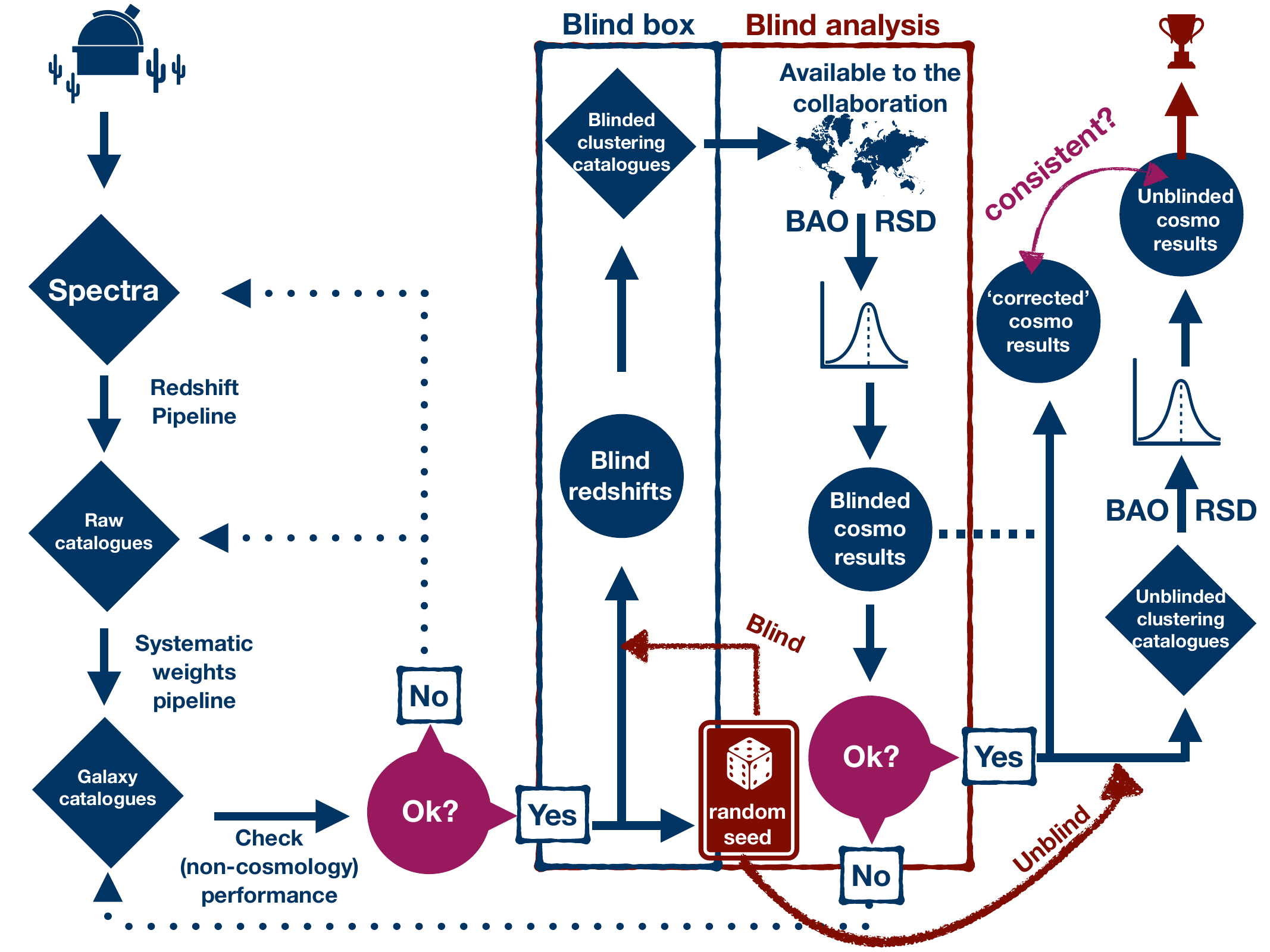}
 \caption{Flowchart of how the blinding fits in the generic workflow of a galaxy redshift survey analysis. See text for details. In this paper we introduce a scheme for the ``blind box", carry out multiple ``blind analyses" on mock and real catalogs and check the consistency between the results on the original catalog and the analytical prediction (fast unblinding). }
    \label{fig:flowchart}
\end{figure}

A flowchart indicating how we envision the blinding scheme to be embedded into a galaxy clustering pipeline is shown in figure~\ref{fig:flowchart}.
All the systematic tests are done on the catalog before blinding, but no comparison with theory or cosmological inference can be done at this stage. Once the catalog is ready for cosmological analysis it goes through blinding (the ``blind box" ) according to the procedure presented in this paper. The blinded catalog is released to the collaboration for cosmological analyses (the ``blind analysis" box). This procedure may be reiterated whenever the methodology for systematic tests or cosmological analysis is changed. Once the analysis is complete, all pre-determined sanity checks are satisfied and the blinded results are obtained, the pipeline should be frozen and unblinding may be done. We propose to unblind in two parallel steps:
\begin{itemize}
\item[1)]The  posterior constraints can be shifted according to analytical expectations, as derived in sections \ref{ch:blinding_AP} and \ref{ch:blinding_RSD}. We call this process ``fast unblinding".
\item[2)] The pipeline can now be re-run on the original catalog for obtaining precise numbers and results.
\end{itemize}

%Structure of the paper
The paper is organised as follows: In section \ref{ch:basics} we summarise the basics of BAO and RSD analyses and introduce the nomenclature and conventions that are used in the following sections. We present our blinding scheme in section \ref{ch:blinding} and discuss how blinding at the catalog level is expected to change the cosmological parameter results (``blind box" in figure \ref{fig:flowchart}).   In section \ref{ch:mocks} we compare this expectation to the results obtained from standard BAO+RSD analyses of unblinded and blinded mocks (``blind analysis" box in figure \ref{fig:flowchart}). The robustness of our results is tested in further detail in appendices \ref{sec:effect-smoothing}, \ref{ch:deviations_reference} where we check for the impact of several changes to our fiducial setup. Results from the application to the BOSS data and a summary of our proposed blinding procedure are presented in section \ref{sec:results}. In particular we are interested in whether the consistency between the two parallel unblinding steps explained before holds. We conclude with section \ref{ch:conclusions}. 
\section{Basics of BAO and RSD analyses}\label{ch:basics}
We start by reviewing the basic background of BAO and RSD analyses from 
the catalog to  cosmological parameter inference.
 The methods and variables presented in the following paragraphs will be used repeatedly throughout the rest of this work, so this section serves also to introduce and define the key quantities used.

\paragraph{The galaxy catalog}\label{ch:basics_cat}
We start with the galaxy catalog, that contains at least three quantities representing the 3D observational position $\mathbf{r}_i = (\mathrm{RA}_i, \mathrm{dec}_i, z_i)$ of each galaxy, parameterised by its right ascension, declination and redshift. These three numbers do not suffice to provide a distance measure though. A reference cosmology is needed to have a unique mapping from redshifts to distances (see next paragraph). Additionally, the catalog may also include relevant quantities such as the average galaxy density $\bar{n}(z)$ and correction weights; for example it is customary to have weights for angular systematics $w_\mathrm{sys}$, fiber collisions $w_\mathrm{fc}$ and redshift failures $w_\mathrm{rf}$ as used for the Baryon Oscillation Spectroscopic Survey\footnote{\url{https://www.sdss.org/surveys/boss/}} (BOSS) and also in the extended Baryon Oscillation Spectroscopic Survey\footnote{\url{https://www.sdss.org/surveys/eboss/}} (eBOSS) in \cite{2012MNRAS.424..564R}, \cite{2014MNRAS.441...24A} and \cite{Dawson:2015wdb}.

\paragraph{Reference cosmology}\label{ch:basics_fid}
As in most of BOSS and  eBOSS analyses, we follow the fixed template-fitting approach, where the cosmology is not varied during posterior exploration, but a reference cosmology $\mathbf{\Omega}\reference$ is chosen for two different purposes:
\begin{enumerate}
\item to convert the catalog redshifts to radial distances $z_i \rightarrow r_i$ 
\item to generate a power spectrum template. 
\end{enumerate}
\noindent In principle, the reference cosmologies for points 1 and 2  do not need to be the same. Here, (both for simplicity and because it is widespread) we use the same reference for both points. We assume a spatially flat model with parameters set
\begin{equation} \label{eq:reference_cosmo}
\begin{aligned} 
\mathbf{\Omega}\reference &= \lbrace \Omega_\mathrm{m}\reference , \Omega_\mathrm{b}\reference , H_0\reference , \sigma_8\reference , n_s\reference , M_\nu\reference , w\reference  , \gamma\reference  \rbrace ~,
\end{aligned}
\end{equation}
where the total matter and baryon densities $ \Omega_\mathrm{m}$ and $ \Omega_\mathrm{b}$ at present,  the Hubble rate evaluated today $H_0$,  the (linear) density fluctuation amplitude $\sigma_8$ (smoothed on spheres of radius $8 \mpcoh$, with $h=H_0/100\, {\rm km/s/Mpc}$)  and the primordial scalar tilt $n_s$ represent the standard $\Lambda$CDM parameter-basis.  In addition, we consider the neutrino mass sum $M_\nu$, the dark energy equation of state parameter $w$ and the growth rate exponent $\gamma$.

For the conversion mentioned in point 1 we need to compute
the angular comoving distance $D_M$ as a function of galaxy redshift $z_i$ and reference cosmology $\mathbf{\Omega}\reference$
\begin{equation}\label{eq:basics_comoving_distance}
\begin{aligned}
r_i = D_M(z_i, \mathbf{\Omega}\reference) =  \int_{0}^{z_i} \! dz \, \frac{c}{H(z, \mathbf{\Omega}\reference)}~,
\end{aligned}
\end{equation}
where $c$ is the speed of light and $H(z, \mathbf{\Omega}\reference)$ is the Hubble expansion rate at a redshift $z$, that in the assumed model is given by
 \begin{equation}\label{eq:basics_hubble}
\begin{aligned}
H(z, \mathbf{\Omega}\reference) =  H_0\reference \sqrt{\Omega_m\reference(1+z)^3  + (1-\Omega_m\reference)(1+z)^{3(1+w\reference)} }~.
\end{aligned}
\end{equation}
The case $w\reference=-1$ corresponds to the standard $\Lambda$CDM model. We call the distance obtained by eq. \eqref{eq:basics_comoving_distance}  the reference distance, which may differ from the underlying one. This fact must be included in the modeling (see \S \nameref{ch:basics_ap}).

The computation of the template of the power spectrum requires to run a Boltzmann code like \texttt{CAMB} \cite{Lewis:1999bs} or \texttt{CLASS} \cite{2011JCAP...07..034B} with the reference cosmological parameters as input. The output linear power spectrum is used in the modelling as explained in \S~\nameref{ch:basics_aniso} et seq.

\paragraph{Power spectrum estimator}\label{ch:basics_ps}
The next step is to compute the summary statistics of choice,  which can be 2-point statistics (power spectrum and 2-point correlation function), 3-point statistics (bispectrum, 3-point correlation function), and even higher-order statistics if needed. As throughout this work we will operate in Fourier space, we introduce the FKP power spectrum estimator \cite{1994ApJ...426...23F}, which relies on computing the field
\begin{align}\label{eq:FKP_pse}
F(\textbf{r}) = \frac{w_\mathrm{FKP}(\mathbf{r})}{I_2^{1/2}} \left[w_\mathrm{c}(\mathbf{r}) n(\mathbf{r}) - \alpha_\mathrm{ran} n_\mathrm{ran}(\mathbf{r})  \right]~,
\end{align}
where $n(\textbf{r})$ and $n_\mathrm{ran}(\mathrm{\mathbf{r}})$ are the number densities of the galaxies and Poisson-sampled randoms, respectively,  $\alpha_\mathrm{ran}$ is the ratio between the sum of weighted data galaxies and the sum of random galaxies, and $w_\mathrm{c}$ denotes the (combined) weight. The standard FKP weight $w_\mathrm{FKP}(r)$~\cite{1994ApJ...426...23F}
is used to minimise the variance of the power spectrum at the BAO scale. The combined weight, $w_\mathrm{c}$, depends on survey characteristic corrections such as redshift failures, fiber collisions and angular systematics.
The normalisation constant $I_2$ is given as 
\begin{align}
I_2 = \int \! d^3\mathbf{r} \, w_\mathrm{FKP}(\mathbf{r}) \left\langle n w_\mathrm{c}(\mathbf{r}) \right\rangle ^2~.
\end{align}
The function defined in eq. \eqref{eq:FKP_pse} is used to build the power spectrum multipoles; for a varying line of sight (LOS), chosen towards one of the galaxies of each pair we follow the Yamamoto approximation to define the power spectrum estimator
\begin{align}
P_\mathrm{meas}^{(\ell)}(k) = \frac{(2 \ell+1)}{2} \int \! \frac{d \Omega}{4 \pi} \, \left[ \int \! d \mathbf{r}_1 \, F(\mathbf{r}_1) e^{i \mathbf{k} \cdot \mathbf{r}_1}
\int \! d \mathbf{r}_2 \, F(\mathbf{r}_2) e^{- i \mathbf{k} \cdot \mathbf{r}_2} \mathcal{L}_\ell (\hat{\mathbf{k}} \cdot \hat{\mathbf{r}}_2) \right]  - P_\mathrm{sn}^{(\ell)} ~,
\end{align}
where $\mathcal{L}_\ell (x) $ is the Legendre Polynomial of order $\ell$ and $P_\mathrm{sn}^{(\ell)}$ is the Poisson shot noise term. Higher order multipoles $\ell >0$ quantify the anisotropic clustering with respect to the LOS. There are two main sources for such an anisotropy, the AP effect and RSD.  

\paragraph{Alcock-Paczynski effect and scaling parameters} \label{ch:basics_ap}
The deviation of the reference cosmology ($\mathbf{\Omega}\reference$)  from the underlying cosmology ($\mathbf{\Omega}\underlying$) leads to an anisotropic clustering signal known as Alcock-Paczynski effect \cite{1979Natur.281..358A}. This and the variation of the acoustic scale with respect to the fiducial template are usually accounted through scaling the wave-vector across $(\perp)$ and along $(\parallel)$ the LOS:
\begin{align} \label{eq:scaling}
k_\perp \longrightarrow \widetilde{k}_\perp = k_\perp / \alpha_\perp~, \quad k_\parallel \longrightarrow \widetilde{k}_\parallel = k_\parallel / \alpha_\parallel~,  
\end{align}
where the scaling parameters are defined as
\begin{equation}\label{eq:alphas}
\begin{aligned}
\alpha_\perp(z_\eff) &= \frac{D_M\underlying(z_\eff)\, r_{\rm d}\reference}{D_M\reference(z_\eff)\, r_{\rm d}\underlying}~, \quad
&
\quad \alpha_\parallel(z_\eff) &= \frac{H\reference(z_\eff)\, r_{\rm d}\reference}{H\underlying(z_\eff)\, r_{\rm d}\underlying}~,
\end{aligned}
\end{equation}
where $r_{\rm d}$ is the sound horizon at radiation drag.  Here, we introduced $D_M\underlying$ and $D_M\reference$ as shorthand notation for $D_M(\mathbf{\Omega}\underlying)$ and $D_M(\mathbf{\Omega}\reference)$ (and same for the Hubble distance $c/H$ and $r_{\rm d}$). The argument $z_\eff$ is the effective redshift of the galaxy sample in a given redshift bin. The ratios between the underlying and reference distances perpendicular $(D_M)$ and parallel $(1/H)$ with respect to the LOS scale the reference mapping from redshifts to distances towards the``correct" one. The ratio of $r_{\rm d}\underlying$ to $r_{\rm d}\reference$ accounts for the different BAO peak position in the reference power spectrum template with respect to the one in the measured power spectrum.

\paragraph{Redshift-space distortions and reconstruction}\label{ch:basics_rsd} 
Driven  by dark matter over-densities, peculiar velocities of galaxies  induce redshift-space distortions in galaxy clustering.  The measured redshift of a galaxy comes from a superposition of its  Hubble flow recession velocity $\mathbf{v_\mathrm{r}}$ and its peculiar velocity $\mathbf{v_\mathrm{p}}$. Hence, the measured redshift-space position $\mathbf{r}$ (found by transforming the observed galaxy redshift to a distance using the reference cosmology) is displaced from the real-space position $\mathbf{x}$ by its peculiar velocity projected on the LOS $(\mathbf{v_\mathrm{p}} \cdot \mathbf{\hat{x}})$, where $\mathbf{\hat{x}}$ is the unit vector in LOS direction. In general, the mapping from $\mathbf{x}$ to $\mathbf{r}$ is given by 

 \begin{equation}\label{eq:mapping_redshift_space}
\begin{aligned}
\mathbf{r}(\mathbf{x}) = \mathbf{x} + \frac{(\mathbf{v_\mathrm{p}}( \mathbf{x}) \cdot \mathbf{\hat{x}})\mathbf{\hat{x}}}{aH(a)}
\end{aligned}
\end{equation}
with scale factor $a$. In the framework of Lagrangian perturbation theory the real-space (Eulerian) position $\mathbf{x}$ of a particle is written as the sum of its initial (Lagrangian) position $\mathbf{q}$ and a displacement vector $\mathbf{\Psi}$
 \begin{equation}\label{eq:mapping_lagrangian}
\begin{aligned}
\mathbf{x}(\mathbf{q},t) = \mathbf{q} + \mathbf{\Psi}(\mathbf{q},t) ~.
\end{aligned}
\end{equation}
At first order, the peculiar velocity field is related to the Lagrangian displacement field by
\begin{equation}\label{eq:mapping_velocity}
\begin{aligned}
\mathbf{v_\mathrm{p}}( \mathbf{x}(\mathbf{q},t)) = a H f   \mathbf{\Psi}(\mathbf{q},t)~,
\end{aligned}
\end{equation}
where
\begin{equation}\label{eq:growth_rate}
\begin{aligned}
f \equiv \frac{d\ln D(a)}{d\ln a} = (\Omega_\mathrm{m}(a))^\gamma
\end{aligned}
\end{equation}
is the logarithmic growth rate, which depends on the scale-independent  linear growth factor $D(a)$, and can be expressed as function of the matter density corresponding to a scale factor $a$ and the growth rate exponent $\gamma$, which for general relativity and $w=-1$ is $\gamma=0.55$ to a very good approximation \cite{PhysRevD.72.043529}. Combining eqs. \eqref{eq:mapping_redshift_space}, \eqref{eq:mapping_lagrangian} and \eqref{eq:mapping_velocity}, and using the fact that $\hat{\mathbf{x}}=\hat{\mathbf{r}}$ yields
 \begin{equation}\label{eq:mapping_recon}
\begin{aligned}
\mathbf{q}(\mathbf{r})= \mathbf{r} - \mathbf{\Psi} - f (\mathbf{\Psi} \cdot \mathbf{\hat{r}})\mathbf{\hat{r}} ~.
\end{aligned}
\end{equation}
Using this equation together with the fact that the number of galaxies is conserved when transforming between real and redshift space, the displacement field is related to the smoothed redshift-space galaxy over-density field $ \delta_g^\mathrm{red} $ as \cite{nusser_prediction_1994}
 \begin{equation}\label{eq:mapping_combined}
\begin{aligned}
\nabla \cdot \mathbf{\Psi} + \frac{f}{b} \, \nabla (\mathbf{\Psi}\cdot \mathbf{\hat{r}} ) \,\mathbf{\hat{r}}  = - \frac{\delta_g^\mathrm{red}}{b} ~,
\end{aligned}
\end{equation}
where $b$ is the scale independent linear galaxy bias relating matter and galaxy over-densities, $\delta_m = b  \delta_g$. 
This is at the basis of the so-called density-field reconstruction procedure, whereby eq.~\eqref{eq:mapping_combined} is used to estimate the displacement field $ \mathbf{\Psi}$ given $\delta_g^\mathrm{red}$ and reference values of $b$ and $f$.  Galaxies, interpreted as test particles for the velocity field, can then be moved back to their initial Lagrangian positions ${\bf q}$ given in eq.~\eqref{eq:mapping_recon}  removing non-linear effects, which sharpens the BAO feature and increases its detection significance (see \cite{eisenstein_robustness_2007}, \cite{padmanabhan_reconstructing_2009} and \cite{burden_efficient_2014} for reference).

\paragraph{BAO anisotropic power spectrum model} \label{ch:basics_aniso}
This model is used to describe the BAO scale in the power spectrum multipoles via template fitting: a linear power spectrum $P_\mathrm{lin}^\mathrm{ref}(k)$ computed at the reference cosmology is used to build a template that is fitted to the oscillatory pattern in the data by varying the scaling factors $\lbrace \alpha_\parallel, \alpha_\perp \rbrace$. The reference template is divided into a smooth part $P_\mathrm{lin,sm}$ and an oscillating part $\mathcal{O}_\mathrm{lin}=P_\mathrm{lin}/P_\mathrm{lin,sm}$ in order to separate the BAO information from the broadband, which is marginalised over. 

Following \cite{Beutler:2016ixs} we model the anisotropic galaxy power spectrum including the non-linear damping of the BAO with two damping scales $\Sigma_\perp, \Sigma_\parallel$ as
\begin{align}\label{eq:Panisofull}
P_\mathrm{aniso}(k,\mu) = P_\mathrm{aniso,sm}(k,\mu) \left[ 1 + \left( \mathcal{O}_\mathrm{lin}(k) - 1 \right) e^{-\frac{1}{2} \left( \mu^2 k^2 \Sigma_\parallel^2 +   (1-\mu^2) k^2 \Sigma_\perp^2 \right)} \right] , 
\end{align}
with the smooth component given by
\begin{align}\label{eq:Panisosmooth}
 P_\mathrm{aniso,sm}(k, \mu) = B^2 (1+\beta \mu^2 R)^2  P_\mathrm{lin,sm}(k)~,
\end{align}
where $B$ is a free amplitude parameter and $\beta$ is a nuisance parameter effectively accounting for linear RSD and galaxy bias. We use the factor $R=1$ before density-field reconstruction and $R=1-\exp{\left( -k^2\Sigma_\mathrm{sm}^2/2 \right)}$ after reconstruction with smoothing scale $\Sigma_\mathrm{sm}$ as in \cite{Beutler:2016ixs} and \cite{Seo:2015eyw}. The coordinates used here are defined as
\begin{align} \label{eq:kmu}
k=|\mathbf{k}|~, \quad \mu = \frac{\mathbf{k}\cdot \mathbf{r}}{kr}
\end{align}
where $\mathbf{k}$ is the Fourier mode and $\mu$ is the cosine of the angle between the Fourier mode and the LOS distance vector $\mathbf{r}$. The components of $\mathbf{k}$ across  and along  the LOS can  be expressed as
\begin{align}
k_\perp = k \sqrt{1-\mu^2} ~, \quad k_\parallel = \mu k\, .
\end{align} 
To include the AP effect, we rescale the wave-vector components as in eq. \eqref{eq:scaling} to scale the observed modes from the reference cosmology to the underlying cosmology. Combining eqs. \ref{eq:scaling} and \ref{eq:kmu} we find the transformation
\begin{align} \label{eq:Panisotrafo}
k \longrightarrow \widetilde{k} = \frac{k}{\alpha_\perp} \left[ 1 + \mu^2 \left( \frac{\alpha_\perp^2}{\alpha_\parallel^2} -1 \right)\right]^{1/2}, \quad \mu \longrightarrow \widetilde{\mu} = \mu \frac{\alpha_\perp}{\alpha_\parallel} \left[ 1 + \mu^2 \left( \frac{\alpha_\perp^2}{\alpha_\parallel^2} -1 \right)\right]^{-1/2}. 
\end{align}
We write the anisotropic power spectrum as a function of the transformed parameters $(\widetilde{k},\widetilde{\mu})$ and 
obtain its multipoles using angle-dependent Legendre polynomials. Also, we add a polynomial expansion in terms of $k$ accounting for nonlinear broadband effects and potential differences between the underlying broadband and the template, that have not been included in the modeling so far:
\begin{align} \label{eq:Panisomultipoles}
P_\mathrm{aniso}^{(\ell)} (k) = \left( \frac{r_{\rm d}^\mathrm{ref}}{r_{\rm d}} \right)^3 \frac{(2\ell+1)}{2\alpha_\perp^2 \alpha_\parallel} \int_{-1}^1 \! P_\mathrm{aniso} (\widetilde{k}(k,\mu), \widetilde{\mu}(\mu)) \mathcal{L}_\ell(\mu) \, d\mu + \sum_{i=1}^{N_i} A_i^{(\ell)} k^{2-i}~.
\end{align}
The extra normalisation factor in front of the integral accounts for the isotropic dilation induced by the change in volume between reference and modelled cosmology. The ratio between the sound horizons is reabsorbed into the overall amplitude parameter $B$. This means that the model consists of the physical parameters $\lbrace \alpha_\parallel, \alpha_\perp \rbrace$ and the nuisance parameters $\lbrace \beta, B_j, A_{i,j}^{(\ell)} \rbrace$ fitted to each redshift bin. The total number of nuisance parameters depends on the order of the polynomial expansion and the number of disconnected patches indexed by $j$ of the sky observed. The dispersion scales $\Sigma_\parallel$ and $\Sigma_\perp$ can either be varied and marginalised over, inferred from theoretical approaches or fitted to mock survey catalogs and then fixed in the data analysis. In this paper we follow the latter approach.

\paragraph{Full shape power spectrum model} \label{ch:basics_fs}
While the BAO anisotropic model is tuned towards detecting the BAO pattern only, the full shape model also extracts information from the broadband  and is not only sensitive to BAO, but  to RSD and non-linear structure formation, too. Again, we use a template-fitting procedure,  but instead of the linear matter power spectrum we use the 2-loop renormalisation perturbation theory extension described in \cite{Gil_Mar_n_2012} to compute the auto- and cross- density and velocity potential power spectra $P_{\delta \delta}$, $P_{\delta \theta}$ and $P_{\theta \theta}$ of the matter field. To go from the matter to the galaxy field we need to assume a galaxy bias model. Here, we use the model proposed by \cite{McDonald_2009}, consisting of four bias parameters $\lbrace b_1,b_2,b_{s^2},b_{3_\mathrm{nl}} \rbrace$. The linear galaxy bias $b_1$ and the non-linear galaxy bias $b_2$ are treated as free nuisance parameters, while the other two are functions of $b_1$, given in \cite{Baldauf_2012} and \cite{Saito:2014qha}. We adopt the exact expressions from appendix B of \cite{Gil-Marin:2014sta} for the galaxy density-density, density-velocity potential and velocity potential-velocity potential power spectra given as
\begin{equation}
\begin{aligned}
P_{g,\delta\delta}(k) ~=~ &b_1^2 P_{m,\delta\delta}(k) +  2b_2 b_1 P_{m,b2\delta}(k) +  2b_{s2} b_1 P_{bs2,\delta}(k) + b_2^2 P_{m,b22}(k) ~ + \\ & 2 b_2 b_{s2} P_{m,b2s2}(k) + b_{s2}^2P_{bs22}(k) + 2 b_1 b_{3nl} \sigma_3^2(k) P_\mathrm{lin}(k) \\
P_{g,\delta\theta}(k) ~=~ &b_1 P_{m,\delta\theta}(k) + b_2 P_{m,b2\theta} (k) + b_{s2} P_{m,bs2\theta}(k) + b_{3nl} \sigma_3^2(k) P_\mathrm{m,lin}(k) \\
P_{g,\theta\theta}(k) ~=~ &P_{\theta\theta}~.
\end{aligned}
\end{equation}
To include the effect of RSD explained in \S \nameref{ch:basics_rsd}
we follow the approach of \cite{Taruya:2010mx} writing the full shape power spectrum as
\begin{equation}
\begin{aligned}
P_\mathrm{FS} (k,\mu) = e^{-\frac{1}{2} \left[ k\mu\sigma_P \right]^2} &\left[ \,P_{g,\delta\delta}(k) + 2 f \mu^2 P_{g,\delta\theta}(k) + f^2 \mu^4 P_{g,\theta\theta}(k) ~+ \right. \\
&\left. ~~b_1^3 A^\mathrm{TNS}(k,\mu,f/b_1)  + b_1^4 B^\mathrm{TNS}(k,\mu,f/b_1) \,\right] ~,
\end{aligned}
\end{equation}
where the functions $ A^\mathrm{TNS}$ and $ B^\mathrm{TNS}$ are defined in \cite{Taruya:2010mx}. The exponential damping term for scales smaller than $\sigma_P$ accounts for the so-called Fingers-of-God (FoG) effect \cite{1972MNRAS.156P...1J}. This effect appears at small scales, where highly non-linear velocities smear out the density field in redshift space, such that the structures appear elongated along the line of sight and the power spectrum is damped. 
Finally, the full shape power spectrum multipoles are given as in eq. \eqref{eq:Panisomultipoles} but without the polynomial term
\begin{align} \label{eq:Pfsmultipoles}
P_\mathrm{FS}^{(\ell)} (k) =  \frac{(2\ell+1)}{2\alpha_\perp^2 \alpha_\parallel} \int_{-1}^1 \! P_\mathrm{FS} (\widetilde{k}(k,\mu), \widetilde{\mu}(\mu)) \mathcal{L}_\ell(\mu) \, d\mu~.
\end{align}
The AP effect is included via the same transformation as in eq. \eqref{eq:Panisotrafo}. This means that there are in total 3 cosmological parameters $\lbrace \alpha_\parallel, \alpha_\perp, f \sigma_8 \rbrace$ per redshift bin and 4 nuisance parameters $\lbrace b_1, b_2, \sigma_P, A_\mathrm{noise}\rbrace$ per redshift bin and per observed patch of the sky. The last nuisance parameter $A_\mathrm{noise}$ is used to model deviations from Poissonian shot noise in the monopole. Note that we measure the growth rate $f$ for a fixed value of $\sigma_8$ of the template. This is why we are only sensitive to the product $f\sigma_8$, which is a template-independent quantity in the wave-vector range of interest (e.g. up to $k=0.2\,\hompc$, see \cite{Percival:2008sh} for reference).

\paragraph{The survey geometry}
In order to compare the model power spectra to the data, we need to take into account the selection function of the survey. This can be achieved by convolving the model power spectrum $P_\mathrm{model} (\mathbf{k})$ with the survey window function $W(\mathbf{k})$ to obtain the ``windowed" power spectrum $P_\mathrm{win, model} (\mathbf{k})$ 
\begin{align}\label{eq:window_convolution}
P_\mathrm{win, model} (\mathbf{k}) = \int \! \frac{\mathrm{d} \mathbf{k}^\prime}{(2\pi)^3} \, P_\mathrm{model} (\mathbf{k}^\prime) \, |W(\mathbf{k}-\mathbf{k}^\prime)|^2~.
\end{align}
To avoid the computation of the expensive 3-dimensional Fourier transform, we follow the framework of \cite{Wilson:2015lup} extended by \cite{Beutler:2016arn} where the window function is obtained as a pair count in configuration space. This is more efficient, because the convolution in Fourier space becomes a multiplication in configuration space. The ``windowed" power spectrum multipoles are then the 1-dimensional Hankel transform of the ``windowed" correlation function multipoles.

The effects of this transformation of the model power spectrum are purely observational, hence it is not particularly interesting for blinding. But, we have to accept, we can only observe the galaxy density fluctuation through a window, which is why this is an important ingredient in BAO and RSD analyses. 

\section{Blinding Scheme} \label{ch:blinding}

\subsection{General Considerations}  \label{ch:blinding_GC}
In this section we introduce the general idea of our blinding scheme and we derive analytic formulae describing its effects. Without introducing blinding, the standard procedure for BAO and RSD analyses is as follows. The galaxy catalog, in terms of angular positions, redshifts and weights, forms the raw data $\mathbf{d}$, the redshifts of which are transformed into distances using a reference cosmology. Subsequently, the power spectrum of this galaxy distribution is measured. Afterwards, a template for the power spectrum is computed at the same reference cosmology to build the models described in section \ref{ch:basics}. Finally, the model parameters that distort the template and are constrained with the data are
\begin{align} \label{eq:observables_definition}
\mathbf{\Theta}(\mathbf{d}, \mathbf{\Omega}\reference) = \lbrace \alpha_\parallel (z_{\eff 1}), ..., \alpha_\parallel (z_{\eff n}),~ \alpha_\perp (z_{\eff 1}), ..., \alpha_\perp (z_{\eff n}),~ f \sigma_8 (z_{\eff 1}), ..., f \sigma_8 (z_{\eff n_\mathrm{bins}}) \rbrace~,
\end{align}
where $n_\mathrm{bins}$ is the number of tomographic redshift bins analysed and $z_{\eff,i}$ denotes the effective redshift of each of them. These are {\it physical  parameters} (two scaling parameters and one growth rate) which capture where the bulk of cosmological information is contained in the data  and can be interpreted in a transparent and  largely model-independent way.  These parameters can also be easily interpreted and modelled within the adopted theory and/or  specific cosmological model.

Next, the constraints on the physical parameters $\mathbf{\Theta}$ are used to infer the underlying cosmological model $\mathbf{\Omega}\underlying$, which is robust to the choice of reference cosmology (especially for BAO-only analyses, see e.g.,~\cite{Carter_BAOtest, Bernal_BAObias}), it only depends on the data:
\begin{align} \label{eq:underlying_cosmo_dependence}
\mathbf{\Omega}\underlying = \mathbf{\Omega}\underlying (\mathbf{\Theta}(\mathbf{d}, \mathbf{\Omega}\reference)) = \mathbf{\Omega}\underlying (\mathbf{d})~.
\end{align}
 In the standard approach to BAO and RSD the derivation of $\mathbf{\Theta}$ can be seen as a necessary intermediate step to compress and extract the signal.  This  makes possible to connect, via a model-independent set of physical parameters, the galaxy catalog to one (or  several) cosmological models in a transparent way.

Since a change in the physical parameters has a very  direct and transparent effect on the commonly used  summary statistics (e.g., power spectra), it is much easier to understand the effect of the proposed blinding procedure, by working with these parameters.

For the purpose of blinding we aim to find a transformation of the raw data
\begin{align} \label{eq:data_shift}
\mathbf{d} \rightarrow \mathbf{d}^\prime = \mathbf{d} + \Delta \mathbf{d}\,(\mathbf{d}, \mathbf{\Omega}\underlying, \mathbf{\Omega}\blind)
\end{align}
fulfilling  the condition that the modified raw data $\mathbf{d}^\prime$ represents a Universe whose cosmological model is $\mathbf{\Omega}\blind$. $\mathbf{\Omega}\blind$ is the target blind cosmology, that we aim for. In what follows we will denote all quantities related to the blinded catalog with a prime in contrast to quantities referring to the original catalog, which are written without superscript. However, it is impossible to find an exact expression for $\Delta \mathbf{d} \,(\mathbf{d}, \mathbf{\Omega}\underlying, \mathbf{\Omega}\blind)$, as by definition we are \textit{blind} towards $\mathbf{\Omega}\underlying$. Hence, in the following we use the reference cosmology $\mathbf{\Omega}\reference$ presented in section \ref{ch:basics} as a starting point and introduce a shifted cosmology $\mathbf{\Omega}\shift$ which depends on the same parameter set, but allowing for a variation $\Delta \mathbf{\Omega}$ with respect to the reference:
\begin{equation} \label{eq:shifted_cosmo}
\begin{aligned} 
\mathbf{\Omega}\shift = \mathbf{\Omega}\reference + \Delta \mathbf{\Omega} ~,
\end{aligned}
\end{equation}
with
\begin{equation} \label{eq:delta_cosmo}
\begin{aligned} 
\Delta \mathbf{\Omega}&= \lbrace \Delta \Omega_\mathrm{m} , \Delta \Omega_\mathrm{b} , \Delta H_0 , \Delta \sigma_8 ,\Delta  n_s , \Delta M_\nu , \Delta w  , \Delta \gamma \rbrace ~.
\end{aligned}
\end{equation}
Then we can write eq. \eqref{eq:data_shift} in terms of the reference and shifted cosmologies:
\begin{align} \label{eq:data_shift_2}
\mathbf{d} \rightarrow \mathbf{d}^\prime = \mathbf{d} + \Delta \mathbf{d}\,(\mathbf{d}, \mathbf{\Omega}\reference, \mathbf{\Omega}\shift)\,.
\end{align}
Note that the shifted cosmology is just a tool to generate the perturbation in the data, so that the results of the analysis correspond to $\mathbf{\Omega}\blind$ instead of $\mathbf{\Omega}\underlying$. Finally, using $\mathbf{\Omega}\reference$ and $\mathbf{\Omega}\shift$, we aim to find a relation between $\mathbf{\Theta}\blind$ and $\mathbf{\Theta}\underlying$.

Here, we propose to shift the position $\mathbf{r}_i$ of galaxies  along the LOS in a redshift- and density-dependent way . We avoid changing the angular positions because in current spectroscopic survey strategies target galaxies are selected based on the photometric surveys. Therefore, perturbing angular positions might add unnecessary complications into e.g. fiber collision corrections and other angular systematics.
Hence, we only shift the individual galaxy redshifts, so we can express eq. \eqref{eq:data_shift_2} as 
\begin{align} \label{eq:shift_general}
z_i^\prime = z_i + \Delta z_i(z_i, \rho(\mathbf{r_i}), \mathbf{\Omega}\reference, \mathbf{\Omega}\shift)~,
\end{align}
where $\rho(\mathbf{r_i})$ denotes the galaxy density in the position of the $i$-th galaxy. From the parameter basis given in eq. \eqref{eq:reference_cosmo} here we choose to blind only for parameters that govern the late-time evolution of the Universe, such that eq. \eqref{eq:delta_cosmo} becomes
\begin{equation} \label{eq:delta_cosmo_2}
\begin{aligned} 
\Delta \mathbf{\Omega}&= \lbrace \Delta \Omega_\mathrm{m},\, 0,\, 0,\, 0,\, 0,\, 0,\, \Delta w ,\, \Delta \gamma \rbrace ~.
\end{aligned}
\end{equation}

Here, we choose not to blind for $H_0$, as its value is absorbed into the unit via the factor $h$. Therefore it has no impact on the galaxy catalog, only on the power spectrum template.
%But in principle it is possible to blind for the parameter combination $H_0 r_s$ by modifying the template.} 
More general blinding, including the effects of other parameters such as $H_0$, $\sigma_8$, $M_\nu$ and $f_{\mathrm{nl}}$, will be presented in a follow up paper \cite{Briedeninprep}. 

The blinding scheme consists of two types of shifts to galaxies' redshifts:
\begin{itemize}
\item A homogeneous shift according to a modified redshift-distance relation (see section~\ref{ch:blinding_AP}). It induces an anisotropy between radial and angular distances equivalent to the AP effect and blinds the cosmological background evolution. We refer to it as AP-like shift, $\Delta^{\rm AP}z$.
\item A shift based on the gradient of the reconstructed galaxy density field along the LOS (see section~\ref{ch:blinding_RSD}). This perturbative shift is designed to imitate RSD and blind the growth of structures, hence we refer to it as RSD-shift, $\Delta^{\rm RSD}z$.
\end{itemize}

In summary $\Delta z_i$ of eq.~\ref{eq:shift_general} is given by the 
combination
of  $\Delta^{\rm AP}z_i$ and $\Delta^{\rm RSD} z_i$.
In the following subsections we describe both shifts in more detail and discuss their effect on the physical parameters:
\begin{align}
\mathbf{\Theta}\underlying \rightarrow \ \mathbf{\Theta}\blind = \mathbf{\Theta}\underlying (\mathbf{d}\blind, \mathbf{\Omega}\reference,\mathbf{\Omega}\shift)
\end{align}

\subsection{AP-like shift} \label{ch:blinding_AP}
We set off to find the redshift dependence of $\Delta z_i$ in Eq.~\eqref{eq:shift_general}, such that it mimics the background expansion of a cosmology different from both the reference and the underlying  ones. In practice, this affects the distances inferred from redshifts using the reference cosmologies, which affects to the measured clustering at all scales modifying the $\alpha_\parallel$ and $\alpha_\perp$ values.

\begin{figure}[t]
    \centering
    \includegraphics[width=\textwidth]{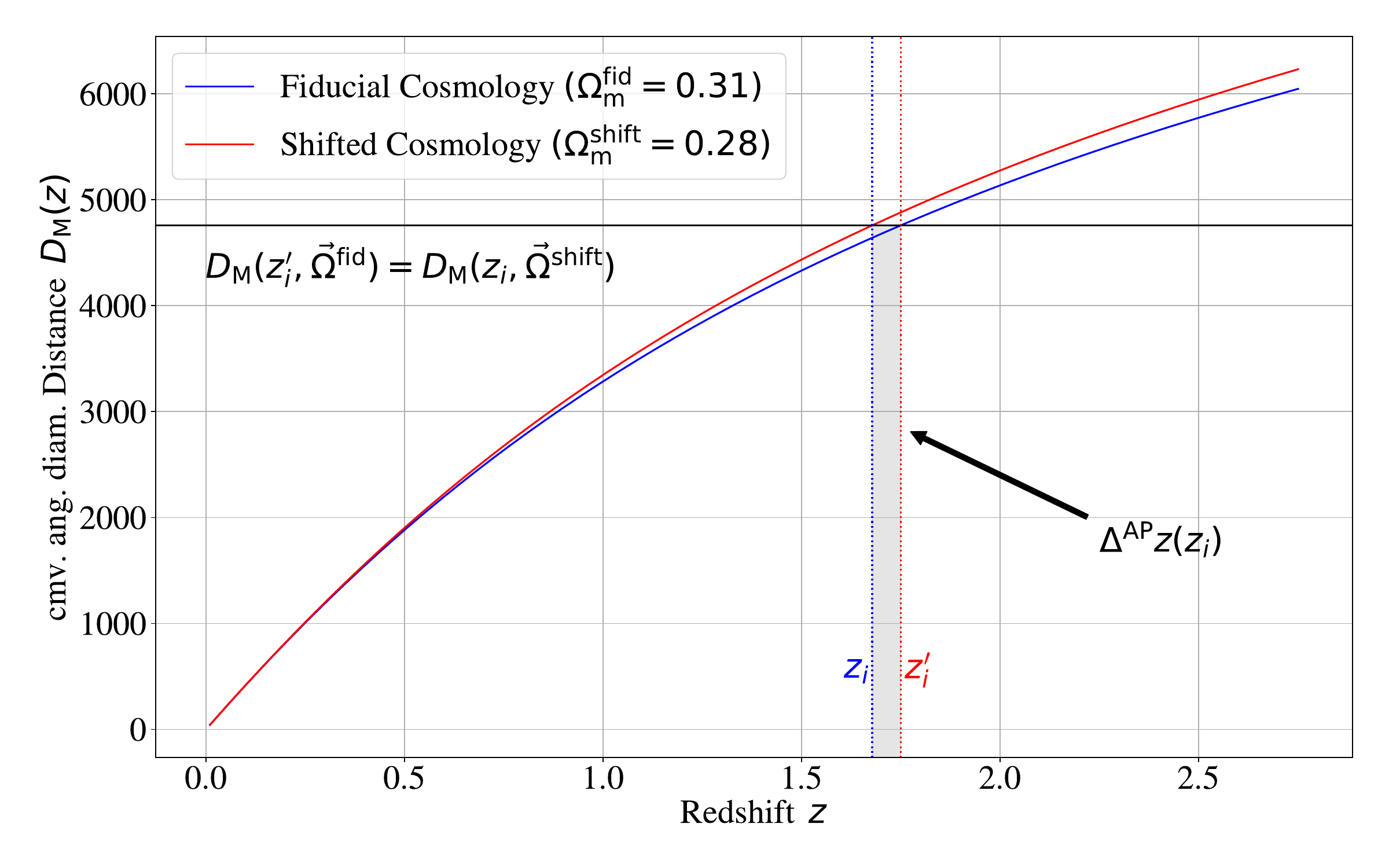}
    \caption{AP-like shift ingredient: determination of   the  shift $\Delta^{\rm AP} z_i(z_i)$ (eq.~\ref{eq:shift_ap}). In this illustration the shifted distance is obtained by a 10\% decrease of the matter density $\Omega_\mathrm{m}$. The blue line is the  comoving angular diameter distance for the reference cosmology, the red  line  is the same quantity for a shifted cosmology where only the matter density parameter has been changed. The black horizontal line  visualises where $D_\mathrm{M}(z_i, \mathbf{\Omega}\shift) = D_\mathrm{M}(z_i^\prime, \mathbf{\Omega}\reference$).  See text for more  details.} 
    \label{fig:diffdistance}
\end{figure}

The shift is obtained by converting the observed redshifts $z_i$ into comoving distances $D_\mathrm{M}(z_i, \mathbf{\Omega}\shift)$ using the shifted cosmology and inverting them back to blinded redshifts $z_i^\prime$ using the reference cosmology as shown in the following sketch:\footnote{The $z_i$ are observed redshifts so they do not have an explicit dependence on the parameters $\mathbf{\Omega}$, which are unknown and we can't change. However, in eq. \ref{eq:shift_sketch} we leave the $\mathbf{\Omega}$ dependence in the argument of $z_i$ to stress that the observed redshifts are drawn from an (unknown) model parameterised by the (unknown) values of the model parameters.}
\begin{align} \label{eq:shift_sketch}
z_i \, (\mathbf{\Omega}\underlying) \xrightarrow{~\mathbf{\Omega}\shift~} D_\mathrm{M}(z_i, \mathbf{\Omega}\shift) = D_\mathrm{M}(z_i^\prime, \mathbf{\Omega}\reference) \xrightarrow{~\, \mathbf{\Omega}\reference~\, } z_i^\prime \, (\mathbf{\Omega}\blind) ~.
\end{align}
In this way we can derive the shift numerically using
\begin{align} \label{eq:shift_ap} \Delta^{\rm AP} z(z_i,  \mathbf{\Omega}\reference, \mathbf{\Omega}\shift) = z_i^\prime (\mathbf{\Omega}\blind)  - z_i(\mathbf{\Omega}\underlying),
\end{align}
 as illustrated in figure \ref{fig:diffdistance}. This equation yields one of the two contributions to the $\Delta z$ to be applied in eq.~\ref{eq:shift_general}.
The blinded redshifts $z_i^\prime$ appear to originate from a  blinded or ``fake" cosmology  $\mathbf{\Omega}\blind$.
Any geometric measurement of the background expansion is affected by this shift. Although only the radial dimension is affected, and not angles between objects, any measurement of a transversal distance on a patch of the sky will also be affected, because the distance to the objects is changed by the shift.

After shifting all galaxies individually, the effective redshift of a given galaxy sample, $z_\eff$, changes  in a way that in general  depends on the details of its initial redshift distribution $\bar{n}(z)$. For simplicity, we only consider the following two cases:

\begin{itemize}
\item \textit{Shifted redshift cuts} (or fixed distance cuts): the redshift range of the blinded sample is changed according to the blinding scheme so that the selection of galaxies remains the same for the pre- and post-blinded catalog. As a consequence, we expect the effective redshift to be shifted according to
\begin{equation} \label{eq:effective_redshift_shifted}
\begin{aligned}
z_\eff^\prime = z_\eff +  \Delta z(z_\eff,  \mathbf{\Omega}\reference, \mathbf{\Omega}\shift) ~.
\end{aligned}
\end{equation}
\item \textit{Fixed redshift cuts} (or shifted distance cuts): the same redshift cuts are applied to the post-blinded catalog as to the pre-blinded one. In this way the effective redshifts of both catalogs remain nearly the same, because  approximately as many galaxies enter the redshift bin of the sample as they leave it after blinding. This is only possible if the redshift cuts lay in regions where the galaxy density is comparable to the mean density $\bar{n}$. In this case we expect
\begin{equation}\label{eq:effective_redshift_fixed}
\begin{aligned}
 z_\eff^\prime \simeq z_\eff~.
 \end{aligned}
\end{equation}
\end{itemize}
We anticipate that we test the validity of eqs. \eqref{eq:effective_redshift_shifted} and \eqref{eq:effective_redshift_fixed} in section \ref{ch:mocks_AP}, where we also motivate our recommendation to use  shifted redshift cuts whenever possible.

We can now derive an expression for the scaling parameters $\alpha_\parallel^\prime$ and $\alpha_\perp^\prime$ that  would  be measured from the blinded catalog. In general they can be written as:
\begin{equation}\label{eq:alphasblinded}
\begin{aligned}
\alpha_\perp^\prime(z_\eff\blind) &\equiv \frac{D_M\blind(z_\eff\blind)\, r_{\rm d}\reference}{D_M\reference(z_\eff\blind)\, r_{\rm d}\blind}~, \quad
&
\quad \alpha_\parallel^\prime(z_\eff\blind) &\equiv \frac{H\reference(z_\eff\blind)\, r_{\rm d}\reference}{H\blind(z_\eff\blind)\, r_{\rm d}\blind}~,
\end{aligned}
\end{equation}
but we need to express them as a function of $\mathbf{\Omega\reference}$, $\mathbf{\Omega\shift}$ and $\mathbf{\Omega\underlying}$.
 Since our blinding scheme does not change the early-time physics,  the sound horizon at the baryon drag epoch is not affected by the blinding, therefore 
\begin{equation}
\begin{aligned}
r_{\rm d}\blind = r_{\rm d}\underlying \,.
\end{aligned}
\end{equation}

Because the distances  in eq. \eqref{eq:alphasblinded} were computed with the shifted cosmology, $\alpha^\prime$ will depend on $D\shift$. But in the blind analysis we assume they were obtained with the reference cosmology and use a power spectrum template  computed from the reference cosmology. 
Hence, the perpendicular and parallel components $\alpha_\perp^\prime$ and $\alpha_\parallel^\prime$  are
\begin{equation}
\begin{aligned}\label{eq:alpharelation}
    \alpha_\perp^\prime(z_\eff^\prime) &=
    \frac{D_M\underlying(z_\eff) r_{\rm d}\reference}{D_M\shift(z_\eff) r_{\rm d}\underlying}~, \quad
&
    \alpha_\parallel^\prime(z_\eff^\prime) &=
    \frac{H\shift(z_\eff)\, r_{\rm d}\reference}{H\underlying(z_\eff)\, r_{\rm d}\underlying}\,.
\end{aligned}
\end{equation}
Finally,  the ratios of the scaling parameters we would measure from the blinded catalog with respect to the unblinded catalog can be obtained by combining eqs. \eqref{eq:alphas} and \eqref{eq:alpharelation} 
\begin{equation}\label{eq:alphasresult}
\begin{aligned}
\frac{ \alpha_\perp^\prime(z_\eff^\prime) }{\alpha_\perp(z_\eff) } &= \frac{D_M\reference(z_\eff)}{D_M\shift(z_\eff)}, \quad 
&
\frac{ \alpha_\parallel^\prime(z_\eff^\prime)  }{\alpha_\parallel(z_\eff) } &= \frac{H\shift(z_\eff) }{H\reference(z_\eff) }~.
\end{aligned}
\end{equation}
This final result eq. \eqref{eq:alphasresult} is valid for both the fixed effective redshift from eq. \eqref{eq:effective_redshift_fixed} and the shifted effective redshift from eq. \eqref{eq:effective_redshift_shifted}. The interpretation of the scaling parameters however is different in the two cases because $z_\eff\blind$ is different. In the case of shifted redshift cuts, the comoving distance measured from the blinded catalog is equal to the unblinded measurement, but interpreted at different redshift. If the measurements are applied to samples with the same effective redshifts, the blinded distance can be obtained from the underlying distance by multiplying it by the correction factor $\alpha^\prime/\alpha$.

\subsection{RSD-shift} \label{ch:blinding_RSD}
 Here we propose to use a similar machinery as the density-field reconstruction procedure to blind the catalog for the growth rate.\footnote{Our numerical implementation is based on the reconstruction code used for eBOSS analysis, which is publicly available at \href{https://github.com/julianbautista/eboss\_clustering}{https://github.com/julianbautista/eboss\_clustering}} First of all, we use the reference cosmology to transform individual redshifts into distances to obtain the original redshift-space position $\mathbf{r}$ for each galaxy. Given that we do not know the underlying values of the galaxy bias and the growth rate in eq. (\ref{eq:mapping_combined}) \textit{a priori}, we choose reference values $b=b\reference$ and $f=f\reference$. Moreover, we need to adopt a method in order to obtain the smoothed density field. Here we choose to apply a gaussian filter with radius $R_\mathrm{sm}$ to the discrete galaxy field. With the reference values and smoothing scale as input, we solve eq. \eqref{eq:mapping_combined} iteratively in Fourier space, using the reconstruction code developed in Ref.~\cite{burden_reconstruction_2015}.  At this point, we diverge from the standard reconstruction procedure and use the estimated displacement field to add a RSD component that mimics a different growth rate $f\shift$.
 In this way we transform the original redshift-space position $\mathbf{r}$ to the blinded redshift-space position $\mathbf{r^\prime}$:
 \begin{equation}\label{eq:mapping_blinded}
\begin{aligned}
\mathbf{r^\prime} = \mathbf{r} - f\reference (\mathbf{\Psi} \cdot \mathbf{\hat{r}})\mathbf{\hat{r}} + f\shift  (\mathbf{\Psi} \cdot \mathbf{\hat{r}})\mathbf{\hat{r}} ~.
\end{aligned}
\end{equation}
Once we have obtained the blinded redshift-space position, we can obtain the blinded redshifts $z^\prime$ using again the reference cosmology. Contrarily to $\Delta^{\rm AP}z$, the RSD-shifts are different for each galaxy, because they depend on the local matter density. Taking this into account, the RSD-shift for the $p$-th galaxy is $\Delta^{\rm RSD}z_p = z^\prime_p - z_p$.

This procedure is equivalent to substitute redshift-space distortions from the underlying cosmology by those of the blinded. Hence, we expect the growth-rate measured from the blinded catalog  $f\blind$ to differ from the underlying value  $f\underlying$ by the difference between the reference and the shifted growth rates, so we write the {\it ansatz}
 \begin{equation}\label{eq:mapping_result}
\begin{aligned}
 f\blind = f\underlying + f\shift -  f\reference   ~.
\end{aligned}
\end{equation}

\begin{figure}[h]
    \centering
    \includegraphics[width=\linewidth]{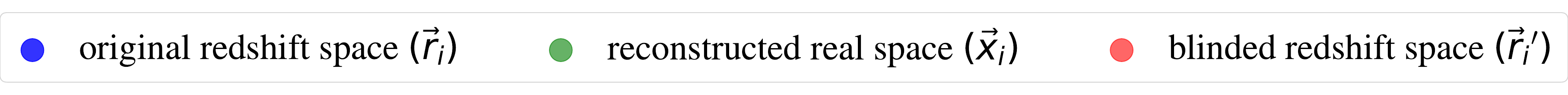}   
    \begin{minipage}[h]{0.495\linewidth}
    \includegraphics[width=\linewidth]{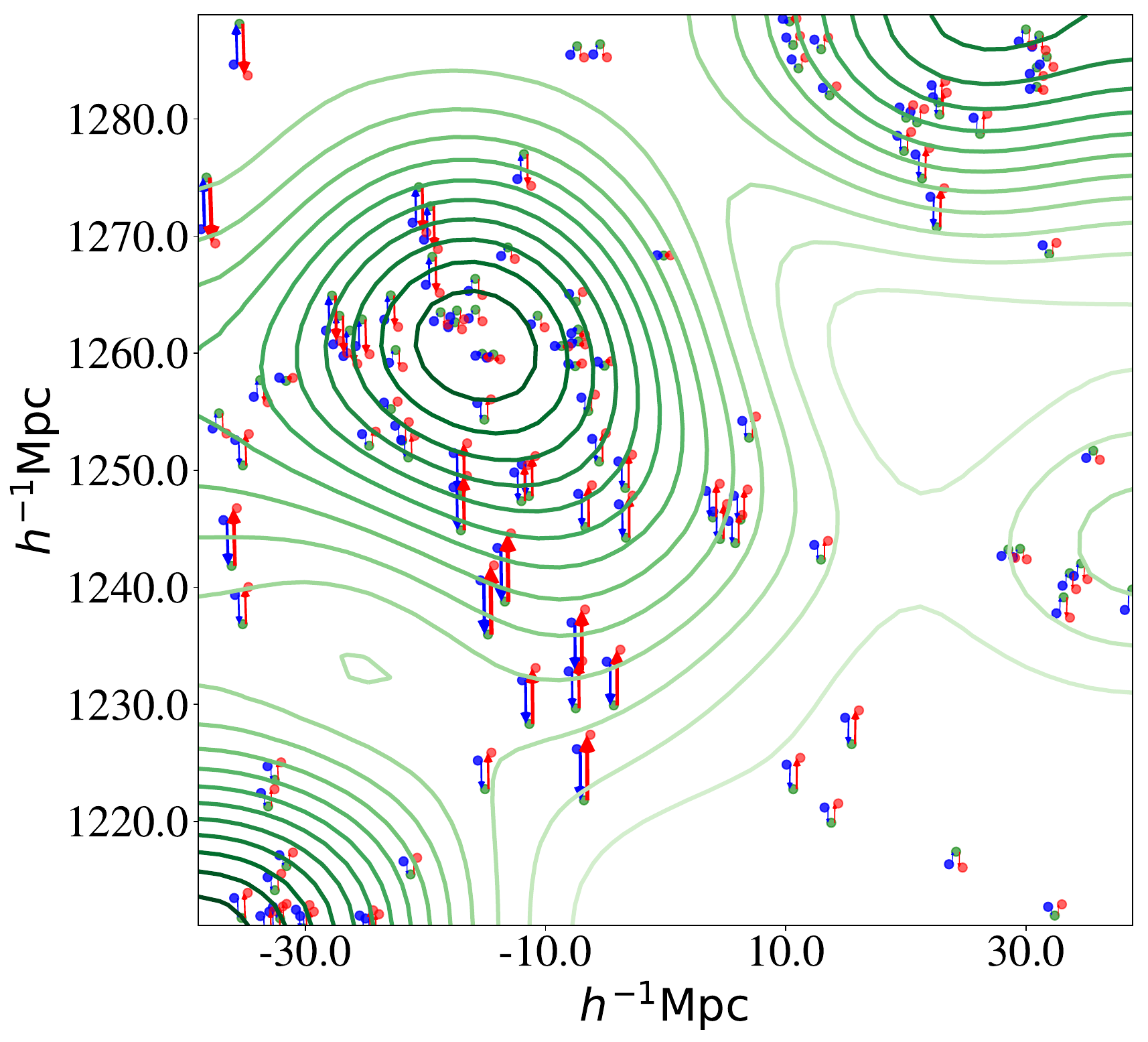}   
    \end{minipage}
    \hfill
    \begin{minipage}[h]{0.495\linewidth}
    \includegraphics[width=\linewidth]{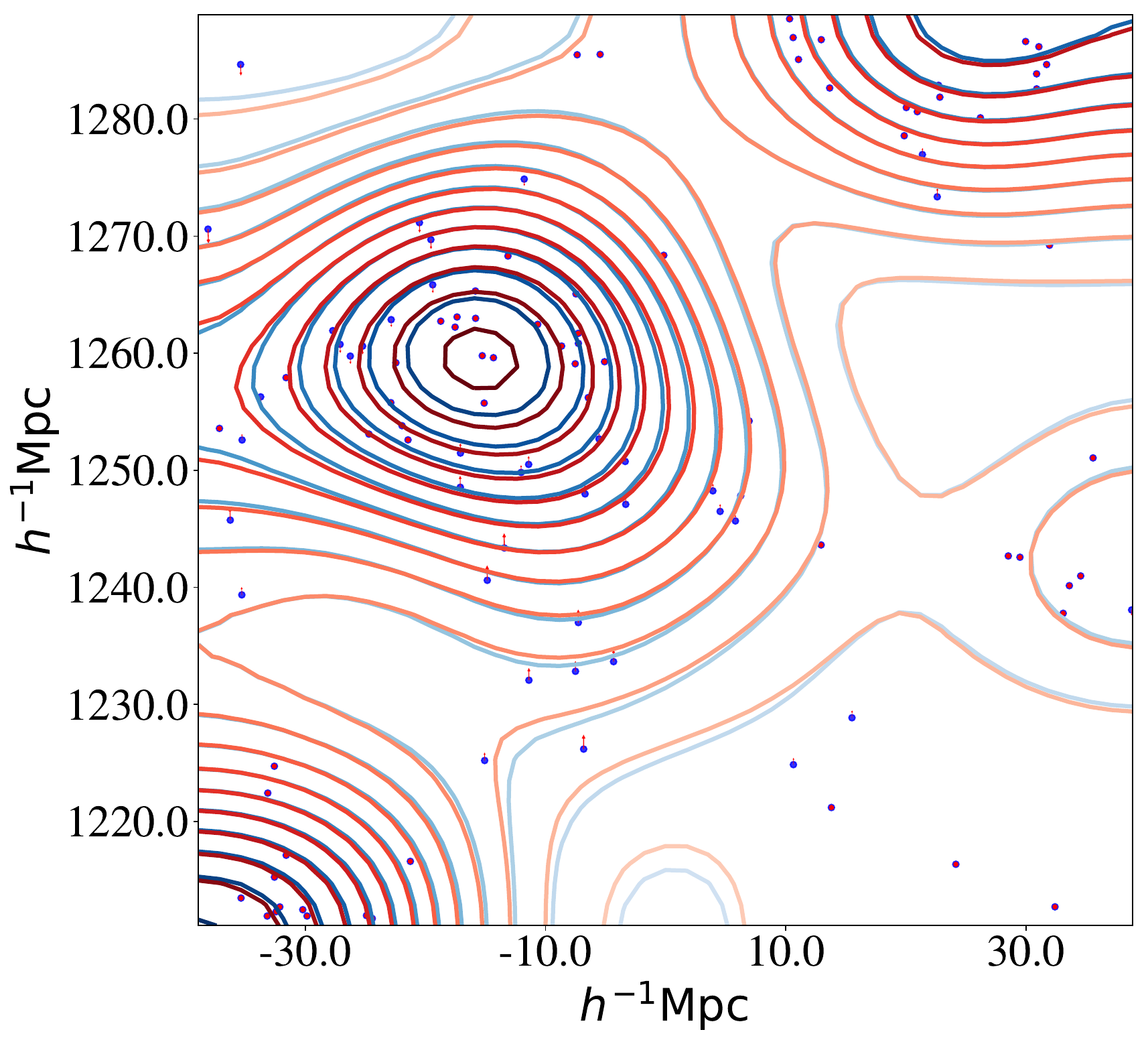}   
    \end{minipage}
    \\   
    \includegraphics[width=\linewidth]{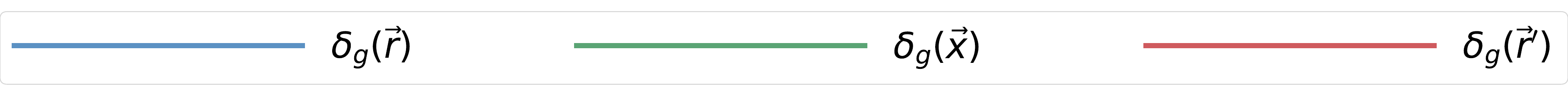}   
	\caption{The two panels show the same slice with sidelength 77 $\mpcoh$ and depth 50 $\mpcoh$ of an original and blinded  realisation of the CMASS North \textsc{MD-Patchy} mocks (see section \ref{ch:mocks_specifications} for more information). The observer is located at position $(0,0)$, which is more than 1200 $\mpcoh$  distant from the slice shown here. We compare the blinded and the original galaxies in two different ways. Left panel: The green contours correspond to the solution of the real-space over-density field smoothed with a gaussian filter of radius $R_\mathrm{sm} = 10 \, \mpcoh$ obtained with the iterative Fourier Space procedure. It is used to displace original galaxy positions (blue dots) along the line of sight (blue arrows) towards their reconstructed real-space positions (green dots) using the reference growth rate $f\reference = 0.78$. From there, the galaxies are displaced back (red arrows) towards their blinded positions (red dots) using a different growth rate $f\shift = 1.0$. To make the effect more visible the red and blue dots are slightly displaced horizontally. Right panel: The same slice, but here we show only the net displacement from the original positions (blue dots) towards the blinded positions (red arrows). Also shown are the original redshift-space over-density field (blue contours) and the blinded one (red contours). For illustrative purposes the shift considered is rather extreme,  so the ``squeezing" effect along the LOS can be appreciated ``by eye''.}    
    \label{fig:rsd2dplot}
\end{figure}

We test the validity of eq. ~(\ref{eq:mapping_result}) in section \ref{ch:mocks_RSD}. The different steps of this procedure are visualised in the left panel of figure \ref{fig:rsd2dplot}. Based on the pseudo-real-space displacement field $\Psi$ computed iteratively with $f\reference$, galaxies are shifted towards their (pseudo-)real-space position (blue arrows) and shifted back using the same displacement field, but  $f\shift$ instead. For illustrative purposes,  in this figure  we show a rather extreme case ($f\reference=0.78$, $f\shift=1.0$) where $f\shift$ differs from $f\reference$ by $\sim 25\%$.
Note that the shift is designed to be reversible: the blinded and original catalog would be exactly the same if $f\reference=f\shift$, which is also evident from eq. \eqref{eq:mapping_result}. The right panel shows the difference between the galaxy over-density fields of the blinded and the original catalog. Although we show a rather extreme case, the net squeezing of the central over-density along the $y$- direction can only be seen in nuances by eye.

For surveys being split in several redshift bins (or covering  different  galaxy samples which span different redshift ranges), the above procedure should be  applied to each redshift bin of the survey. Hence, for each bin, different values for the input parameters $(b\reference,f\reference)$ as well as the shifted growth rate $f\shift$ need to be chosen. It is possible to treat each bin independently, such that the overall growth history among samples cannot be modelled before unblinding. However, a more useful approach in practice is to blind coherently among samples by choosing a reasonable parameterisation of the redshift evolution of the growth rate as for example eq. (\ref{eq:growth_rate}).

\section{Test on Mocks}\label{ch:mocks}
Having described the theory of our blinding scheme, we now validate it on the (mock) galaxy catalogs from BOSS. Their detailed specifications are described in section \ref{ch:mocks_specifications}. We apply the individual blinding shifts $\Delta^{\rm AP}z_i$ , $\Delta^\mathrm{RSD} z_i$ and their combination on the mock galaxy catalogs. We treat each of the blinded mock catalogs as if it were an actual blinded measurement and run the analysis pipeline described in section \ref{ch:basics} with a fiducial configuration specified in section \ref{ch:mocks_methodology}. Finally, we quantify the performance of the studied blinding schemes by comparing the physical parameters measured from the blinded catalog with their theoretical expectation given by eqs. \eqref{eq:alphasresult} and \eqref{eq:mapping_result}.

Note that throughout this paper we work in Fourier space and do not extend our analysis to configuration space for simplicity. This is, because a correlation function analysis basically relies on the same information content as a power spectrum analysis and our blinding scheme affects both analysis types in the same way. We leave a detailed comparison of the blinding performance for different estimators for future work.

\subsection{Original and Blinded Mock Catalogs} \label{ch:mocks_specifications}
Mock galaxy catalogs play an essential role in interpreting cosmological data. They are calibrated on N-body simulations at a given cosmological model and correspond to independent realisations of the galaxy distribution as observed by the galaxy survey.
We use  the MultiDark-Patchy BOSS DR12 (\textsc{MD-Patchy}) mocks provided by \cite{kitaura_clustering_2016} and \cite{rodriguez-torres_clustering_2016} which include a low redshift sample LOWZ ($z_* = 0.33$) with roughly 350,000 galaxies and a higher redshift sample CMASS ($z_* = 0.60$) with roughly 800,000 galaxies.\footnote{The effective redshifts used in this work are different from the actual BOSS samples because the redshift cuts are different.}
The cosmological parameters used to generate the initial conditions of the mocks, as well as the reference cosmology adopted to analyse the original and the blinded catalogs, are given in table \ref{tab:cosmoparams}.  
Note a difference of  $0.1\%$ on $r_{\rm d}$ values predicted by these two cosmology models. This shift in the BAO scale is reflected by the factor $(r_{\rm d}\underlying/r_{\rm d}\reference)$ in the scaling parameters, $\alpha_\parallel$ and $\alpha_\perp$, and it is not modified during the blinding procedure.

\begin{table}[h]
    \centering
    \begin{tabular}{|c||c|c|c|c|c|c|c||c|}
    \hline
        Cosmology & $\Omega_m$ & $\Omega_b h^2$ & $h$ & $\sigma_8$ & $n_s$ & $M_\nu\,[\mathrm{eV}]$ & $r_{\rm d} \,[\mathrm{Mpc}]$ \\ \hline
       Reference & 0.31 & 0.022  & 0.676 & 0.8288 & 0.9611 & 0.06 & 147.78 \\
       \textsc{MD-Patchy} & 0.307115  & 0.02214 & 0.6777 & 0.8288 & 0.9611 & 0 & 147.66 \\ \hline
    \end{tabular}
    \caption{Values of cosmological parameters and sound horizon at radiation drag for the reference and the \textsc{MD-Patchy} mocks cosmology. In both cases a flat-$\Lambda$CDM cosmology and GR are assumed, so $w=-1$ and $\gamma=0.55$.}
    \label{tab:cosmoparams}
\end{table}

We employ the blinding scheme presented in section \ref{ch:blinding} to produce different sets of blinded catalogs. For one of the sets we only change the background evolution (AP only), for another we only apply the density-dependent shift (RSD only), and finally we combine the two (Combined). The values for the shifted physical parameters $\Omega\shift$ as well as the expectation for the cosmological parameters $\mathbf{\Theta}$ at each of the two redshift bins LOWZ ($z_\eff=0.33$) and CMASS ($z_\eff=0.60$) are given in table \ref{tab:shiftedparams}. In addition,  the parameter values of the blinded cosmology $\Omega\blind$, that we expect to infer from the blinded physical parameters, are also reported. Note that these cosmological parameters differ from the shifted cosmology used to generate the blinded catalog. As in our blinding scheme, the background parameters $\Omega_\mathrm{m}\shift$ and $ w\shift$ are basically used to replace the reference cosmology, $\Omega_\mathrm{m}\reference$ and $ w\reference$, so for those parameters we expect the deviation between $\mathbf{\Omega}\blind$ and $\mathbf{\Omega}\reference$ to be the same as the one between $\mathbf{\Omega}\reference$ and  $\mathbf{\Omega}\shift$, but in the opposite direction. Thus, for $\mathbf{\Omega}\shift - \mathbf{\Omega}\reference = \Delta \mathbf{\Omega}$ we expect $\mathbf{\Omega}\blind - \mathbf{\Omega}\reference = - \Delta \mathbf{\Omega}$.

We apply relative shifts on $\Omega_{\rm m}$ and  $w$ with respect to their reference values of order of $10\%$. We find that this corresponds approximately to  $1\sigma$ deviation when fitting $w$ and $\Omega_{\rm m}$ to BAO LOWZ and CMASS data with a Planck prior on $\Omega_bh^2$.
For the RSD-only blinding we perform a very extreme shift with $\gamma=0$ or $f(z)=1$ to fully exploit the capabilities of our method and to make the effect of RSD-blinding more visible. For the combined shift we choose a relatively moderate value. 

In practice, the choice of how much the shifted values can deviate from the reference parameters should depend on the forecasted sensitivity of the data and probably not exceed deviations of $3\sigma$. Too large deviations can spoil the robustness of the analysis, because some of the usually made approximations, e.g. a fixed covariance matrix computed at the reference cosmology, might not hold.

\begin{table}[h]
    \centering
    \resizebox{\textwidth}{!}{
    \begin{tabular}{|c|c||c|c|c||c|c|c|c|c|c|}
    \hline
       \multicolumn{2}{|c||}{Cosmology/} & \multirow{2}{*}{$\Omega_{\rm m}$} & \multirow{2}{*}{$w$} & \multirow{2}{*}{$\gamma$} & \multicolumn{2}{c|}{$\alpha_\parallel(z_\eff)$} & \multicolumn{2}{c|}{$\alpha_\perp(z_\eff)$} & \multicolumn{2}{c|}{$f(z_\eff)$} \\ \cline{6-11}  
       \multicolumn{2}{|c||}{Physical parameter at $z_\eff$} & & & & 0.33 & 0.60 & 0.33 & 0.60 & 0.33 & 0.60 \\ \hline
       \multicolumn{2}{|c||}{Reference $\Omega_\mathrm{m}\reference$} & 0.31 & \text{-}1.0  & 0.55 & 1.000 & 1.001 & 0.999 & 1.000 & 0.69 & 0.78  \\ \hline
       \multirow{2}{*}{AP only} & Shifted $\Omega_\mathrm{m}\shift$ & 0.279 & \text{-}1.1  & 0.55 &  \multirow{2}{*}{0.964} &  \multirow{2}{*}{0.980} &  \multirow{2}{*}{0.950} &  \multirow{2}{*}{0.970} &  \multirow{2}{*}{0.69} &  \multirow{2}{*}{0.78} \\
       & Blinded $\Omega_\mathrm{m}\blind$& 0.341 & \text{-}0.9  & 0.55 &  & & & & &\\ \hline
       \multirow{2}{*}{RSD only} & Shifted $\Omega_\mathrm{m}\shift$ & 0.31 & \text{-}1.0  & 0.0 & \multirow{2}{*}{1.002} & \multirow{2}{*}{1.002} & \multirow{2}{*}{1.003} & \multirow{2}{*}{1.002} & \multirow{2}{*}{1.0} & \multirow{2}{*}{1.0} \\
       & Blinded $\Omega_\mathrm{m}\blind$ & 0.31 & \text{-}1.0  & 0.0 &  & & & & &\\ \hline
       \multirow{2}{*}{Combined} & Shifted $\Omega_\mathrm{m}\shift$ & 0.279 & \text{-}1.1  & 0.825 & \multirow{2}{*}{0.964} & \multirow{2}{*}{0.980} & \multirow{2}{*}{0.950} & \multirow{2}{*}{0.970} & \multirow{2}{*}{0.58} & \multirow{2}{*}{0.70} \\
        & Blinded $\Omega_\mathrm{m}\blind$ & 0.341 & \text{-}0.9  & 0.825 &  & & & & & \\ \hline
    \end{tabular}
    }
    \caption{In the first line of this table we provide the values of the reference cosmological parameters that we aim to blind for $\left\lbrace \Omega_{\rm m}, w, \gamma \right\rbrace$, together with the corresponding derived values  $\left\lbrace \alpha_\parallel, \alpha_\perp, f \right\rbrace$ in each redshift bin. In the following lines the cosmological parameters and expected derived parameters are shown for each of the 3 blinding cases we investigate. The shifted cosmology is used to generate the blinded catalog, whereas the blinded cosmology is the one that we expect to measure.}
    \label{tab:shiftedparams}
\end{table}

\subsection{Analysis on mocks: set up and approach} 
 \label{ch:mocks_methodology}
Here we present the fiducial setup of the BAO and RSD analyses.
\textit{A priori}, the blinded data should be treated as if it were the original data, or in other words, the analyses should not depend on the blinding. Therefore the methodology presented here is applied to the original and the blinded catalogs in the same way.
The general procedure is already explained in section \ref{ch:basics}, but in the following we provide the numerical details.

For the conversion from redshifts to distances we always use the reference cosmology given in table \ref{tab:cosmoparams}. These cosmological parameters are also used to compute the linear matter power spectrum template with \texttt{CLASS} \cite{2011JCAP...07..034B}.

To estimate the original and the blinded power spectra we follow the steps explained in the \S ~ \nameref{ch:basics_ps} of section \ref{ch:basics}. Using the distance-redshift relation assuming the reference cosmology, we place the LOWZ and CMASS galaxies in boxes of length $L_\mathrm{box} = 2500\,\mpcoh$  (LOWZ) and $L_\mathrm{box} = 3700\,\mpcoh$ (CMASS) with $512^3$ grid cells each. We compute the multipole power spectra in Fourier space using the Yamamoto method \cite{Yamamoto:2005dz, bianchi_measuring_2015}. We interpolate galaxies on the grid by using the triangular-shaped-cloud scheme and obtain the monopole and quadrupole in the interval $0<k\,[\hompc]<0.30$ for linearly binned wave vectors $k$ with $\Delta k = 0.01\,\hompc$. For both the BAO anisotropic and the full-shape analysis we restrict our fits to $0.02 < k\, [\hompc] < 0.20 $. 

We fit the BAO and RSD models described in \S  ~ \nameref{ch:basics_aniso} and \S~ \nameref{ch:basics_fs} of section~\ref{ch:basics} to the original and blinded power spectra using Monte-Carlo Markov Chains (MCMCs).  
We use the standard Metropolis Hastings algorithm, where sequences (called chains) starting at different points  sample parameter space so that the density of points sampled is proportional to the posterior. Convergence is assessed by  the Gelman-Rubin criterion \cite{gelman1992} and the stopping criterion adopted for the MCMC chains  is $1-R<0.01$  for all parameters.
For the full-shape RSD model we have 3 physical parameters and $(4 N_j)$ nuisance parameters, where $N_j$ is the number of disconnected patches of the sky observed. 
For the BAO anisotropic model we have 2 physical parameters and $[1+(1+N_i N_\ell) N_j ]$ nuisance parameters per redshift bin, where $N_\ell$ is the number of multipoles and $N_i$ the number of coefficients in the broadband polynomial.
Regarding the dispersion scales $\Sigma_\parallel$ and $\Sigma_\perp$ of eq. \eqref{eq:Panisofull},
 we fit the mean of 1000 mock catalog power spectra and find 
\begin{equation}
\begin{aligned}
      \Sigma_\parallel^\mathrm{LOWZ} &= 13.85 \, \mpcoh~, \qquad &
      \Sigma_\perp^\mathrm{LOWZ} &= 9.03 \, \mpcoh ~, \\
      \Sigma_\parallel^\mathrm{CMASS} &=12.18 \, \mpcoh~, \qquad  &  
      \Sigma_\perp^\mathrm{CMASS} &= 7.31  \, \mpcoh~.
\end{aligned}
\end{equation}

For both models we perform a joint monopole and quadrupole fit ($N_\ell=2$) for the north and south galactic caps of LOWZ and CMASS ($N_j=2$). For the polynomial broadband expansion we go up to order $N_i=3$. Hence, we have 17 free parameters for the BAO anisotropic model and 11 for the full-shape RSD model per redshift bin. 
We refer to these parameters as ${\boldsymbol \theta}=\{\boldsymbol{\Theta},$ \rm{nuisance parameters}\}. The prior ranges of all these parameters are given in table \ref{tab:priors} and they are the same for all redshift bins.

\begin{table}[h]
    \centering
    \begin{tabular}{c c}
      \hline \hline
       Parameter & Prior range \\ \hline
       $\alpha_\parallel$  & $[0.6, 1.4]$  \\
      $\alpha_\perp$ & $[0.6, 1.4]$  \\ \hline
      $\beta$ & $[0, 30]$ \\
      $B$ & $[0, 20]$ \\
      $A_i ~ [10^3 	(\mpcoh)^{5-i} ]$ & $[\text{-} 20, 20]$ \\  \hline
      $f$ & $[0, 3]$ \\
      $b_1$ & $[0, 10]$ \\
      $b_2$ & $[\text{-} 10, 10]$ \\
      $\sigma_P \, [\mpcoh]$ & $[0, 20]$ \\
      $A_\mathrm{noise} $ & $[\text{-} 5, 5]$ \\ \hline
    \end{tabular}
    \caption{Uniform prior ranges for parameters used in anisotropic BAO and FS analysis. The $\alpha$'s in the top section are used in both BAO and FS analyses, parameters in the second and third sections are used only in BAO and FS analysis, respectively. The $A_i$ are the three parameters to model the broadband.}
    \label{tab:priors}
\end{table}

The likelihood function for the parameters ${\boldsymbol \theta}$ is defined as  
\begin{align}
\mathcal{L}(\boldsymbol{\theta}) \propto e^{-\chi^2(\boldsymbol{\theta})/2}~, 
\end{align}
where the $\chi^2$ is obtained by a matrix multiplication of the difference-vector between data and model with the inverse covariance:
\begin{align}
\chi^2(\boldsymbol{\theta}) = \sum_{\ell,\ell^\prime}^{\ell_\mathrm{max}} \sum_{k,k^\prime}^{k_\mathrm{max}} \left( P_\mathrm{model}^{(\ell)}(k,\boldsymbol{\theta}) - P_\mathrm{data}^{(\ell)}(k) \right) C_{(\ell k)(\ell^\prime k^\prime)}^{-1}  \left( P_\mathrm{model}^{(\ell^\prime)}(k^\prime,\boldsymbol{\theta}) - P_\mathrm{data}^{(\ell^\prime)}(k^\prime) \right) ~.
\end{align}
Both data and model vectors contain $(k_\mathrm{max}-k_\mathrm{min})/\Delta k$ entries for monopole and quadrupole each. The covariance matrix $C$ is assumed to be fixed, and estimated from 1000 independent realisations per patch of the  \textsc{MD-Patchy} mock catalogs without blinding. We do not modify the covariance matrix when analyzing blinded catalogs (e.g. by estimating it from 1000 blinded mocks), because the blinded data should be treated as if we did not know that it was blinded. When inverting $C$ we apply the Hartlap correction \cite{Hartlap:2006kj} to take into account the finite number of mock catalogs. The north and south galactic cap are assumed to be fully independent, and therefore, their likelihoods are multiplied to obtain the combined one, $\mathcal{L}_{\rm NGC+SGC}=\mathcal{L}_{\rm NGC}\times\mathcal{L}_{\rm SGC}$.

\begin{figure}[t]
    \centering
    \includegraphics[width=\textwidth]{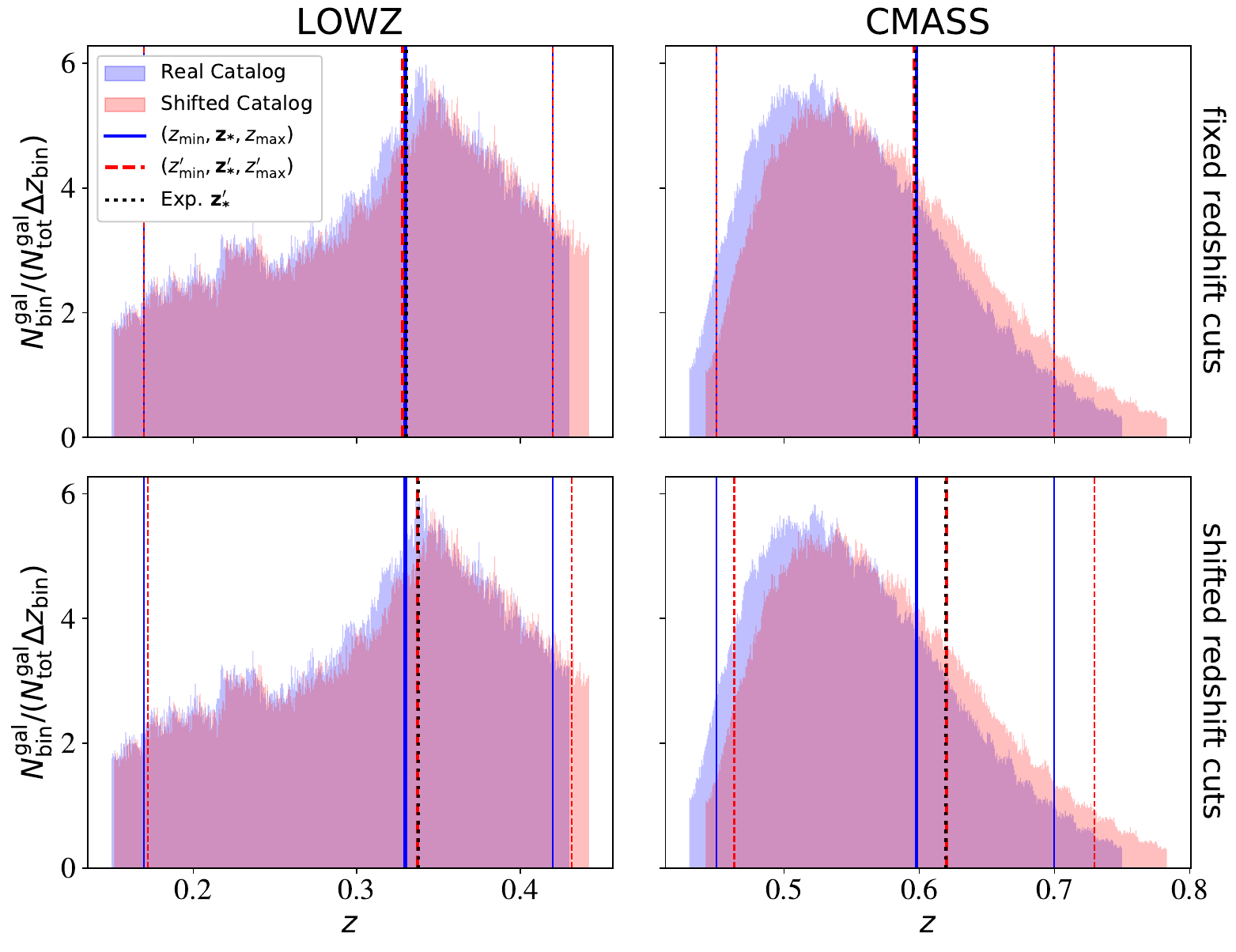}
    \caption{Normalised galaxy redshift distributions of LOWZ (left panels) and CMASS (right panels). The blue area represents the original catalog galaxies and the red area the blinded catalog of a scenario where $\Omega_{\rm m}$ and $w$ were reduced by 10\%. Overlapping regions are shown in purple. The distribution of the upper and lower panels are the same, but with different cuts in redshift. If we keep the cuts fixed (upper panels), the effective redshift does not change significantly after blinding. By shifting the cuts according to the blinding scheme (lower panels) the effective redshift is shifted in the same way as expected.}
    \label{fig:mocksnofz}
\end{figure}

\subsection{AP-like shift Performance}\label{ch:mocks_AP}

We now focus on the performance of the geometrical blinding shift $\Delta^\mathrm{AP} z$ (section \ref{ch:blinding_AP}). We apply the methodology explained in section \ref{ch:mocks_methodology} to the set of blinded mocks characterised by the parameter values of  row ``AP only" in table \ref{tab:shiftedparams}. In this section, the blinding and the analysis are performed on pre-reconstruction catalogs i.e., on the original ones. Results relative to the  combination of the two  effects including reconstruction procedure are presented in section~\ref{ch:mocks_comb}.

We use  the shifted cosmology specified in  row ``AP only"of  table \ref{tab:shiftedparams} to convert the original  catalog redshifts to distances, then transform them to blinded redshifts using the reference cosmology (yielding the blind catalog). This changes the number density distribution $\bar{n}(z)$, as shown in figure \ref{fig:mocksnofz}. The  $\bar{n}(z)$ of the original catalog (blue histogram) is transformed to the shifted distribution (orange histogram). The overlap area appears in purple. The shifted parameters are lowered by $10\%$ with respect to the reference ones, thus the (shifted) distances for individual galaxies are larger, and consequently the blinded redshifts are (slightly)  increased by $\Delta^\mathrm{AP} z$.

This naturally leads to a shift of the effective redshift of the sample, whose implications for the analysis are discussed below.

The top panels of the figure \ref{fig:mocksnofz} illustrate the ``fixed redshift cuts" convention and the bottom panel the  ``shifted redshift cuts". In the  ``fixed redshift cuts" case,  the cuts (thin dashed lines) are decided and fixed before blinding the catalog. If the cuts are not located at the edge of the distribution -- i.e,. there is some padding-- the effective redshift does not change considerably after blinding,  because  while some galaxies leave the selected region on one side, new galaxies enter the region from the other side.  Here, we cut the original catalogs at $z=0.17$ and $z=0.42$ for LOWZ, and $z=0.45$ and $z=0.7$ for CMASS. In both redshift bins, after blinding, approximately $5\%$ of galaxies leave the selected range at the top edge and about $5\%$ enter from the bottom edge. The effective redshifts of the original and blinded catalogs agree within $0.1\%$.  However, the change of effective redshift highly depends on the details of the redshift distribution, as well as the position of the cuts, so the agreement we find between $z_\eff$ and $z_\eff^\prime$ in the case of LOWZ and CMASS should not be generalised to other galaxy samples.

In the ``shifted redshift cuts" case on the other hand,  
the selection cuts are made in such a way that the objects belonging to the sample are the same pre- and post-blinding.
Hence, the cuts on the blinded catalog (thin red dashed lines) are displaced from the original cuts (thin blue continuous lines) according to $\Delta^\mathrm{AP} z$; the measured effective redshifts of the samples (thick lines) are also displaced, matching exactly the shift predicted by $\Delta^\mathrm{AP} z$ (black dotted line). 

\begin{figure}[t]
    \begin{minipage}[h]{0.5\textwidth}
    \centering
    \includegraphics[width=\textwidth]{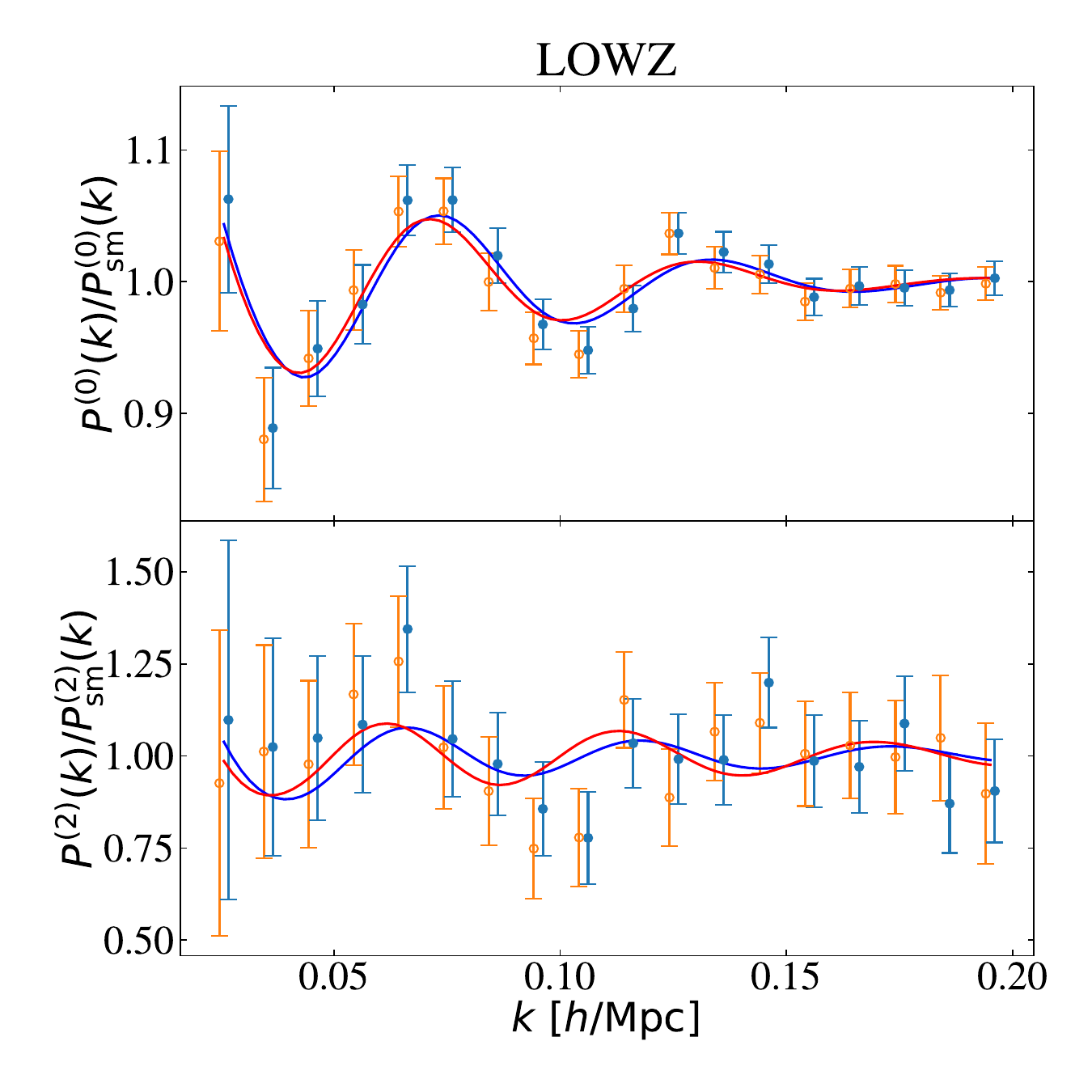}
    \end{minipage}
    \begin{minipage}[h]{0.5\textwidth}
    \centering
    \includegraphics[width=\textwidth]{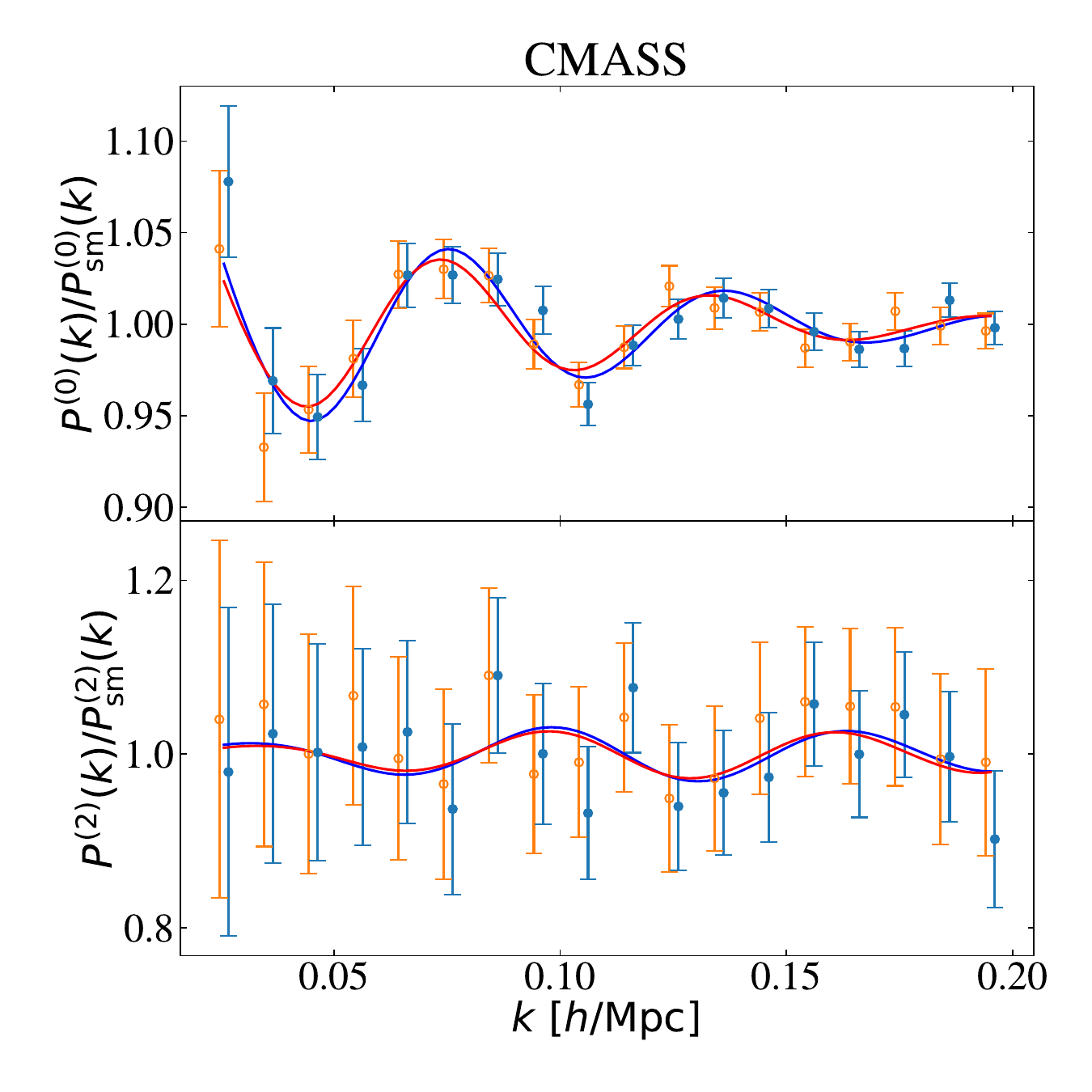}
    \end{minipage}  
    \\
    \includegraphics[width=\textwidth]{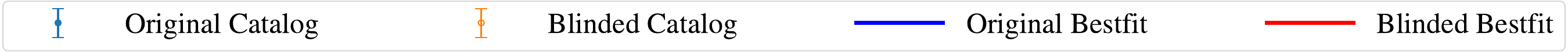}
    \caption{Measured power spectra of one LOWZ (left) and one CMASS (right) mock catalog (both original and blinded) with the best-fit anisotropic BAO model (blue) in comparison with its blinded counterpart (red) using the ``shifted redshift cuts'' convention.  Data points are slightly displaced by $\Delta k = \pm 0.001 \, \hompc$ for better visibility. In all cases, both data and model prediction are divided by the corresponding best-fit prediction of the broadband.}
    \label{fig:BAOmodelvsdatadiffz}
      
\end{figure}

In the next step, we measure the monopole and quadrupole power spectra of the original and the blinded catalogs and fit to them our anisotropic BAO and full-shape RSD models as described in section \ref{ch:mocks_methodology}. As an example, in figure \ref{fig:BAOmodelvsdatadiffz} we show the  power spectra together with the best-fit anisotropic BAO model for a pair of original and blinded mock with the ``shifted redshift cuts'' convention. We show the oscillatory part of the power spectrum only. After blinding, the oscillation peaks are shifted towards smaller $k$ leading to a decrease of the measured $\alpha_\parallel$ and $\alpha_\perp$ with respect to the original catalog. This is consistent with our expectation given in table \ref{tab:shiftedparams}.

\begin{figure}[t]
    \begin{minipage}[h]{0.5\textwidth}
    \centering
    \includegraphics[width=\textwidth]{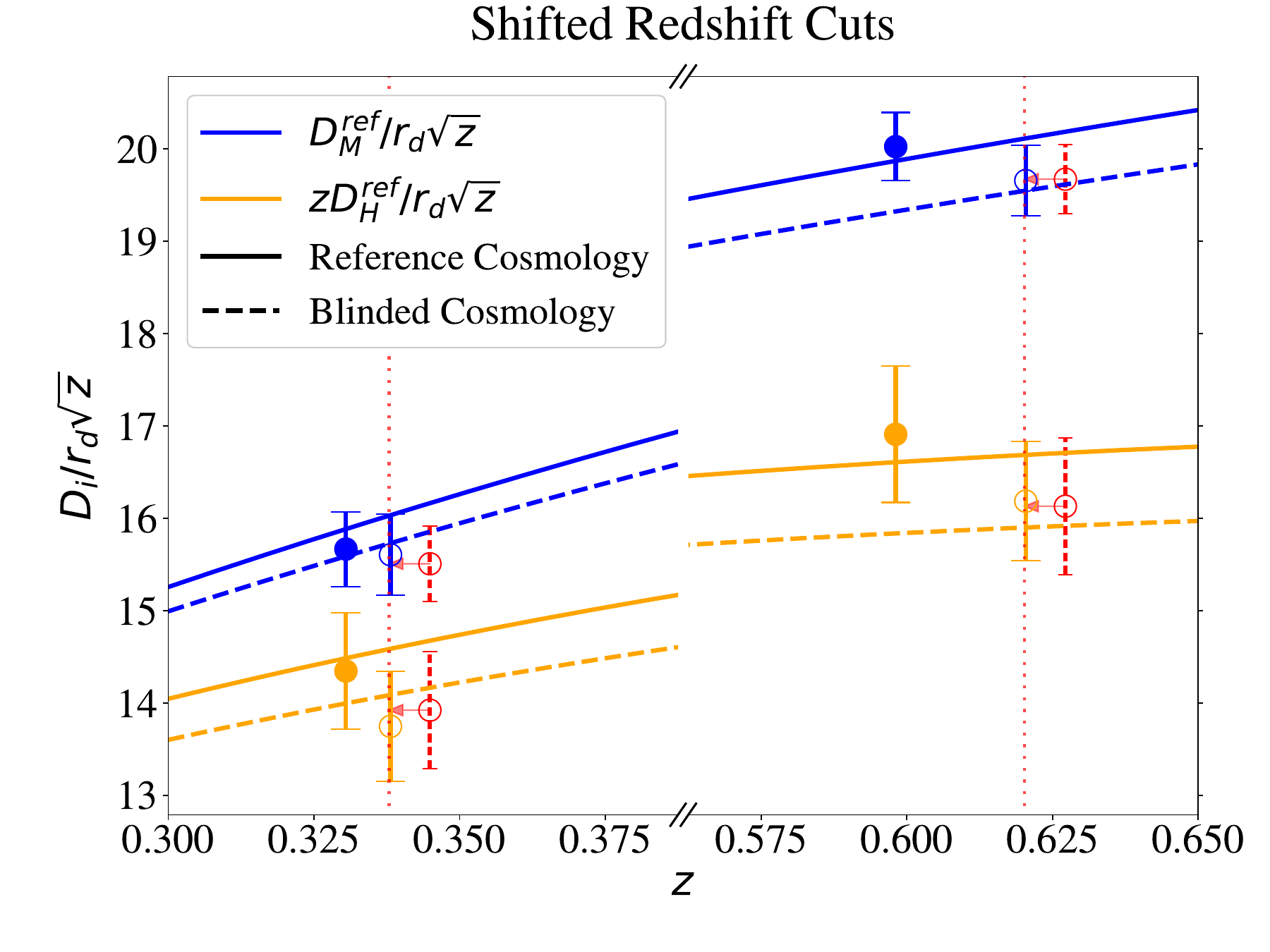}
    \end{minipage}
    \begin{minipage}[h]{0.5\textwidth}
    \centering
    \includegraphics[width=\textwidth]{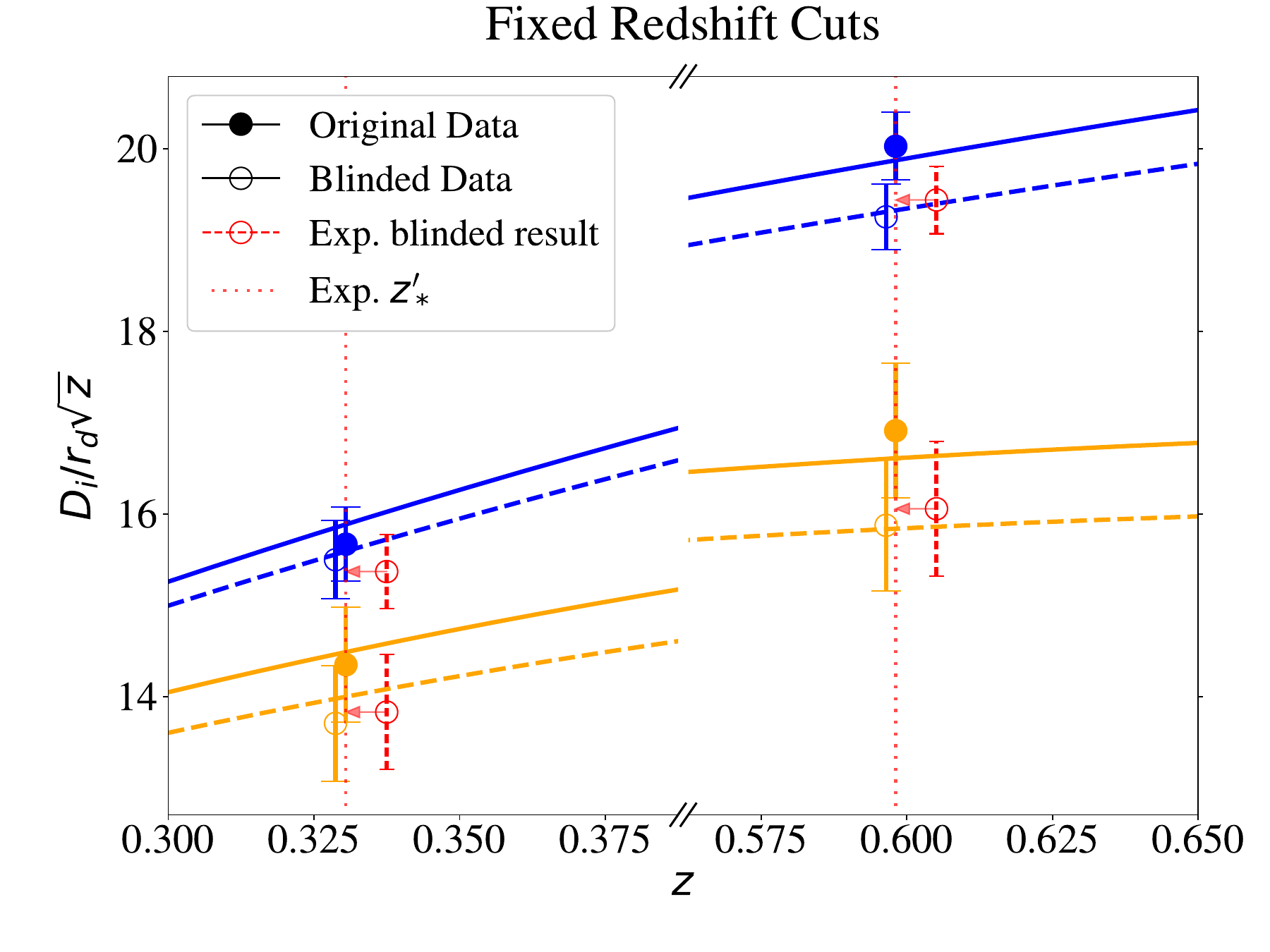}
    \end{minipage}
    \caption{Comparison of the anisotropic distance measurement between one of the original mock realisations (filled dots) and its blinded counterpart (empty dots). The effective redshift positions are different in the case of shifted redshift cuts (left panel), while they are almost the same ($<0.5\%$ difference) in the case of fixed redshift cuts (right panel). Red data points mark the expected blinded result obtained by applying eq. \eqref{eq:alphasresult} to the original data. They are slightly displaced to avoid overlap; their true redshift positions are indicated by red arrows and vertical dotted lines. The rescaling factor $\sqrt{z}$ improves the visibility of the different data points and has no physical meaning.}
    \label{fig:BAOdistances}
\end{figure}

In  figure \ref{fig:BAOdistances} we show a comparison between the scaling parameters corresponding to the best-fit anisotropic BAO model of figure \ref{fig:BAOmodelvsdatadiffz} and the expected values given in table \ref{tab:shiftedparams}. We directly show the corresponding distances perpendicular (comoving angular diameter distance $D_M(z)$) and parallel (Hubble distance $D_H=c/H(z)$) to the LOS rescaled by $r_{\rm d}$. The left panel shows the results using ``shifted redshift cuts" convention, where  blinding changes the effective redshift of the sample.

On the right panel the ``fixed redshift cuts" convention is shown, where the effective redshift does not change. In both cases the expectation represented by the red points matches the blinded distance measurement very well. The error bars of the expected blinded result are chosen to have the same size as the original error bars. As we can see, their size is similar to the error bars obtained from the blinded data. In addition, figure \ref{fig:BAOdistances} shows the distance-redshift relation of the reference cosmology $\mathbf{\Omega\reference}$, that is very close to the \textsc{MD-Patchy} mock underlying cosmology, and the blinded cosmology, which is the one that we would expect to measure.

\begin{figure}[t]
    \begin{minipage}[h]{0.5\textwidth}
    \centering
    \includegraphics[width=\textwidth]{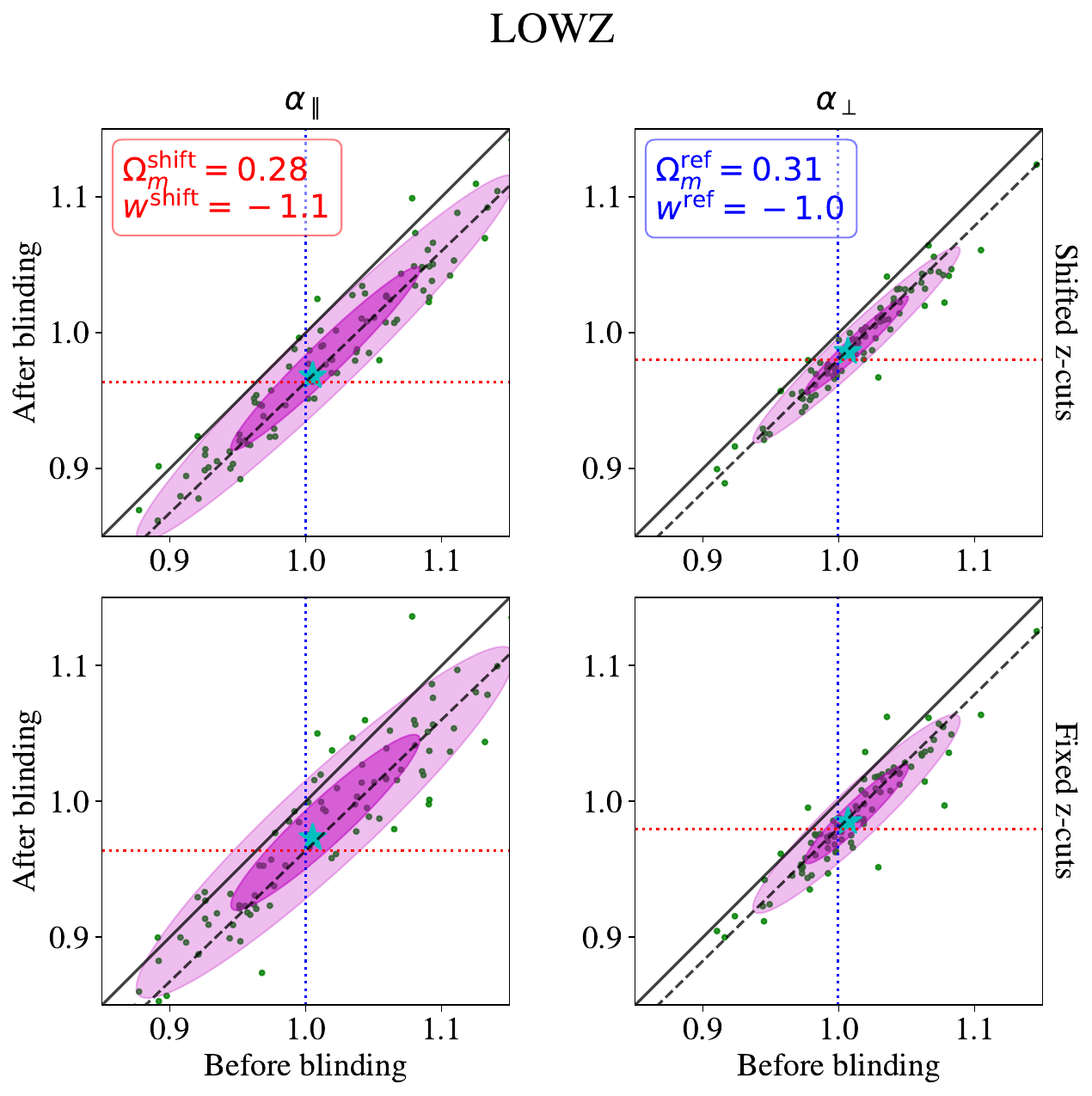}
    \end{minipage}
    \begin{minipage}[h]{0.5\textwidth}
    \centering
    \includegraphics[width=\textwidth]{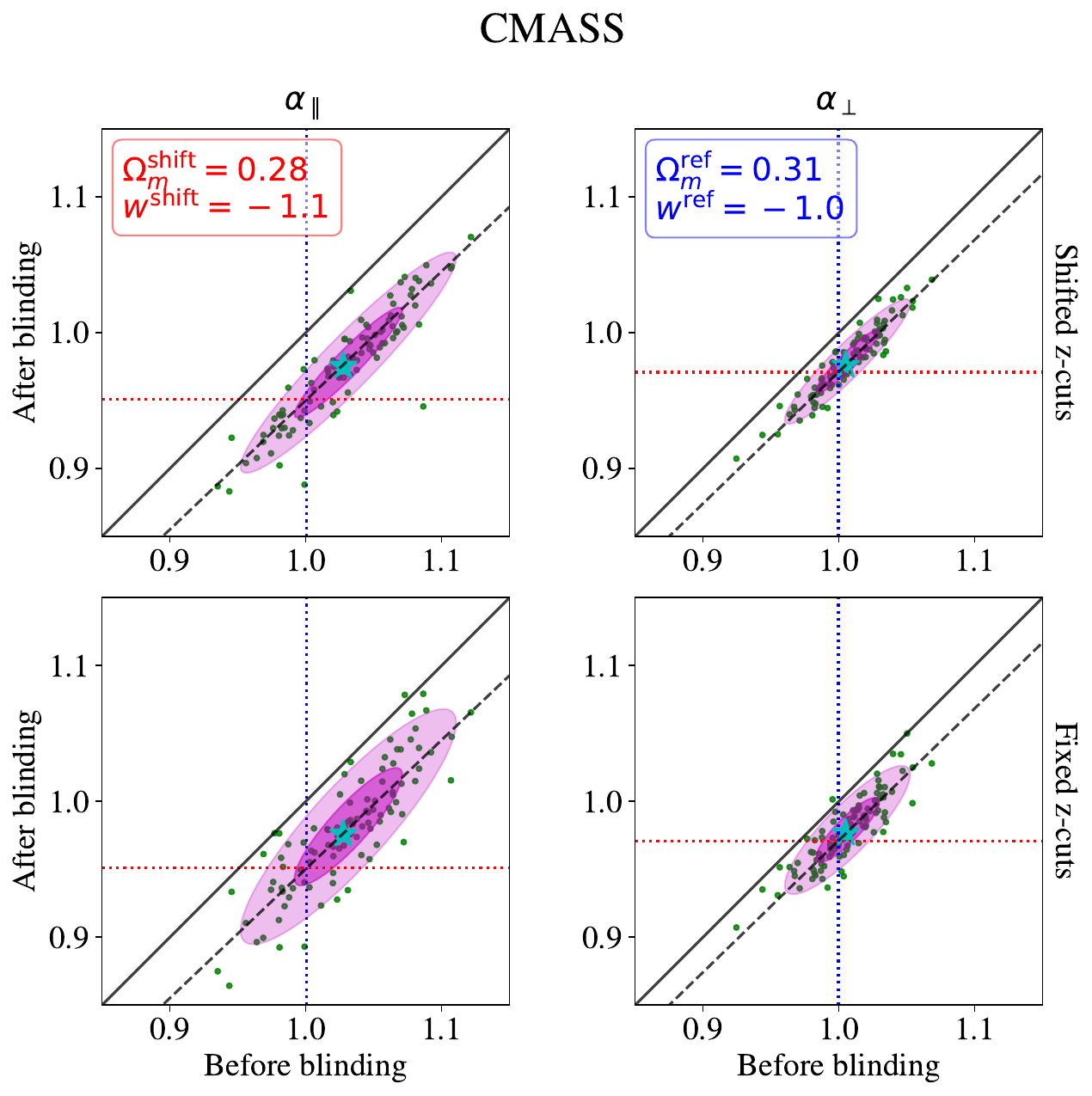}
    \end{minipage} 
 \includegraphics[width=\textwidth]{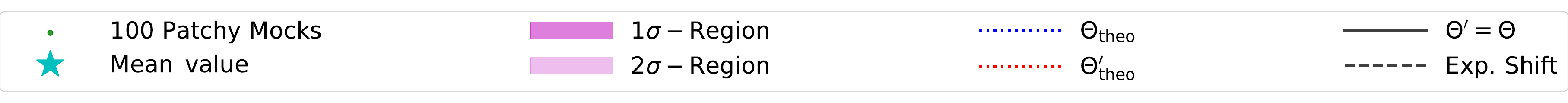}    
    \caption{Effect of  AP-only shifts blinding on pre-reconstruction BAO-only analysis. The panels show $\alpha_\parallel$  and $\alpha_\perp$ best-fitting results for the pre-reconstruction BAO-only analysis on \Nmocks mock realisations (green dots) and their mean values (cyan stars) for the underlying cosmology and the blinded cosmology (using only AP-like shifts). Left (right) panels display the measurements from the LOWZ (CMASS) sample. Upper panels correspond to the ``shifted redshift cuts" and lower panels to the ``fixed redshift cuts" convention. Purple regions indicate the 68\% and 95\% confidence level regions. Dashed lines represent the expected shift of the mock distribution from the diagonal due to blinding. Blue dotted lines represent the underlying values corresponding to the \textsc{MD-Patchy} mocks cosmology, whereas red dotted lines show the ``underlying'' values of the blinded catalogs.}
    \label{fig:alphascatterBAOdiffz}   
\end{figure}

We repeat this analysis for \Nmocks realisations (both original and blinded) pairs of mock catalogs and test the agreement between measurement and expectation of the scaling parameters for each  pair.\footnote{Note that for each mock catalog  6 MCMC are run in parallel (we refer to these as sets of chains). In total in this paper we are presenting the results of 2216 sets of MCMC chains}

The overall performance is summarised in figure \ref{fig:alphascatterBAOdiffz}. Each dot represents the measured values of a parameter from the blinded and original catalogs, shown in the $y$- and $x$-axis, respectively. The scatter of the mocks is visualised by the purple 68\% and 95\% confidence level contours, that were obtained by weighting each point with its corresponding uncertainties for each catalog. 
The spread of the points parallel to the diagonal $y=x$ indicates the intrinsic scatter of the mocks with respect to the given parameter; in case the original and the blinded catalog were the same, all points would be located along that diagonal. However, by using the shifted cosmology given in table \ref{tab:shiftedparams} we expect the points to scatter around the dashed line described by the formula
\begin{align} \label{eq:dashed_line}
y(x) = \frac{\alpha_\mathrm{theo}\blind}{\alpha_\mathrm{theo}\underlying} \, x~,
\end{align} 
where $\alpha_\mathrm{theo}\underlying$ and $\alpha_\mathrm{theo}\blind$ correspond to the theoretical values given in table \ref{tab:shiftedparams}. The scatter of points perpendicular to the line parameterised by eq.~\eqref{eq:dashed_line} indicates how much the measurement of a single blinded mock catalog typically deviates from the expectation given the measurement of the original mock. This deviation is purely statistical, as the overall cloud scatters around the line predicted by eq.~\eqref{eq:dashed_line}. 
 For all cases there is a good agreement between the mock distribution and the prediction of the blinded result represented by eq. \eqref{eq:dashed_line}.

The mean values of the mocks (cyan stars) agree very well with the theoretical values of table \ref{tab:shiftedparams} given by the coloured dotted lines in figure \ref{fig:alphascatterBAOdiffz}. Also, our results for the original mocks are in concordance with previous analysis as in \cite{Beutler:2016ixs}.

From figure~\ref{fig:alphascatterBAOdiffz} we can clearly appreciate that for the  fixed redshift cuts case (right panels) the scatter in direction perpendicular to the diagonal is slightly enhanced with respect to the shifted redshift cuts case (left panels). This is because in the former case the galaxy sample changes after blinding and thus the Fourier modes sampled, leading to a cosmic variance effect. This is avoided in the latter case, where the blinded catalog contains the same galaxies (and samples exactly the same modes) as the original catalog.

We conclude that the shifted redshift cuts case gives a blinding procedure that is a more accurate prediction of the behaviour of each individual mock catalog. For this reason we recommend whenever possible to adopt the ``shifted redshift cuts" convention. In the following we always refer to the ``shifted redshift cuts" convention  unless explicitly mentioned otherwise.

\subsection{RSD shift Performance}\label{ch:mocks_RSD}

Here we focus on the perturbation-level blinding shift $\Delta^\mathrm{RSD} z$ (section \ref{ch:blinding_RSD}). Starting from the original catalogs, we generate the set blinded mocks using the parameter values of  row ``RSD only" in table \ref{tab:shiftedparams} and a reference galaxy bias value of $b\reference=1.85$.  Then we apply the methodology explained in section \ref{ch:mocks_methodology} to the set of blinded mocks.

\begin{figure}[t]
    \begin{minipage}[h]{0.5\textwidth}
    \centering
    \includegraphics[width=\textwidth]{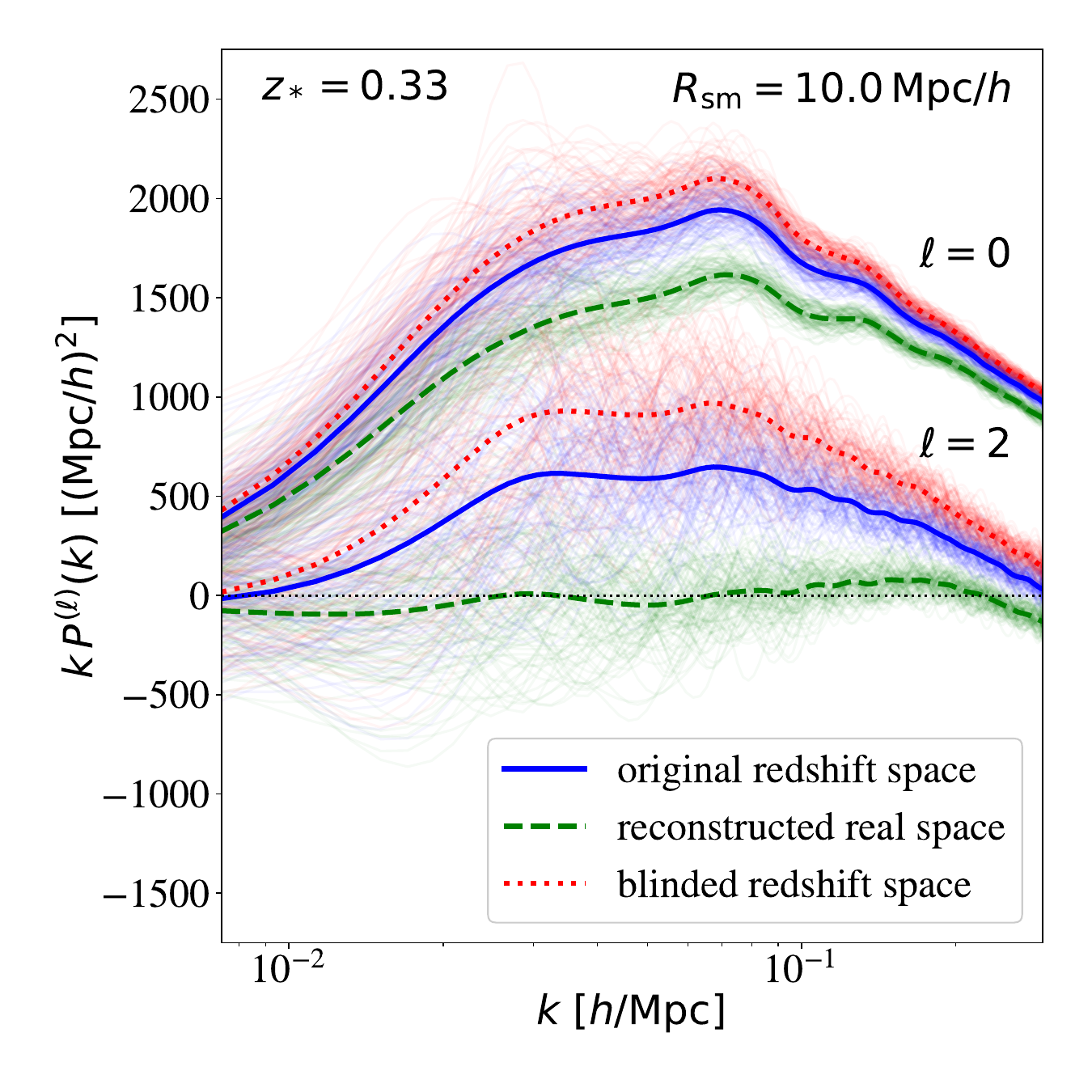}
    \end{minipage}
    \begin{minipage}[h]{0.5\textwidth}
    \centering
    \includegraphics[width=\textwidth]{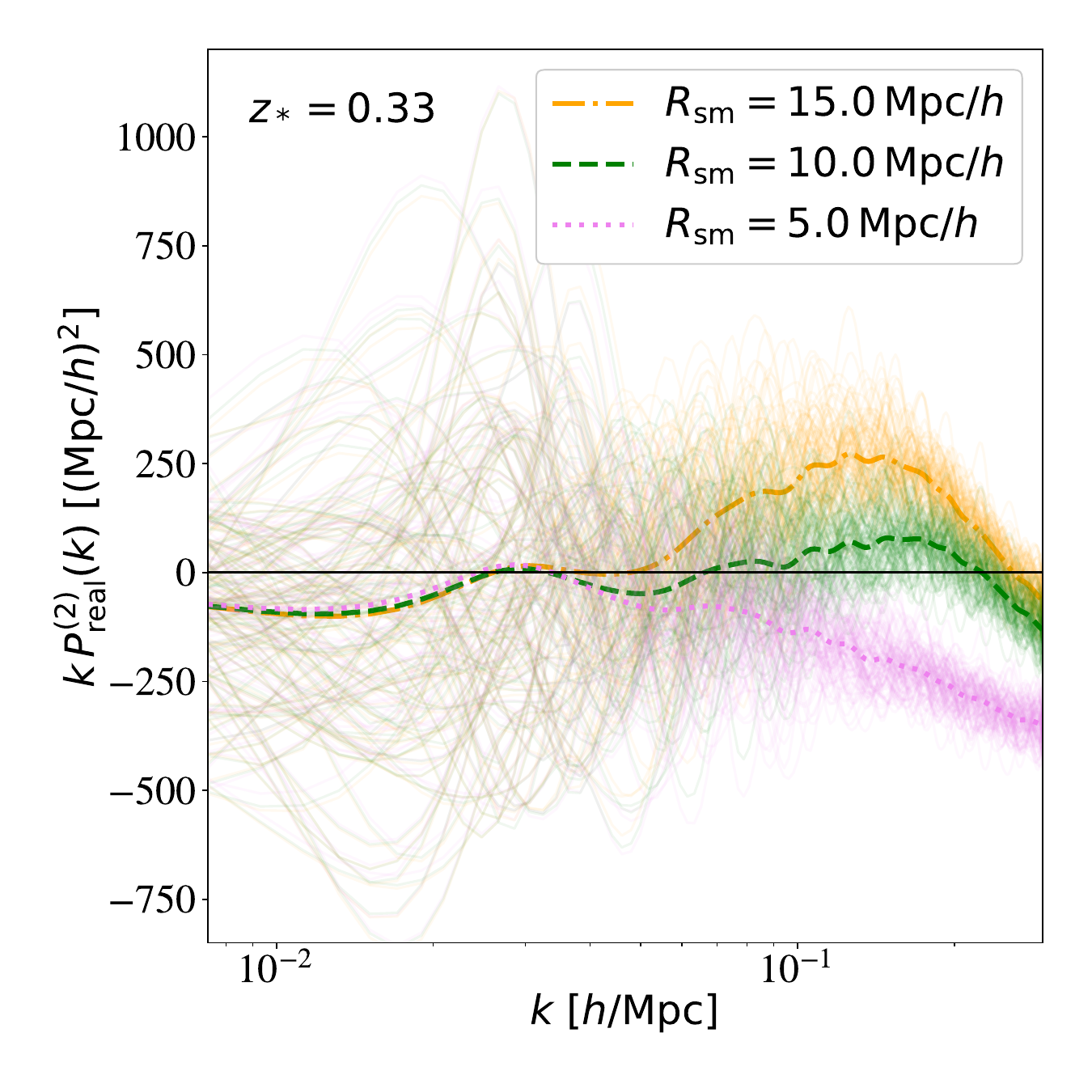}
    \end{minipage}  
    \caption{Left panel: measured monopole and quadrupole from \Nmocks LOWZ SGC mocks. Thin lines correspond to individual measurements, thick lines to their mean. Right panel: reconstructed real-space quadrupoles for different choices of smoothing scales adopted in the RSD-blinding procedure.}
    \label{fig:powerspectraRSD}      
\end{figure}

The left panel of figure \ref{fig:powerspectraRSD}  shows the measurements obtained from \Nmocks LOWZ mocks for three different cases. The blue (solid) lines correspond to  the multipoles measured from the original catalogs. The green (dashed) lines correspond to the catalogs obtained after removing the redshift-space component as an intermediate step during the reconstruction procedure using the reference growth rate and galaxy bias. The quadrupole is almost zero, which demonstrates that reconstruction efficiently removes the RSD-induced anisotropy on large scales. As explained in section \ref{ch:blinding_RSD}, for the blinding procedure we use the displacement field obtained from the reconstructed real space positions to add a new RSD  contribution corresponding to the shifted growth rate. The multipoles measured from  the  blinded catalogs are shown in red (dotted) lines. 

One important ingredient of the blinding procedure is the  choice of the smoothing scale $R_\mathrm{sm}$ used to filter the discrete galaxy over-density field. For too-large value of $R_\mathrm{sm}$ the small scale perturbations are washed out, and thus ignored. On the other hand, for a too-small value of $R_\mathrm{sm}$, the obtained field is dominated by nonlinear velocities, which cannot be efficiently modelled. Hence, the choice of smoothing scale implies a trade-off between these effects. 

The right panel of  figure \ref{fig:powerspectraRSD} shows the reconstructed  real space power spectra obtained for three different smoothing scales. For $R_\mathrm{sm}=15 \, \mpcoh$ we see a bump for $0.05 <k\, [\, \hompc]<0.20 $ indicating that anisotropies due to RSD are not properly removed. In the same $k$-range, a smoothing scale of $R_\mathrm{sm}=5 \, \mpcoh$ leads to a negative quadrupole, suggesting that it over-predicts the linear RSD anisotropies because of the superimposed nonlinear velocities. We find that a value of  $R_\mathrm{sm}=10 \, \mpcoh$ optimally reduces both effects (for LOWZ and CMASS); thus we  adopt this choice for the rest of this work. This is also in agreement with previous findings as in Ref.~\cite{burden_efficient_2014}. The choice of smoothing scale is further discussed in  appendix \ref{sec:effect-smoothing} where we show that the procedure presented here is robust to the choice of the smoothing scale.

\begin{figure}[t]
    \centering
    \includegraphics[width=\textwidth]{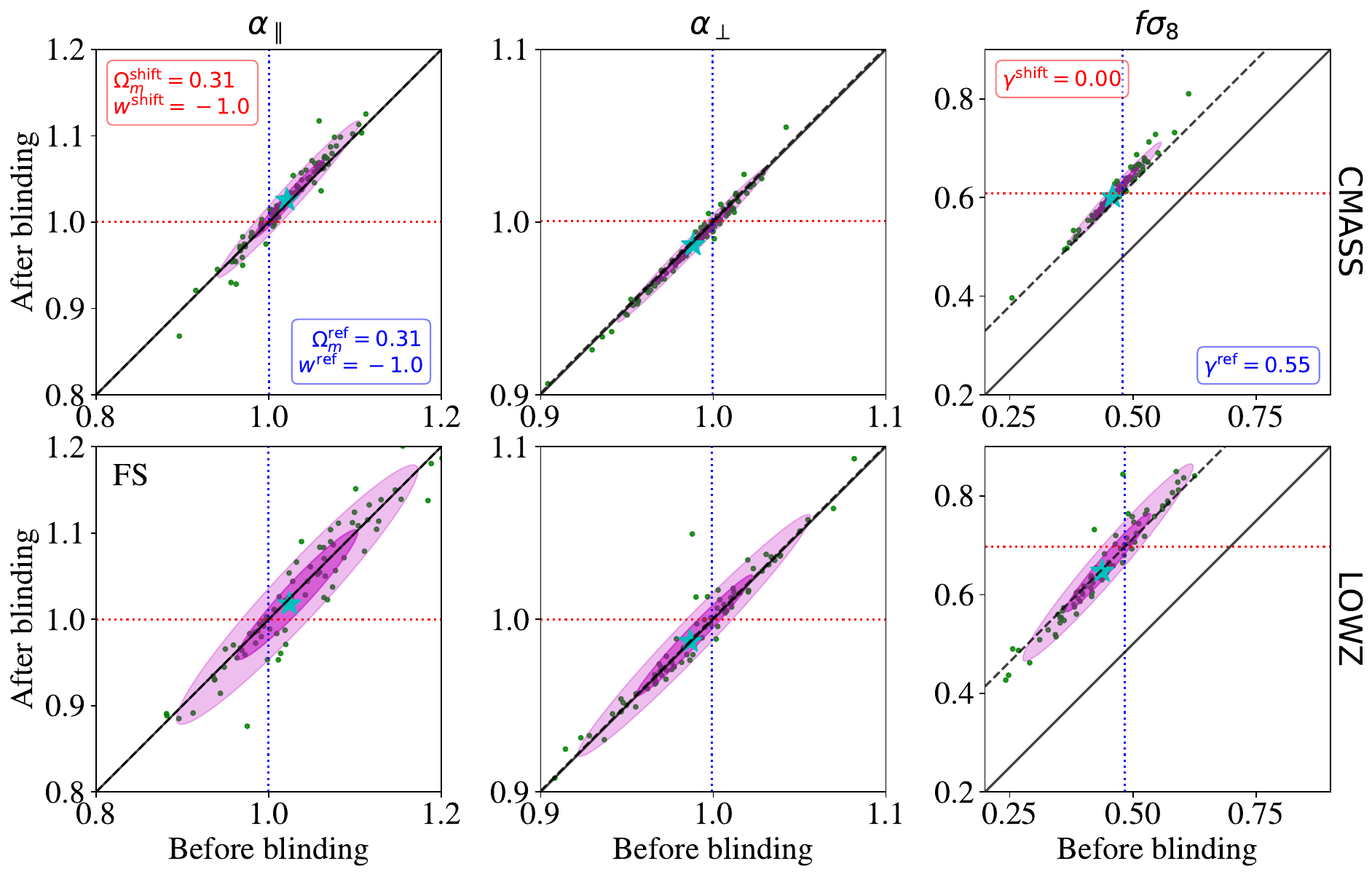}
    \includegraphics[width=\textwidth]{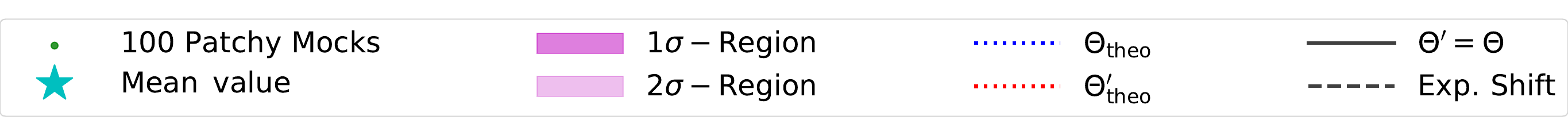}
    \caption{Effect of RSD-only shift blinding on full shape analysis results. Best-fit physical parameters $\alpha_\parallel$ (left panels), $\alpha_\perp$ (middle panels) and $f \sigma_8$ (right panel) from \Nmocks mock realisations (green dots) and their mean values (cyan stars) for the underlying (reference) cosmology and the blinded cosmology. Upper panels show the measurements from the CMASS mocks, lower panels the LOWZ mocks and the purple regions give the $1-2\sigma$ spread of the mocks. The dashed line shows the expected shift of the mock distribution from the diagonal due to blinding. The blue dotted lines represent the underlying values corresponding to the \textsc{MD-Patchy} mocks cosmology, whereas red dotted lines show the ``underlying'' values of the blinded catalogs.} 
    \label{fig:alphascatterFSdiffz}
\end{figure}

We apply the full shape model introduced in \S~\nameref{ch:basics_fs} to the blinded mocks  and estimate the relevant parameters.
The results for the recovered parameters $\mathbf{\Theta} = \lbrace \alpha_\parallel, \alpha_\perp, f\sigma_8 \rbrace$ %of the mcmc's applied to  
for each individual mock are summarised in figure~\ref{fig:alphascatterFSdiffz}.

Since the AP-like shift is not included here, we expect no deviation due to blinding for $\lbrace \alpha_\parallel, \alpha_\perp \rbrace$. As we can see from the two left columns, on average  this is indeed the case: statistically, the recovered parameters from  the blinded mocks follow this expectation. This shows  that  RSD-like shifts along the LOS are not degenerate with the AP-like shifts: the induced change of  $f$ due to blinding does not propagate into a systematic bias on $\alpha_\parallel$,  $\alpha_\perp$.

The expected change  due to blinding of $f \sigma_8$ (see right columns) is given by the dashed line parameterised as
\begin{align} \label{eq:dashed_line_RSD}
y(x) = x + (f\shift - f\reference)\sigma_8\reference ~,
\end{align} 
corresponding to Eq.~\eqref{eq:mapping_result} multiplied by the matter density fluctuation amplitude of the reference template  $\sigma_8$.\footnote{Note that the  method presented here does not allow one to  blind for the value of $\sigma_8$ itself, as we only shift radial and not angular positions.}
This can be understood by considering that, in standard RSD analyses, $f$ and $\sigma_8$ are completely anti-correlated, and the measurements are thus sensitive  to their product only.
As stated in section \ref{ch:blinding_RSD}  here we decide to blind for $f$ coherently following Eq.~\eqref{eq:growth_rate}, but this choice is not 
unique and could be replaced by a relation that also takes into account the different evolution of $\sigma_8$ with $\gamma\shift$, e.g., by rescaling $f\shift$ accordingly.

Note  that here we have deliberately chosen to show a very extreme shift towards $\gamma\shift=0$. As presented in table \ref{tab:shiftedparams}, this corresponds to a $20\%$ change of the growth rate parameter $f$ for CMASS and almost $50\%$ for LOWZ galaxies (recall that effective redshifts are $z_*=0.33$ for  LOWZ and $z_*=0.6$ for CMASS). 

This consideration is key to understand the different behaviour of the mocks in figure~\ref{fig:alphascatterFSdiffz}. While the blind CMASS catalogue matches the expectation very accurately,  the value of $(f\sigma_8)\blind$ is over predicted by $\sim 1\sigma$ for the LOWZ sample.
The chosen value of $\gamma\shift=0$, and hence $\Delta \gamma=0.55$ indicates the upper limit for an accurate prediction for the growth rate after blinding. 
This extreme case, while very illustrative, should not be applied if possible  in any realistic blinding scenario. The resulting catalog would still be blind, but the condition to be able to ``unblind parameter values without the need to rerun the analysis pipeline" would not be satisfied.

The predictability of the blinding procedure proposed here for shifted values deviating by \sout{(at least)} up to $20\%$ should be more than enough given the anticipated precision level of future galaxy surveys at all redshifts.

Another important observation from figure \ref{fig:alphascatterFSdiffz} is that even mocks that are distant from the dashed blue line follow eq. \eqref{eq:dashed_line_RSD} very well. Therefore we conclude that even for mocks with different best fit parameters with respect to the reference ones,  the blinding shift can be well predicted.

\subsection{Combined shift Performance}\label{ch:mocks_comb}

Having shown the performance of the AP-like and the RSD shift individually, here we investigate the performance when both shifts are combined. This provides a  more realistic blinding scenario, closer to the way we envision it to be applied to real data.

The first question to be addressed is how both shifts shall be combined.  While in principle the shifts can be applied in any order, in practice each operation on the catalog relies on different inputs, different assumptions and has different optimisation requirements, and thus the order does matter.

In particular, the AP-like shift $\Delta^{\mathrm{AP}} z$ is cosmology-independent , as it  does not depend  on $\mathbf{\Omega}\reference$, but only  on the input $\Delta \mathbf{\Omega}\shift$ propagating directly into the $\alpha$-parameters which describe generically the expansion history. 
While for this application we have decided to adopt a  $w$CDM-type expansion history, it could incorporate even more general models such as varying dark energy or modified gravity.

On the other hand the RSD shift $\Delta^{\mathrm{RSD}} z$ relies on additional assumptions. The RSD blinding algorithm requires, as input, reference values for $\lbrace \Omega_\mathrm{m}\reference, w\reference \rbrace$ to convert redshifts to comoving distances, and $\lbrace f\reference, b\reference, R_\mathrm{sm}\rbrace$ to find a solution for the reconstructed smoothed real-space over-density. Although the impact of an unphysical reference cosmology on the obtained blinded catalog is small, it is important to recognise this conceptual difference with respect to the AP-like shift, which is independent of $\mathbf{\Omega}\reference$. 

In this case, if we were to  first apply to the catalogue $\Delta^{\mathrm{AP}} z$, followed by $\Delta^{\mathrm{RSD}} z$, the intended deviation from the reference cosmology induced by $\Delta^{\mathrm{AP}} z$ would propagate into the calculation of $\Delta^{\mathrm{RSD}} z$. We find that this leads to small but unnecessary biases contributing to the computation of the final blinded catalog. Therefore, a more natural  and recommended choice consists in applying the shifts in reversed order. First, run the reconstruction-like algorithm with input parameters tuned to catalogs generated from a reference cosmology
that match key properties of the sample. Second, take the RSD-blinded catalog and convert redshifts to distances using a shifted cosmology (AP-like shift). 
Any biases that might occur during the first step (RSD shift) do not propagate into the second step (AP-like shift), because the latter is purely geometrical, homogeneous and model-independent. Another point in favour of this order is, that it nicely matches
 the order of steps in which the BAO and RSD analyses is carried out. We \textit{first} convert redshifts to distances using a reference cosmology and \textit{then} measure the summary statistic of choice. The idea of our scheme is to carry out these steps ``backwards" in order to mimic a different cosmology. Therefore, we advocate to use the order ``RSD shift first, AP-like shift second" for the combined blinding shift and we adopt that strategy in what follows.

We blind the mocks according to  the last entry of  table \ref{tab:shiftedparams}. We carry out two different types of analysis: We fit the \nameref{ch:basics_aniso} both to pre-reconstruction (BAO prerecon fit) and post-reconstruction (BAO postrecon fit) mocks to obtain constraints on the scaling parameters in section \ref{ch:comb_bao}.  In section \ref{ch:comb_fs} we show the results obtained for the \nameref{ch:basics_fs} which is used to jointly fit the scaling parameters and growth rate amplitude (FS fit).

\begin{figure}[t]
    \centering
    \begin{minipage}[h]{0.495\textwidth}
    \centering
    BAO prerecon
    \includegraphics[width=\textwidth]{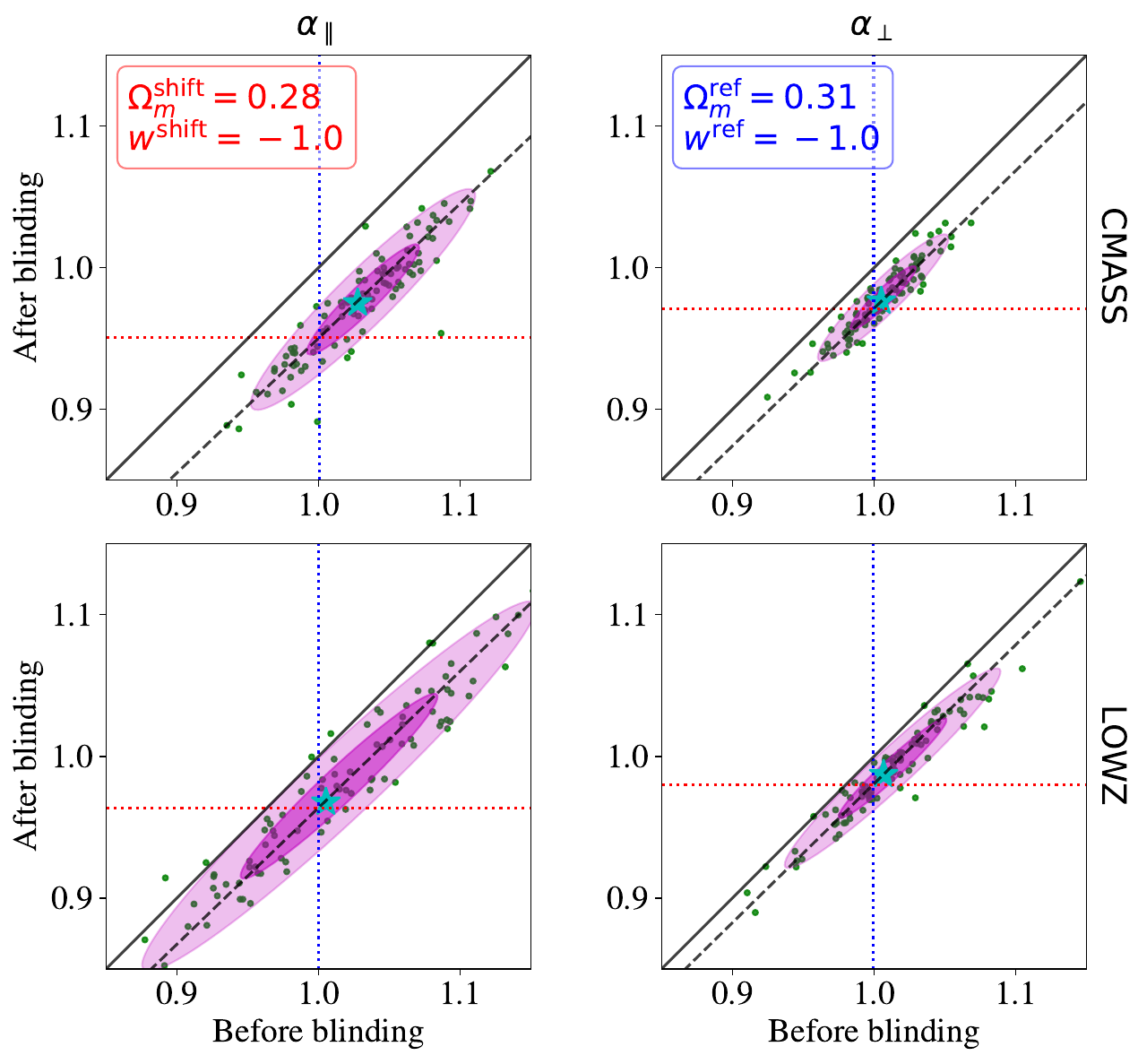}
    \end{minipage}
    \begin{minipage}[h]{0.495\textwidth}
    \centering   
    BAO postrecon
    \includegraphics[width=\textwidth]{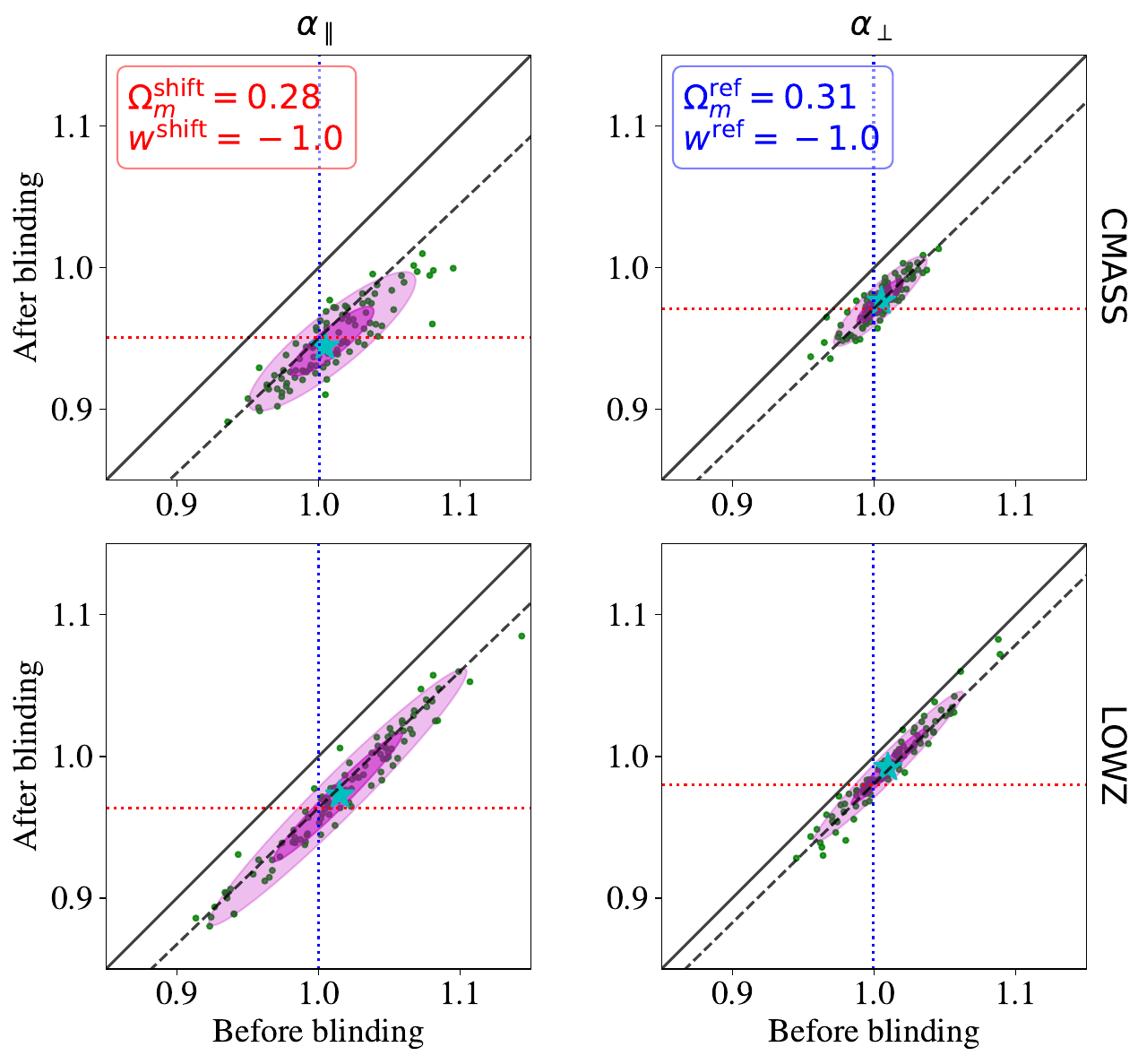}  
 \end{minipage}
 \includegraphics[width=\textwidth]{plots/aniso/leg_alpha_scatter.pdf}  
    \caption{Effect of AP-like+RSD shift blinding  on prerecon (left panels) and postrecon (right panels) BAO-only analyses. The panels show $\alpha_\parallel$  and $\alpha_\perp$ best-fit results for \Nmocks mock realisations (green dots) and their mean values (cyan stars) for the underlying cosmology and the blinded cosmology (using AP only shifts). Upper (lower) panels display the measurements from the CMASS (LOWZ) sample. Purple regions indicate the 68\% and 95\% confidence level regions. Dashed lines represent the expected shift of the mock distribution from the diagonal due to blinding. Blue dotted lines represent the underlying values corresponding to the \textsc{MD-Patchy} mocks cosmology, whereas red dotted lines show the ``underlying'' values of the blinded catalogs.}
    \label{fig:alphascatterBAOdiffzCOMB}   
\end{figure}

\begin{figure}[t]
    \begin{minipage}[h]{0.495\textwidth}
    \centering
    BAO prerecon
    \includegraphics[width=\textwidth]{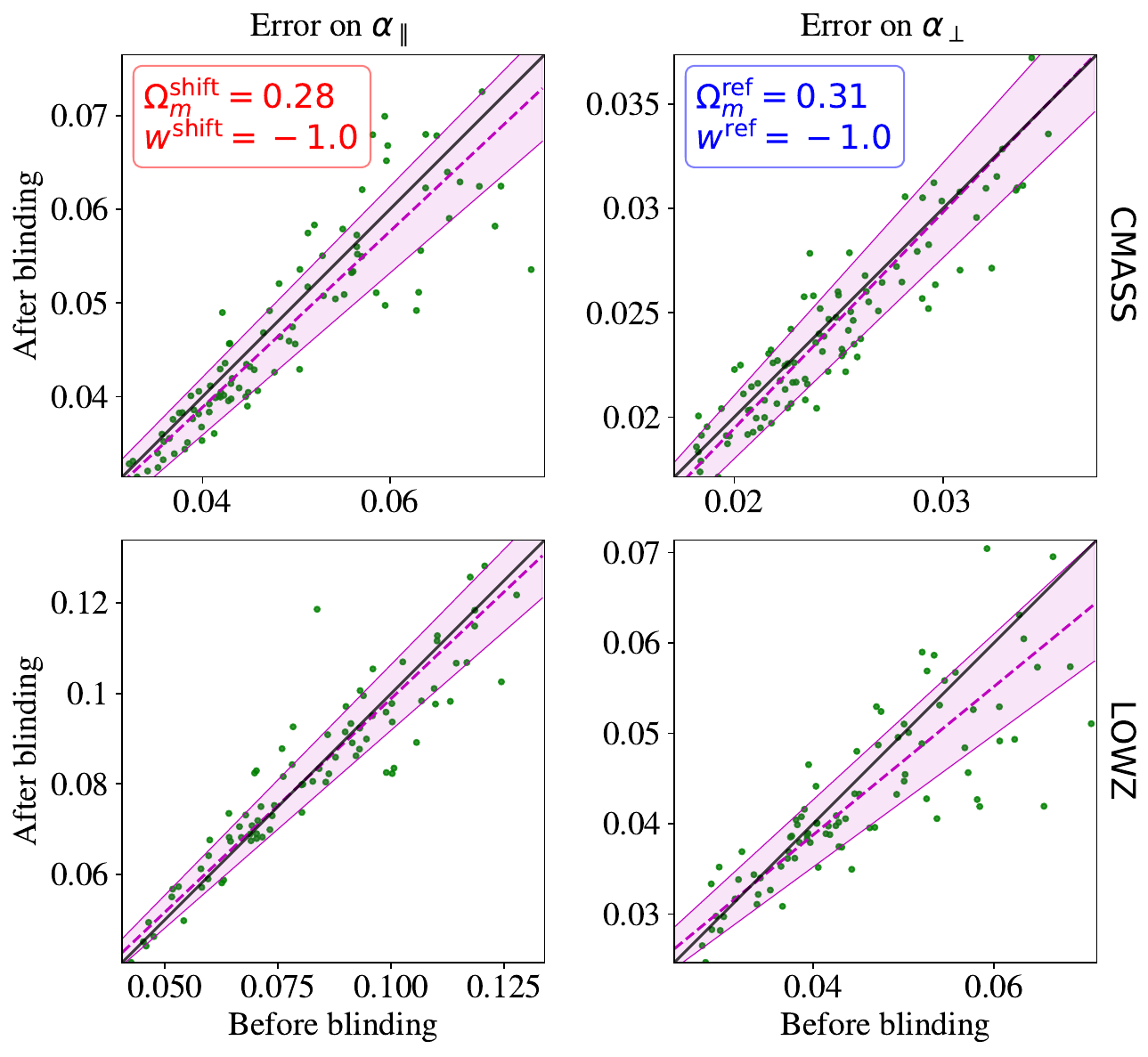}
    \end{minipage}
    \begin{minipage}[h]{0.495\textwidth}
    \centering
        BAO postrecon
    \includegraphics[width=\textwidth]{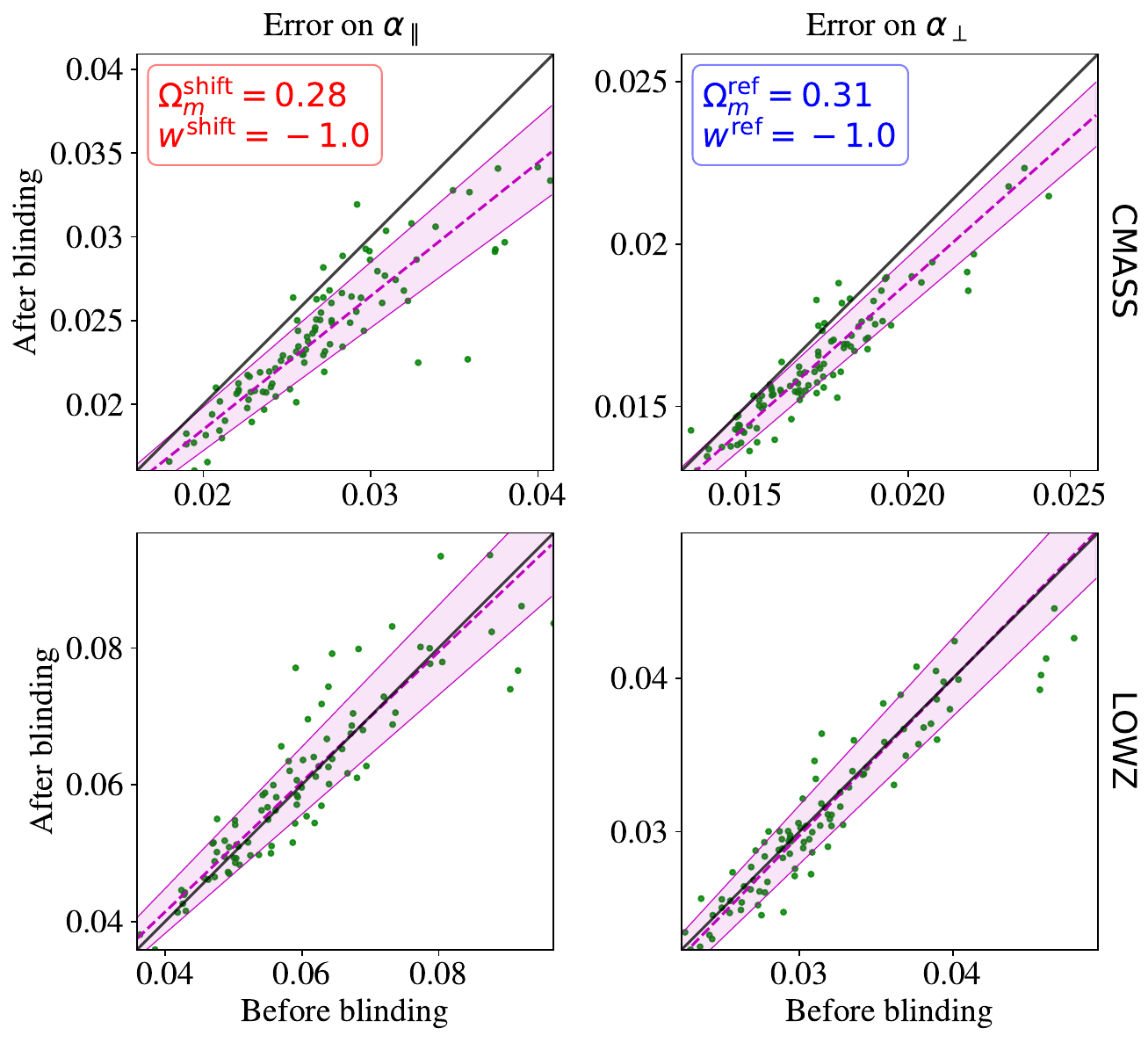}
    \end{minipage}
    \includegraphics[width=\textwidth]{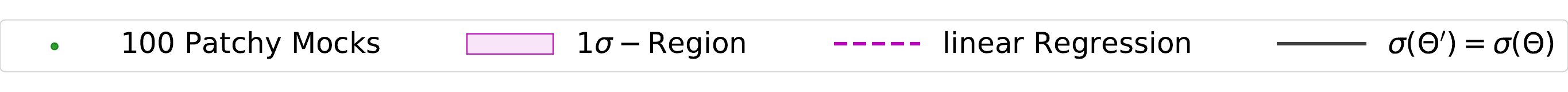}
    \caption{Effect of AP+RSD shift blinding on error-bars obtained from prerecon (left panels) and postrecon (right panels) BAO fits. The shaded regions describe the $1\sigma$-region when fitting a linear regression to the mock distribution.
    }
    \label{fig:sigmaalphascatterBAOdiffzCOMB}
\end{figure}

\subsubsection{BAO analysis results}\label{ch:comb_bao} 

Here we show the results of our BAO prerecon and postrecon fits. For reconstruction, we follow the procedure explained in \S~\nameref{ch:basics_rsd} using an input growth rate $f\reference(z)$ corresponding to $\Lambda$CDM $(\gamma\reference=0.55)$, a galaxy bias of $b\reference=1.85$, and a smoothing scale of $R_\mathrm{sm}=10\,\mpcoh$ both for the blinded and the original mock catalogs. The impact of choosing the same growth rate in both cases is very small as shown in more detail in section \ref{sec:results}.

All BAO fit results obtained from the original and the blinded mocks generated using the combined shift are shown in figure \ref{fig:alphascatterBAOdiffzCOMB}. For the prerecon case (left panels) we see that the agreement between the mock distribution and the prediction (dashed line) is as accurate as that obtained for the AP-only shift in figure \ref{fig:alphascatterBAOdiffz}. This suggests, that the prerecon BAO analysis results are stable against the additional RSD shift included in this section. In other words: modifying the RSD signal at the catalog level leaves unaltered the information extracted from the anisotropic BAO signal, confirming the discussion of the previous subsection. The right panels show that reconstruction reduces the intrinsic scatter along the diagonal (note that the x-axis and y-axis scales are the same for all subplots). This is expected, as reconstruction enhances the BAO peak, which increases the signal-to-noise ratio. Furthermore, reconstruction brings the mean of the mocks (cyan star) closer to the theoretical prediction given by the blue and red lines, in particular for CMASS. 

A potential worry with this procedure is that, as shown in Ref \cite{Sherwin:2018wbu}, using input parameters for the reconstruction procedure significantly different from the underlying ones, may lead to biases in postreconstruction BAO analyses. Here, we find that also in the BAO postrecon case the agreement between mock distribution and expectation is encouragingly good.

In addition to the best-fit results, we are particularly interested in whether the blinding scheme leads to a change of error bars, and hence the significance of the measurement.  Therefore, we show in figure \ref{fig:sigmaalphascatterBAOdiffzCOMB} the errors on the scaling parameters on a mock-to-mock basis with the same setup as in figure \ref{fig:alphascatterBAOdiffzCOMB}. Black lines indicate where the errors of blinded and original mocks coincide, whereas purple dashed lines show a linear regression to the mocks represented by green dots. Shaded contours show the $1\sigma$-region of the linear regression. The scatter of the error bars before and after blinding is consistent with the identity line. Only for the CMASS postrecon case we observe a small trend towards smaller errors after blinding. Again, this could be related to a suboptimal choice of input parameters for reconstruction as for example the galaxy bias. Also note that blind measurements make use of a covariance matrix obtained from unblind mocks, which could propagate into an over- or underestimation of the error bars, depending on the shifted cosmology used to generate the blinded data. But as figure \ref{fig:sigmaalphascatterBAOdiffzCOMB} suggests, this effect is very small. Therefore we argue that the overall agreement is promising and conclude that the blinding procedure does not lead to additional biases corrupting the significance of the blind measurements. 

\begin{figure}[t]
    \centering
    \includegraphics[width=\textwidth]{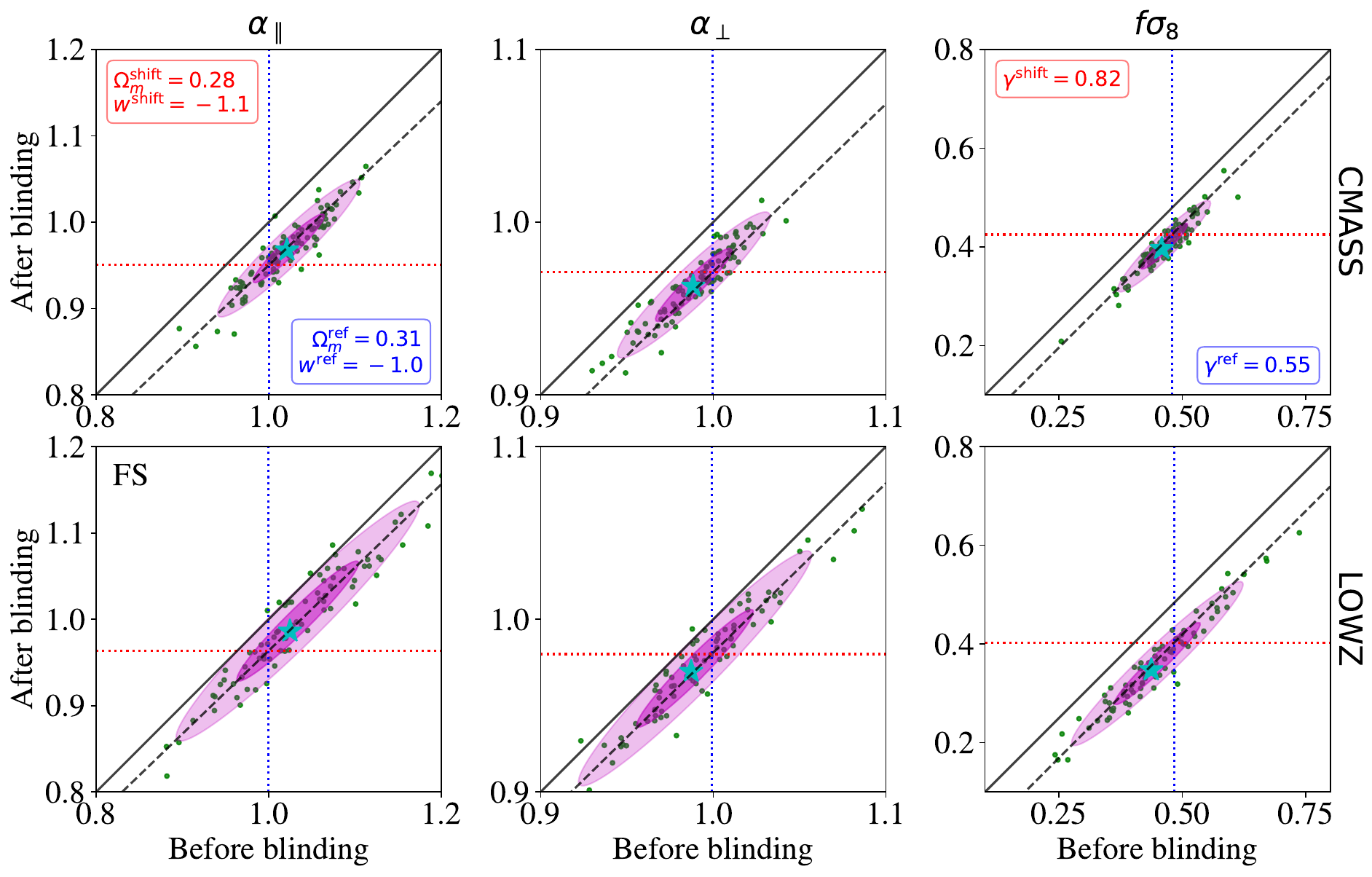}
    \includegraphics[width=\textwidth]{plots/fs/leg_alpha_scatter.pdf}    
    \caption{``Full shape" fits on mocks with combined shift blinding. Comparison of recovered values for $\alpha_\parallel$ (left panels), $\alpha_\perp$ (middle panels) and $f \sigma_8$ (right panel) from \Nmocks mock realisations (green dots) and their mean values (cyan stars) for the underlying (reference) cosmology and the blinded cosmology.  Upper panels show the measurements from the CMASS mocks, lower panels the LOWZ mocks and the purple regions give the $1-2\sigma$ spread of the mocks. The dashed line shows the expected shift of the mock distribution from the diagonal due to blinding. The blue dotted lines represent the underlying values corresponding to the \textsc{MD-Patchy} mocks cosmology, whereas red dotted lines show the `underlying' values of the blinded catalogs.} 
    \label{fig:alphascatterFSdiffzCOMB}
\end{figure}

\begin{figure}[t]
    \centering
    \includegraphics[width=\textwidth]{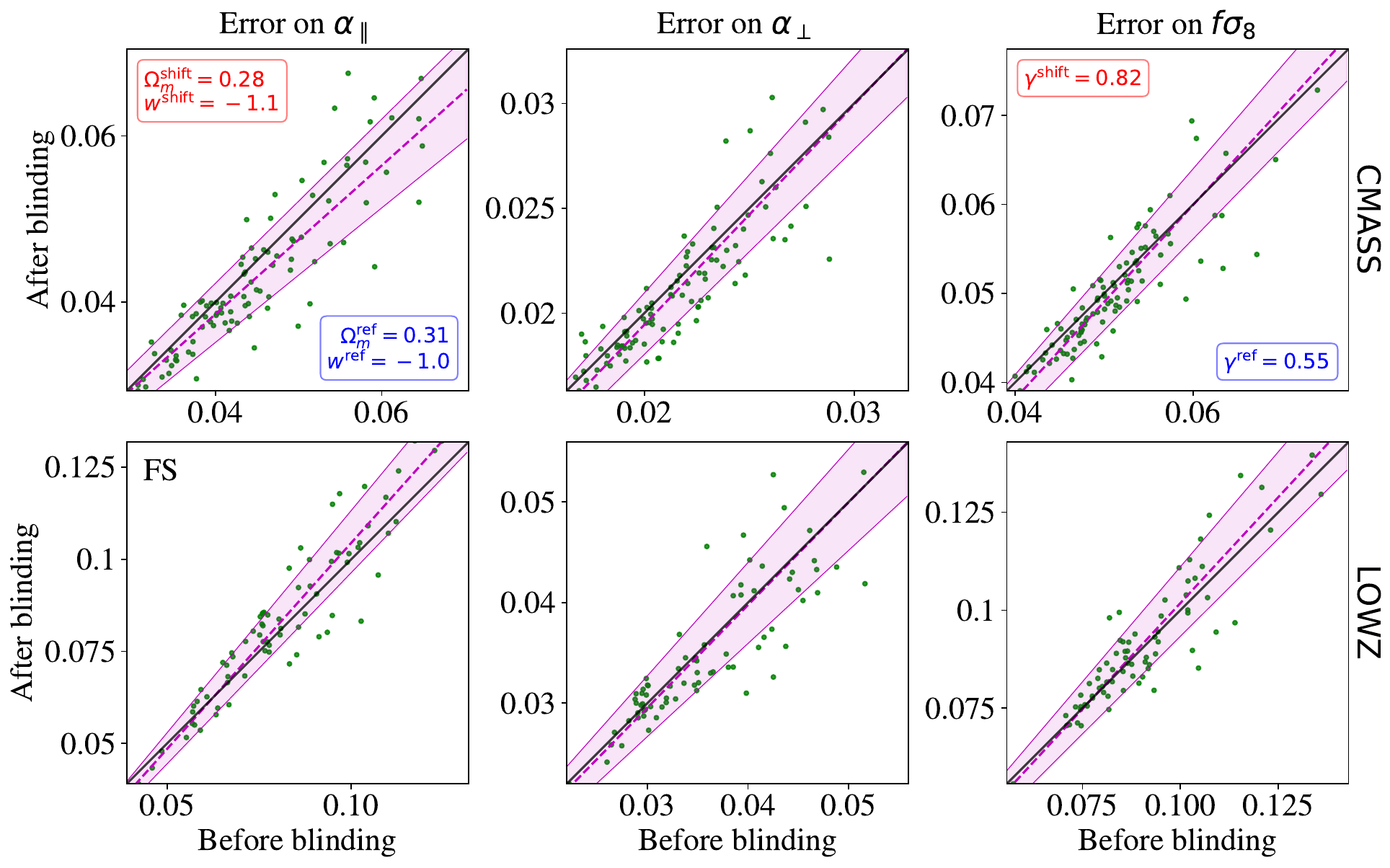}
    \includegraphics[width=\textwidth]{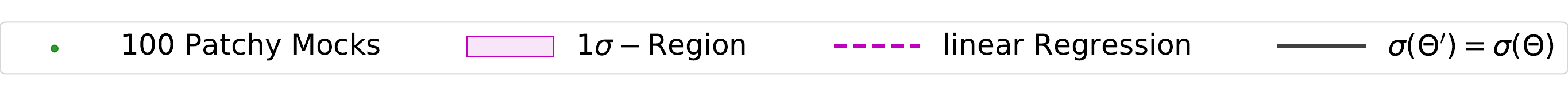}    
    \caption{Effect of  combined  AP+RSD shifts blinding for full shape analysis on the errors-bars.  Errors on $\alpha_\parallel$, $\alpha_\perp$ and $f \sigma_8$ from \Nmocks mock realisations (green dots) for the underlying (reference) cosmology and the blinded cosmology (last entry of table~\ref{tab:shiftedparams}). The errors before and after blinding are centred around the diagonal indicating no systematic change. The dashed purple line is a linear regression to the points and the shaded region its $1\sigma$ errors. }
    \label{fig:sigmaalphascatterFSdiffzCOMB}
\end{figure}

\subsubsection{FS analysis results}\label{ch:comb_fs} 

The results for the FS fits are shown in figure \ref{fig:alphascatterFSdiffzCOMB}. 
As in figure \ref{fig:alphascatterBAOdiffzCOMB}, the green points correspond to the different realisations and the cyan stars represent the mean  of the mocks.
 It is possible to appreciate that the  inferred  parameters from the blinded  mocks  are nicely aligned with the dashed line, which is the expected ``locus".  The fact that the cyan star does not coincide with the intersection of the red and blue dotted line is not due to the blinding procedure (in fact the systematic affects both determinations before and after blinding); it is possibly caused by the modeling itself applied to ``fast" mocks rather than full N-body and to the relatively small number of realisations used. 

Here, we also quantify the effect of the combined blinding on the recovered parameters errors. This is illustrated in figure \ref{fig:sigmaalphascatterFSdiffzCOMB}, for the same set up as in figure \ref{fig:alphascatterFSdiffzCOMB} (combined shifts, FS fit). Again, the black diagonal line indicates where errors before and after blinding coincide and the dashed purple line is a linear regression to the points with $1\sigma$ uncertainty given by the  shaded region.  Similarly to figure~\ref{fig:sigmaalphascatterBAOdiffzCOMB}, the errors before and after blinding are centered around the diagonal indicating no systematic change.
This demonstrates that blinding does not affect significantly the error-bars.

\begin{figure}[t]
    \begin{minipage}[h]{0.31644\textwidth}
    \centering
    %BAO prerecon
	\includegraphics[width=\textwidth]{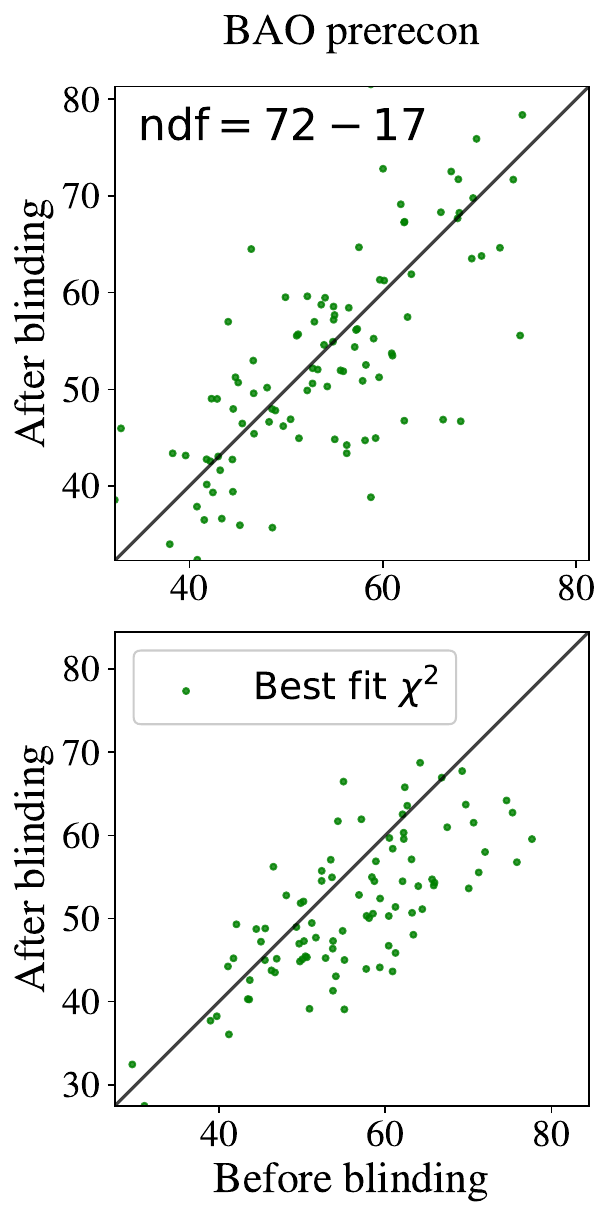}
	\end{minipage}
    \begin{minipage}[h]{0.31644\textwidth}
    \centering
    %BAO postrecon
	\includegraphics[width=\textwidth]{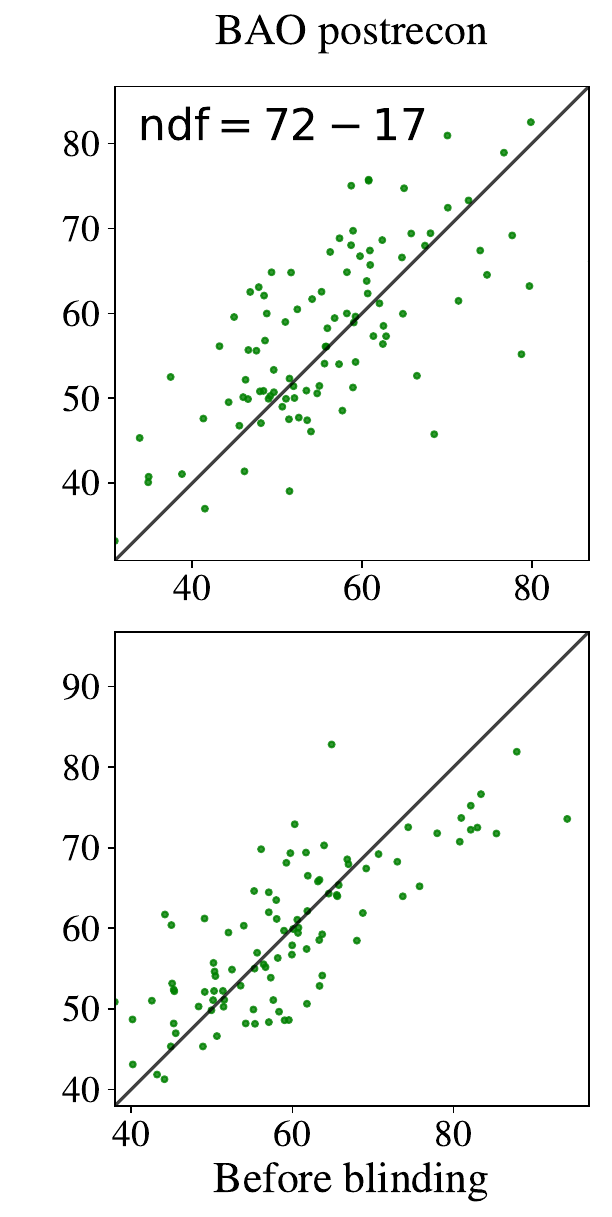}    \end{minipage}
        \begin{minipage}[h]{0.31644\textwidth}
    \centering
    %FS
	\includegraphics[width=\textwidth]{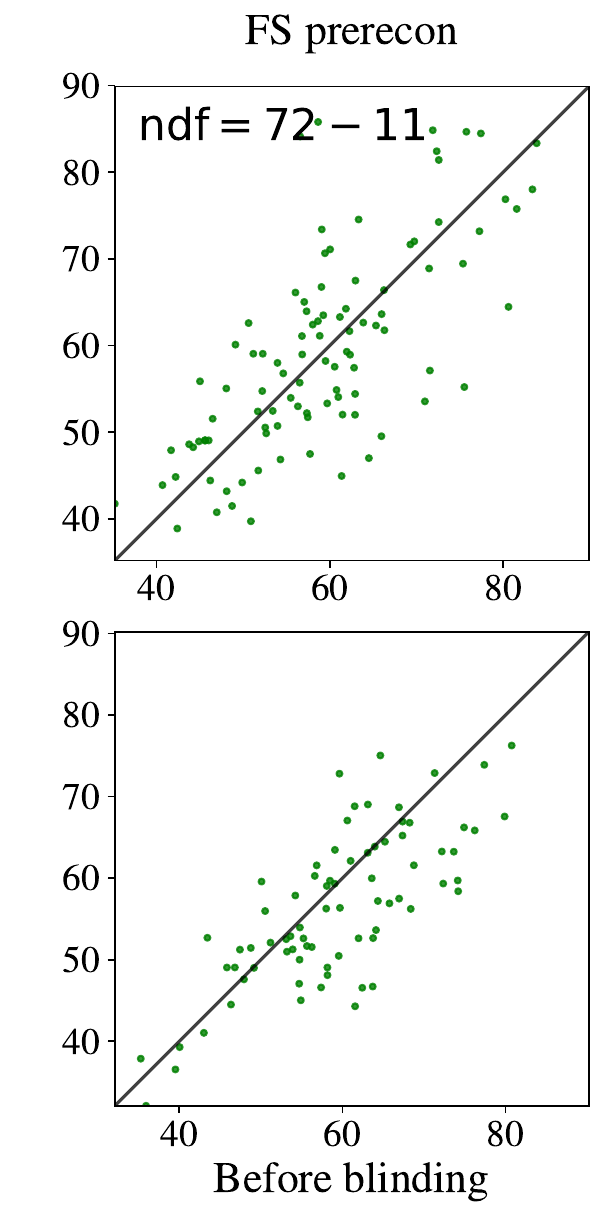}    
    \end{minipage}
    \caption{Effect of AP+RSD shift blinding on $\chi^2$ values from \Nmocks mock realisations (green dots) for all types of analysis considered here.  The left, middle and right panels show the best fit $\chi^2$ for BAO prerecon, BAO postrecon and FS respectively. The top row corresponds to CMASS and the bottom row to LOWZ fits.}
    \label{fig:chiscatterBAOdiffzCOMB}
\end{figure}

\subsubsection{$\chi^2$ before and after blinding} \label{ch:comb_chi} 
Another important consistency check is the comparison of best fit  $\chi^2$ before and after blinding. This is shown in figure \ref{fig:chiscatterBAOdiffzCOMB} for all cases discussed in sections \ref{ch:comb_bao} and \ref{ch:comb_fs}. Again, each pair of mocks is represented by a green dot and the black diagonal shows where the values before and after blinding coincide. We note  that there is an intrinsic scatter of $\chi^2$ among the mocks in a range of about $0.6<\chi^2/\mathrm{ndf}<1.4$. In general we see that the mock distribution is aligned with the black diagonal, meaning that blinding does not systematically change the $\chi^2$. However, there is some scatter around the diagonal, meaning that the blinding shift can either improve or worsen the goodness of fit. Since this scatter is much smaller than the intrinsic scatter of mocks, we argue that it is not relevant. Overall, the alignment along the identity line of $\chi^2$ before and after blinding proves our blinding procedure to be stable enough to be applied to spectroscopic galaxy survey data.

\section{Worked example: application to BOSS DR12 data}\label{sec:results}

We use BOSS DR12 catalog \cite{BOSSDR12} to present  a practical example of applying the blinding procedure proposed in this paper  to real data. 
The Baryon Oscillation Spectroscopic Survey (BOSS) \cite{BOSS1} is part of the  Sloan Digital Sky Survey III \cite{EisensteinBOSS2011}. BOSS measured spectroscopic redshifts  \cite{Bolton12,smee13} for more than 1 million galaxies in an effective volume of $V_\mathrm{eff} = 7.4\,\mathrm{Gpc}^3$. The galaxy survey includes the  LOWZ sample  with 361 762 galaxies \cite{alamdr12} covering the redshift range of $0.15<z<0.43$  and the CMASS sample, with 777 202 galaxies  at $0.43<z<0.70$.  

We proceed by applying all the steps presented in sections \ref{ch:blinding_AP} and \ref{ch:blinding_RSD} to create a blinded catalog. We assume the same reference cosmology as given in table \ref{tab:cosmoparams} and use the shifted cosmology of the ``Combined" case given in table \ref{tab:shiftedparams}. We use the shifted redshift cuts convention explained in section \ref{ch:mocks_AP}. Hence, the blinded catalog contains $334\,053$ LOWZ galaxies in the redshift range $0.17<z<0.43$ and $746\,889$ CMASS galaxies in the range $0.46<z<0.73$; this corresponds to an effective redshift of $0.34$ for LOWZ and $0.62$ for CMASS. Once obtained the blinded catalog, we create a synthetic random catalog matching its number density distribution with redshift but with 50 times the number of objects. 
This random catalog is used to calculate the blinded survey window function. For the power spectrum measurement we proceed similarly to section \ref{ch:mocks_methodology} with the only difference being  that we use a finer grid with $1024^3$ cells for the same volume.

The catalogs contain a certain set of weights to account for several observational effects: a redshift failure weight ($w_\mathrm{rf}$), a fibre collision weight  ($w_\mathrm{fc}$) and a weight  ($w_\mathrm{sys}$) incorporating angular systematics such as seeing conditions and stellar foreground \cite{2012MNRAS.424..564R}. These weights are common to the original and blinded catalogs. The combined weight for each individual target galaxy is given as
\begin{align}
w_c = w_\mathrm{sys} (w_\mathrm{rf} + w_\mathrm{fc} -1)~.
\end{align}
For the power spectrum measurements we take the combined weight $w_c$  into account as described in \S~\nameref{ch:basics_ps}.

\begin{figure}[b]
    \centering
    \begin{minipage}[h]{0.495\textwidth}
    \centering
    \textbf{LOWZ} \\
    \includegraphics[width=\textwidth]{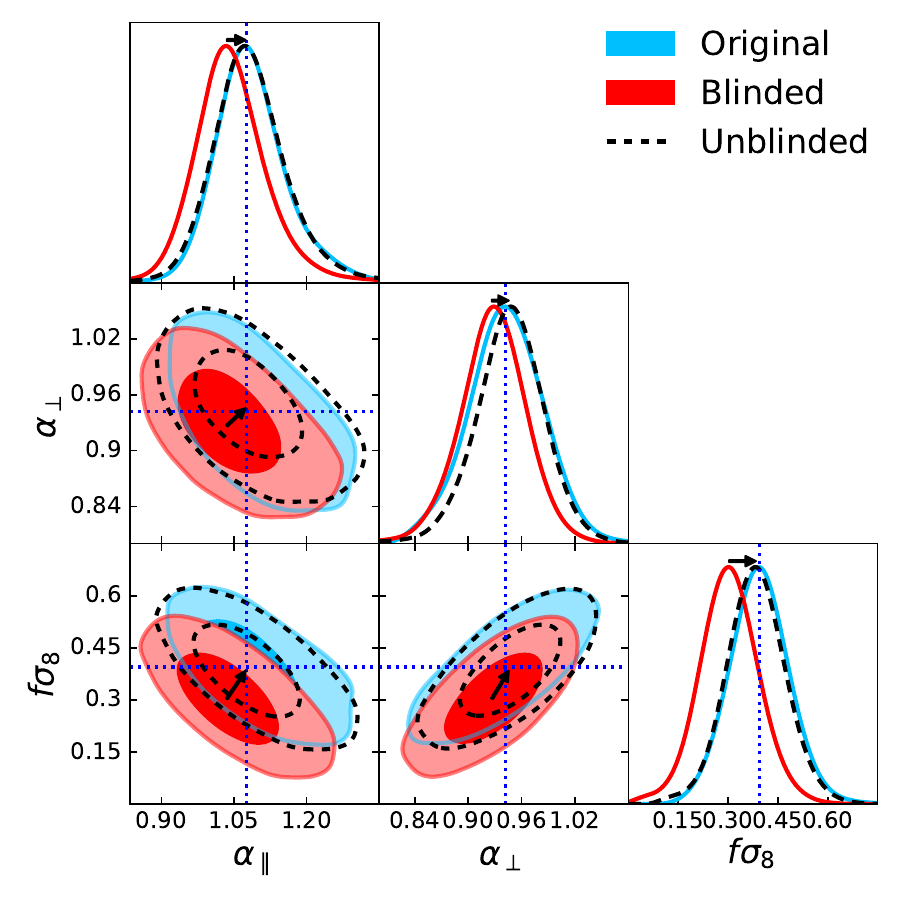} 
    \end{minipage}
     \hfill
    \begin{minipage}[h]{0.495\textwidth}
    \centering
    \textbf{CMASS} \\
    \includegraphics[width=\textwidth]{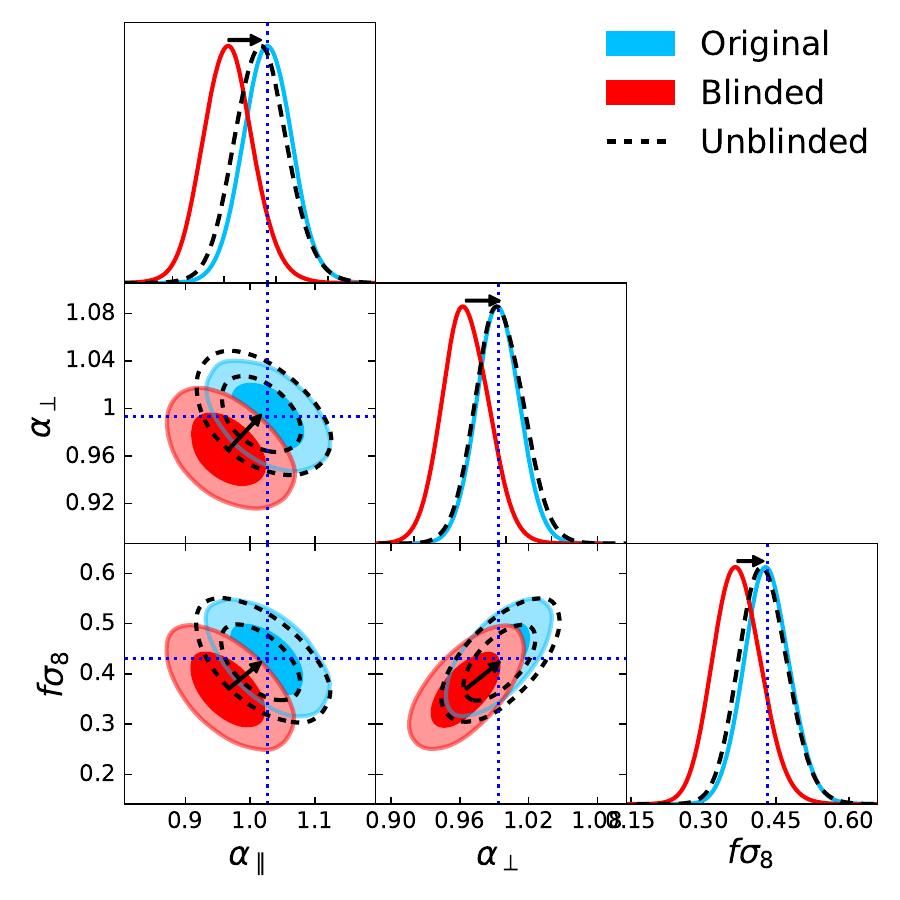} 
    \end{minipage}    
    \caption{Full shape fit to the combined shift blinding applied to the BOSS DR12 data. 68\% and 95\% confidence level marginalised posteriors of the physical parameters $\alpha_\parallel$, $\alpha_\perp$ and $f \sigma_8$ obtained when fitting the FS model to the blinded (red contours) and the original catalog (blue contours) from LOWZ (left) and CMASS (right) galaxy samples. The dashed contours are obtained by correcting the blinded posteriors by eqs. \eqref{eq:dashed_line} and \eqref{eq:dashed_line_RSD} (fast unblinding) as indicated by the arrows. Blue dotted lines mark the best-fit parameters obtained after rerunning the FS analysis on the unblinded catalog.}
    \label{fig:triangleFS}
\end{figure}

For the fits we use a wave number range of $0.02<k\,[\hompc]<0.20$ for the monopole and $0.04<k\,[\hompc]<0.20$ for the quadrupole. The cut on the largest scales is chosen  to be more conservative than in the case of the \textsc{MD-Patchy} mocks due to the known presence of  systematic effects in the quadrupole for scales larger than $k=0.04\, \hompc$ as shown in \cite{gil-marin_clustering_2016-1}.

For the purpose of testing the effect of blinding on reconstruction we apply the reconstruction algorithm described in \S~\nameref{ch:basics_rsd} on the blinded catalogs. We use the reference cosmology $\mathbf{\Omega}\reference$ to convert redshifts to distances, a galaxy bias of $b\reference=1.85$, and a smoothing scale of $R_\mathrm{sm}=10\,\mpcoh$. We create two sets of reconstructed catalogs using two different reference cosmologies. In one set we use the reference value of the growth rate $f(z) = f\reference (z)$ corresponding to $\gamma=0.55$ while in the other set  we use the shifted values (see table \ref{tab:shiftedparams}) corresponding to a growth history with $\gamma=0.825$,  which is that of our target blind  cosmology.  We expect the choice of the reference cosmology to deliver a more precise measurement on the reconstructed catalogs, as it would correctly  ``undo" the RSD signal induced by blinding.
To quantify the effect of a sub-optimal choice of $f$ we present results for these two representative values.

\begin{figure}[t]
    \centering
    \begin{minipage}[h]{0.495\textwidth}
    \centering
    \textbf{Reference growth rate} \\
    \includegraphics[width=\textwidth]{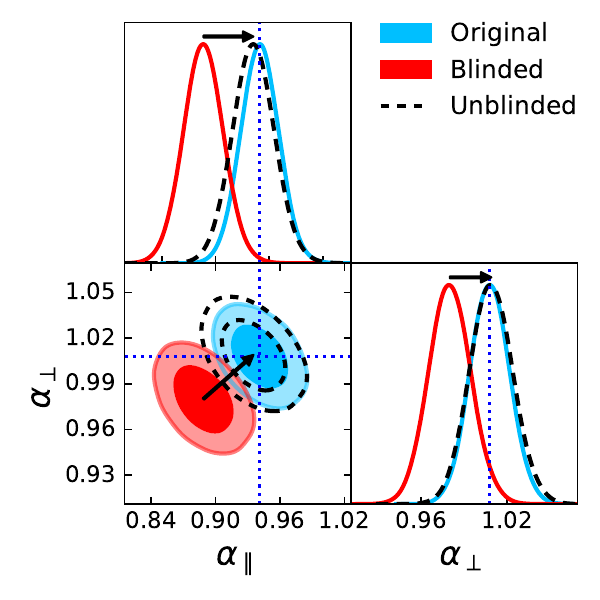} 
    \end{minipage}
     \hfill
    \begin{minipage}[h]{0.495\textwidth}
    \centering
    \textbf{Shifted growth rate} \\
    \includegraphics[width=\textwidth]{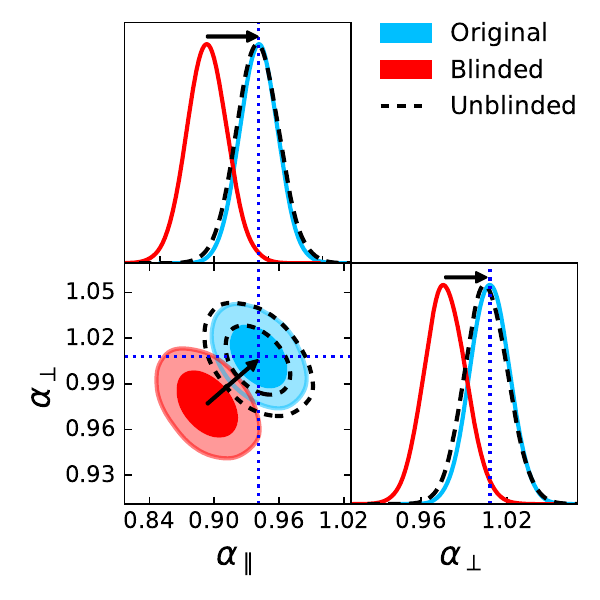} 
    \end{minipage}    
    \caption{Post-reconstruction anisotropic BAO fit to the combined shift blinding applied to the BOSS DR12 data. The figure shows 68\% and 95\% confidence level marginalised posteriors of $\alpha_\parallel$ and $\alpha_\perp$ obtained when fitting the BAO anisotropic model to the blinded (red contours) and the original CMASS catalog (blue contours) post-reconstruction. In the left panel the same reference growth rate $f=0.784$ was used, while in the right panel we performed reconstruction on the blinded catalog with $f=0.695$ corresponding to the growth history we aimed to blind for. The dashed contours are obtained by correcting the blinded posteriors by eqs. \eqref{eq:dashed_line} and \eqref{eq:dashed_line_RSD} (fast unblinding) as indicated by the arrows. Blue dotted lines mark the best-fit parameters obtained after rerunning the BAO analysis on the unblinded catalog.}
    \label{fig:triangleBAO}
\end{figure}

We proceed by fitting the FS model to the blinded catalogs and the BAO anisotropic model to the reconstructed blinded catalogs with the same parameters and prior ranges as given in table \ref{tab:priors}.  
Sometimes in the BOSS literature, these get referred to as ``pre-recon" and ``post-recon" analyses,  respectively. We fit the BAO anisotropic model to the mean of 1000 reconstructed original mock realisations with varying nonlinear damping scales. As a result we find
\begin{equation}
\begin{aligned}
      \Sigma_\parallel^\mathrm{LOWZ} &=  8.6  \, \mpcoh \qquad &
      \Sigma_\perp^\mathrm{LOWZ} &= 5.8 \, \mpcoh ~, \\
      \Sigma_\parallel^\mathrm{CMASS} &=  4.3  \, \mpcoh \qquad &  
      \Sigma_\perp^\mathrm{CMASS} &= 4.3  \, \mpcoh~.
\end{aligned}
\end{equation}
We keep these values fixed when analyzing the data, both for the blinded and the original catalogs. Again, this is to follow the good practice of treating the blinded data in the same way as if we would not know that it is blinded.

In figure \ref{fig:triangleFS} we present blinded two-dimensional marginalised constraints (red contours)  from the  FS analysis for  model parameters $\alpha_\parallel$, $\alpha_\perp$ and $f \sigma_8$ for the LOWZ (left panel) and CMASS samples (right panel). In a similar fashion, we show the results for the BAO anisotropic model parameters $\alpha_\parallel$ and $\alpha_\perp$ fitted to reconstructed CMASS catalogs in figure \ref{fig:triangleBAO}.  On the left we use the reference value of the growth rate $f\reference$ and on the right the shifted value $f\shift$ for reconstruction. 

Next, we unblind in two steps as anticipated in figure \ref{fig:flowchart}. First, from our knowledge of $\Delta \mathbf{\Omega}$ we can directly use eqs. \eqref{eq:alphasresult} and \eqref{eq:mapping_result} to predict the expected underlying results:

\begin{equation}\label{eq:combinedresult}
\begin{aligned}
\alpha_\perp(z_\eff) &= \frac{D_M\shift(z_\eff)}{D_M\reference(z_\eff)} \alpha_\perp^\prime(z_\eff^\prime) ~, \\
  \alpha_\parallel(z_\eff) &= \frac{H\reference(z_\eff) }{H\shift(z_\eff) } \alpha_\parallel^\prime(z_\eff^\prime) ~, \\
 f\underlying &= f\blind - f\shift +  f\reference  ~,
 \end{aligned}
\end{equation}
where the primed quantities are measured from the blinded data.
These shifts are indicated by the black arrows and yield the dashed contours in figures \ref{fig:triangleFS} and \ref{fig:triangleBAO}. Then we repeat the analysis on the unblinded (original) catalogs with the same settings as for the blinded catalogs. The results are represented by the blue contours in figures \ref{fig:triangleFS} and \ref{fig:triangleBAO}, where the position of the corresponding best fit parameters is represented by the  vertical dotted blue lines.

  LOWZ and  CMASS samples have been blinded coherently,   but LOWZ is at lower redshift and  covers a smaller effective volume; as a result  the blinding-induced shift is less significant in LOWZ than in CMASS. 

Figure \ref{fig:triangleFS} shows an excellent agreement between the best-fit scaling parameters  obtained  from the full shape analysis of the original  catalog and those  obtained  from the blinded catalog  after  correction of the estimated posterior.  For  $f\sigma_8$ we observe a larger difference between the measured and predicted, but this is a small fraction of a standard deviation, which in any case is consistent with the effect of sample variance, as observed in the mocks (see fig. \ref{fig:alphascatterFSdiffzCOMB}). The size and shape of the confidence regions is also well recovered by the  blinded posterior after  correction.

Figure \ref{fig:triangleBAO}  highlights the  agreement of  mean values and confidence regions of the scaling parameters obtained post reconstruction from the anisotropic BAO analysis of the original  catalog and of the blinded catalog after  correction. The mean values of scaling parameters are recovered  particularly accurately after unblinding (keeping in mind that reconstruction step has been implemented on the blinded catalog).
The figure also shows that the choice of the adopted value of the growth rate parameter for reconstruction is not important: the difference between the underlying and blinding values, $\Delta f\sim 0.09$ does not lead to a significant change for the mean values nor for the error bars. We do observe that the choice of the shifted growth rate delivers a slightly better match between the corrected and the original posteriors (especially regarding the shape and not so much the position of the peak), but this can be subject to statistical fluctuations.

The results of all fits described in this section are presented in table \ref{tab:dataresults}. As already shown in figures \ref{fig:triangleFS} and \ref{fig:triangleBAO}, the recovered mean values by applying the ``correction" of eq. \eqref{eq:combinedresult} to blinded catalogs agree very well with the actual measurements from the original catalogs. Furthermore, we see that they do not significantly change the error bars. We observe differences of the best-fit $\chi^2$ between blinded and original catalogs,  but there is no preference towards increasing or decreasing it: for LOWZ it decreases after unblinding, while for CMASS it increases. We argue, that this is likely to be a statistical fluctuation, that we also observed for pairs of blinded/original mock catalogs (see for instance figure \ref{fig:chiscatterBAOdiffzCOMB}). All results obtained from the original catalogs are within $1-\sigma$  agreement with previous analyses such as \cite{gil-marin_clustering_2016b}. Though, small difference sourced from the different choice of redshift cuts is expected.

Finally, table \ref{tab:dataresults} as well as figures \ref{fig:triangleFS}  and \ref{fig:triangleBAO} demonstrate that the blinding procedure presented here is suitable to be applied to a real survey and it comfortably meets  all  the criteria  we initially  required for a good blinding scheme.

\begin{table}[h]
\centering
    \resizebox{\textwidth}{!}{
\begin{tabular}{ccc|ccc|c}
Fit                                                                                                       & Sample                 & Case  & $\alpha_\parallel$ & $\alpha_\perp$ & $f \sigma_8$ & $\chi^2/\mathrm{ndf}$   \\ \hline \hline 
\multirow{6}{*}{FS}                                                                                         & \multirow{3}{*}{LOWZ}  & blinded   & $1.046 \pm 0.079$  & $0.927 \pm 0.042$ &$0.303 \pm 0.091$ &  $45.3/(68-11)$ \\ 
                                                                                                                    &                        & unblinded & $1.085 \pm 0.082$  & $0.945 \pm 0.043$ &$0.383 \pm 0.091$ &  $45.3/(68-11)$ \\ 
                                                                                                                     &                        & original  & $1.086 \pm 0.077$  & $0.943 \pm 0.043$ &$0.398 \pm 0.092$ &  $46.5/(68-11)$ \\ 
                                                                                                                     & \multirow{3}{*}{CMASS} & blinded   & $0.966 \pm 0.040$  & $0.965 \pm 0.020$ &$0.370 \pm 0.050$ &  $38.8/(68-11)$ \\ 
                                                                                                                     &                        & unblinded & $1.017 \pm 0.042$  & $0.995 \pm 0.021$ &$0.424 \pm 0.050$ &  $38.8/(68-11)$ \\ 
                                                                                                                     &                        & original  & $1.027 \pm 0.038$  & $0.994 \pm 0.019$ &$0.431 \pm 0.047$ &  $57.8/(68-11)$ \\ \hline 
\multirow{6}{*}{\begin{tabular}[c]{@{}c@{}}BAO \\ + \\ post- \\ recon \\ ($f\reference$)\end{tabular}}     & \multirow{3}{*}{LOWZ}  & blinded   & $0.977 \pm 0.042$  & $0.997 \pm 0.033$ & - &  $61.3/(68-17)$ \\ 
                                                                                                                    &                        & unblinded & $1.014 \pm 0.047$  & $1.018 \pm 0.036$ & - &  $61.3/(68-17)$ \\ 
                                                                                                                    &                        & original  & $0.997 \pm 0.039$  & $1.036 \pm 0.032$ & - &  $48.9/(68-17)$ \\ 
                                                                                                                    & \multirow{3}{*}{CMASS} & blinded   & $0.889 \pm 0.019$  & $0.980 \pm 0.015$ & - &  $45.2/(68-17)$ \\ 
                                                                                                                    &                        & unblinded & $0.935 \pm 0.021$  & $1.010 \pm 0.016$ & - &  $45.2/(68-17)$ \\ 
                                                                                                                    &                        & original  & $0.941 \pm 0.018$  & $1.008 \pm 0.014$ & - &  $59.6/(68-17)$ \\ \hline 
\multirow{6}{*}{\begin{tabular}[c]{@{}c@{}}BAO \\ + \\ post- \\ recon \\ ($f\shift$)\end{tabular}} & \multirow{3}{*}{LOWZ}  & blinded   &  $0.979 \pm 0.042$  & $0.995 \pm 0.033$ & - &  $64.8/(68-17)$ \\ 
                                                                                                                    &                        & unblinded & $1.015 \pm 0.045$  & $1.015 \pm 0.034$ & - &  $64.8/(68-17)$ \\ 
                                                                                                                    &                        & original  & $0.997 \pm 0.039$  & $1.036 \pm 0.032$ & - &  $48.9/(68-17)$ \\ 
                                                                                                                    & \multirow{3}{*}{CMASS} & blinded   & $0.893 \pm 0.019$  & $0.977 \pm 0.015$ & - &  $44.9/(68-17)$ \\ 
                                                                                                                    &                        & unblinded & $0.940 \pm 0.021$  & $1.007 \pm 0.016$ & - &  $44.9/(68-17)$ \\ 
                                                                                                                    &                        & original  & $0.941 \pm 0.018$  & $1.008 \pm 0.014$ & - &  $59.6/(68-17)$ \\ 
\end{tabular}

}
\caption{Parameter results, errors and best-fit $\chi^2$ for all the fits on BOSS LOWZ and CMASS data presented in this section. The full chains, bestfit values and the blinded catalogs can be found at \href{https://github.com/SamuelBrieden/BlindingCatalogs}{https://github.com/SamuelBrieden/BlindingCatalogs}. The first rows correspond to the FS fits on the blinded, original and corrected catalogs. The second and third groups of rows refer  to the correction given by eq. \eqref{eq:combinedresult} and visualized in figures \ref{fig:triangleFS} and \ref{fig:triangleBAO}. The following rows represent the BAO anisotropic fits on reconstructed catalogs, where reconstruction of blinded catalogs was carried out either with the reference growth rate or the shifted growth rate. Note that for the original catalogs we always used the reference growth rate.} \label{tab:dataresults}
\end{table}

\section{Conclusions and discussion} \label{ch:conclusions}

\tikzstyle{block} = [rectangle, draw, fill=blue!15, 
    text width=20em, minimum width = 20em, rounded corners, minimum height=4em]
\tikzstyle{line} = [draw, line width = 0.5mm, -latex']
\tikzstyle{cloud} = [draw, ellipse,fill=red!15, node distance=6.5cm, text width=5em, text centered,
    minimum height=2em]

\begin{figure}[t]
\begin{tikzpicture}[node distance = 3cm, auto]
    % Place nodes
    \node [block] (init) {1. Pick a reference cosmology with parameters (cosmological and physical) 
$\Omega^{\rm ref}$. Pick a set of shift parameters $\Delta \Omega$ so that $\Omega^{\rm shift}=\Omega^{\rm ref}+\Delta \Omega$ as described in section~\ref{ch:blinding_GC}. Note that given  $\Omega^{\rm shift}$ and  $\Omega^{\rm ref}$, $\Omega^{\rm blind}$ is known.};
    \node [block, below of=init] (transform) {2. Given a set of galaxies with observed redshifts $z_i$ transform these into distances using the reference cosmology.};
    \node [block, below of=transform] (RSD) {3. Apply RSD shift from section \ref{ch:blinding_RSD}, eq.~\eqref{eq:mapping_blinded}.};
    \node [block, below of=RSD] (AP) {4. Apply AP-like shift of eq. \eqref{eq:shift_ap}.};   
    \node [cloud, left of=RSD] (inputs) {Inputs: measured redshifts};
     \node [cloud, right of=RSD] (outputs) {Outputs: Blind catalog + reference cosmology};
    \node [block, below of=AP] (end) {5. This produces the blinded catalog. Distribute blinded catalog and reference cosmology.  Keep original catalog, $\Delta \Omega$ and  $\Omega^{\rm blind}$ hidden until ``unblinding".};
   
    % Draw arrows
    \path [line] (init) -- (transform);
    \path [line] (transform) -- (RSD);
    \path [line] (RSD) -- (AP);
    \path [line] (AP) -- (end);
    \path [line,dashed] (inputs) |- (init);
    \path [line,dashed] (end) -| (outputs);
\end{tikzpicture}
 \caption{Summary of the proposed blinding procedure presented in this paper. In case that blinding for the growth rate is not necessary, step 3 can be skipped.} \label{fig:cheatsheet}
\end{figure}
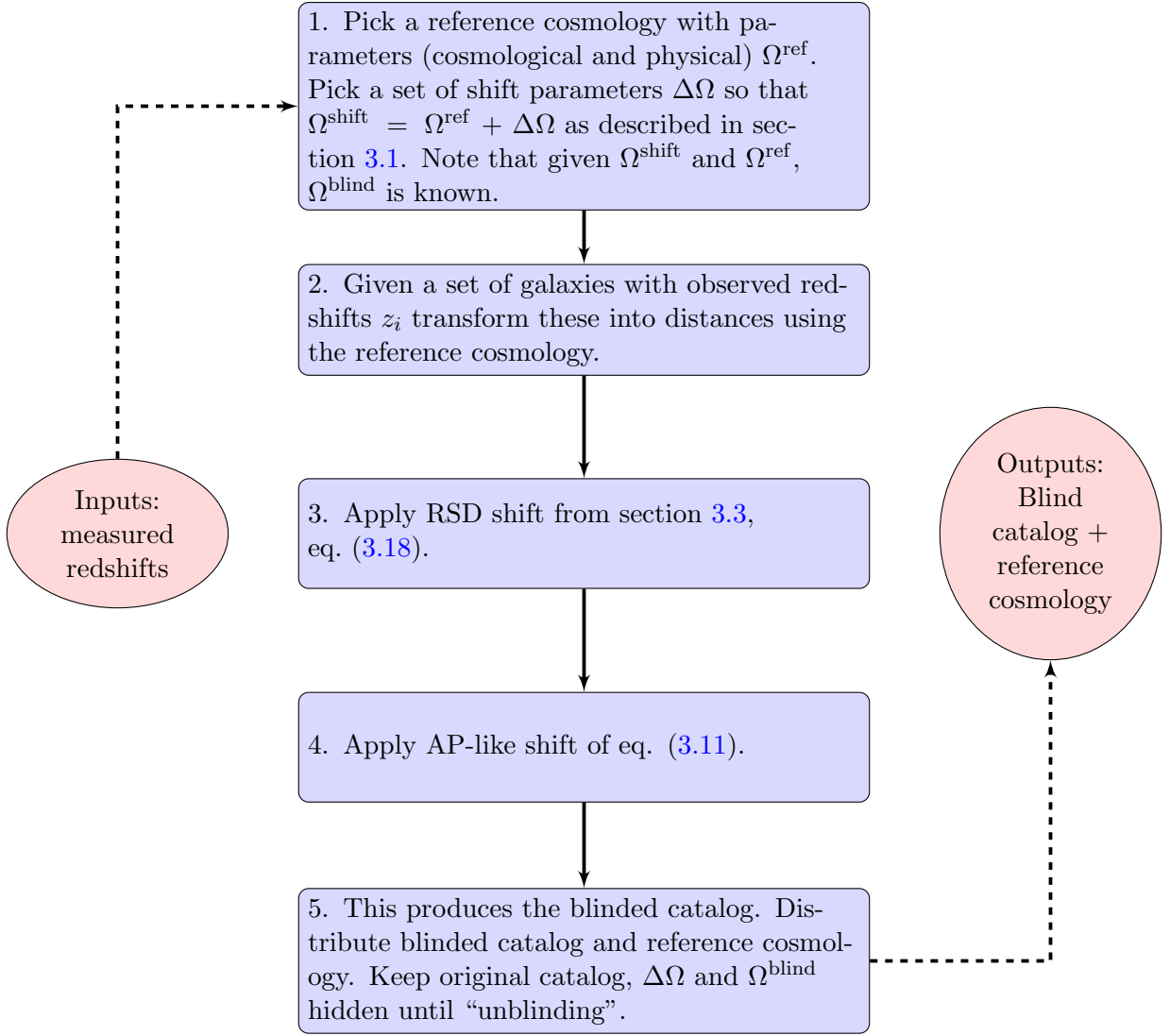

We have presented a novel catalog-level blinding scheme for galaxy redshift surveys.  We have shown how it is possible to shift the position of galaxies in a redshift- and density-dependent way along the line of sight as to implement designed modifications to the  measurements of expansion history and growth rate. In particular it is possible to modify in a fully controlled way, the physical parameters $\alpha_\parallel,  \alpha_\perp, f \sigma_8$, where the  first two (scaling parameters) describe  the background expansion history and the latter  the growth of perturbations.   These parameters can be shifted  coherently across redshift and galaxy samples according to a chosen cosmological model, but shifts can also  be  applied in a much more flexible  way. 
By not altering the galaxies' angular positions,  this approach avoids adding  unnecessary complications into the treatment of angular systematics and  makes  accidental unblinding extremely unlikely. In addition, the blinding technique does not significantly affect the shape of the marginalised posterior distributions of the measured parameters when the covariance matrix appropriately describe the blinded statistics. 

An executive summary of our blinding scheme is shown in figure \ref{fig:cheatsheet}. It can be understood as a zoom into the ``blind box" of figure \ref{fig:flowchart}, that shows our proposal for how to embed blinding in standard clustering analyses.  

By applying the proposed blinding scheme to a suite of mock survey catalogs, we have shown (either explicitely or by construction) that it satisfies all the requirements for a good blinding procedure: {\it i)} the  effect on observables and best-fit parameters can be predicted easily; {\it ii)} it is difficult to unblind accidentally;  {\it iii)} it does not interfere  with key analysis aspects (especially regarding systematics); {\it iv)} it  does not change significantly the error bars;  {\it v)} it is not too numerically intensive  {\it vi)} it is easy to infer  unblinded parameter  constraints with enough accuracy prior to rerun the analysis pipeline again on the original catalog.

In particular, we have verified and quantified  the robustness and accuracy of the proposed blinding scheme regarding points {\it iv)} and  {\it vi)}; the parameters posterior  obtained for the blind catalog  can be shifted back analytically to recover to a good approximation the posterior for the original catalog. This can be done for example,  directly from the output of  MCMCs. In addition, the proposed blinding methodology fulfills points \textit{i)} and \textit{v)} by design. The most numerically intensive step is the RSD-blinding step which takes as much time as applying the reconstruction algorithm to the catalog. In our configuration using BOSS catalogs this takes 5-7 minutes on a dual core CPU depending on the Sample it is applied to. This is comparably mild given that the blinding code needs to be applied to the data once, and not to thousands of mock realisations.

The optimal performance of the blinding (regarding points \textit{iv)} and \textit{vi)}) depends on  the magnitude of  the blinding shift applied: the larger the shift the less robust the approach becomes;   too large deviations between the reference and the shifted cosmology can spoil the predictability of the blinding scheme, because some of the usually made approximations may fail (fixed covariance matrix, adoption of reference model etc.). The choice of how much the shifted values can deviate from the reference parameters should depend on the forecasted sensitivity of the data, and should  probably not exceed deviations of $3\sigma$. We have demonstrated how a large shift leads to a (graceful) degradation of performance by shifting the growth rate parameter by 50\%; even in this extreme case, only point {\it vi)} slightly looses accuracy.  

To satisfy all the above criteria  to a high level of fidelity (especially point  {\it iv)}) the approach can  be  tuned (beyond  setting reasonable shifts amplitudes).   For example,  the smoothing scale intrinsic to the reconstruction procedure may  be adjusted to reduce  noise, but a sub-optimal  choice does not affect the recovered parameter values. 
Similarly, the choice of the linear bias parameter and the growth rate $f$ adopted for the reconstruction step, may affect the  error-bars estimated from the (blinded or original) catalog, if they do not match the ``underlying values". However, in this work we have found that the effect of the latter is very small.

 Another  point, also  common to the  traditional (non-blind) analyses, is the calibration of the mock survey catalogs used to estimate the covariance matrix.  It is well known that  to yield a correct estimate of the parameters errors,  the amplitude of clustering in the mocks used to estimate the covariance must match the clustering amplitude of the data. But the clustering amplitude is not well known when the mocks are being created (before the  data are taken and/or because the data are blind). Usually this calibration is done at the halo occupation distribution level and on small non-cosmological scales, and in this case one has the freedom to calibrate the mocks to the blinded data or to the original catalog. Should the mocks be calibrated on the blind catalog,  one possibility  is  that,  at the  unblinding step, the galaxies positions in the  mocks can  be shifted ``back" with a procedure  similar to the blinding done on the data but in the opposite direction. In this way the mocks would match more closely the  clustering properties of the original  data.

Finally we have applied  the procedure to the BOSS DR12 data release, which serves also as an illustrative example of how the  whole process (from original data, to build catalog, analysis,  unblinding procedure until final results) could be implemented in practice. Even for BOSS data,  we have demonstrated that the blinding procedure recovers  to high accuracy the official (original) results. 

However, our idea laid out in figure \ref{fig:flowchart} of how to implement the blinding procedure in the analysis pipeline can be further adapted depending on specific needs of the collaboration/survey. Other ways are possible: instead of creating a single blinded catalog only, one can create several ones to be analyzed by different subgroups of the collaboration or circulate the blinded catalogs together with the original catalog. We prioritize our proposal to create a single blind catalog to be used by all collaboration members, because this maintains the possibility to cross-check between subgroups. Also we prefer to leave the original catalog hidden until unblinding, because it makes blinding more robust avoiding potential accidental or willing unblinding.

For the magnitude of the blinding shift with respect to the anticipated (Fisher-forecasted) sensitivity we suggested to choose values up to $3\sigma$ %as a compromise between not spoiling the analysis with too large shifts and ensuring that the blinded catalog might return values far away from our expectation
to ensure that the assumption of a parameter-independent (fixed) covariance does still hold. Again, this number is not set in stone and should be chosen carefully depending both on the sensitivity of the experiment and the ``state-of-the-art" knowledge of the parameters in question. We have worked here under the assumption that the covariance matrix is computed only once and is not changed with the blinding/unblinding procedure. However, if computationally feasible, the covariance matrix could be initially  calibrated on the blinded  catalogs and then re-calibrated on the unblinded one. This approach would support larger shifts.

We have not discussed  the criteria for freezing the analysis pipeline and going ahead with unblinding because this may also be very experiment-specific. We have only commented here that any criteria based on the value of the best-fit $\chi^2$ (whether can be deemed acceptable or not)  should take into consideration the spread in best-fit $\chi^2$ values observed in fig.~\ref{fig:chiscatterBAOdiffzCOMB}. The spread in best-fit $\chi^2/\mathrm{ndf}$ values for both the original catalog  and the blinded one range from $\sim 0.5$ to $\sim 1.5$. The spread induced by blinding is a fraction of that scatter and there is no systematic trend: blinding does not  statistically alter the best-fit $\chi^2$ values, but the criteria for defining an ``acceptable" $\chi^2$ should be calibrated on a suite of survey mocks.

While we have only tested explicitly the full shape and anisotropic BAO analyses, there is no reason to believe that the proposed scheme would not work for non-template-based  analysis methods such as  EFTofLSS \cite{DAmico:2019fhj}, \cite{Ivanov:2019pdj} . Moreover, being only a line-of-sight shift,   it is probably  possible to generalise  the scheme to other  tracers such as Lyman-$\alpha$, quasars or line-intensity mapping. 
 
To conclude, we have presented a fast, robust and reliable catalog-level blinding scheme for galaxy redshift catalogs. Although this work is motivated to be applied to spectroscopic surveys, we do not see any obvious limitation for it to be used in future photometric surveys as well.
We envision that it will be of value  to increase the robustness  of cosmological results from  forthcoming LSS surveys and making their cosmological implications free from possible confirmation bias and experimenter's bias. We make publicly available at \href{https://github.com/SamuelBrieden/BlindingCatalogs}{https://github.com/SamuelBrieden/BlindingCatalogs} the following products: BOSSDR12 data and 100 PatchyMock catalogs, that have all been blinded using the ``Combined" case given in table \ref{tab:shiftedparams} as well as associated power spectrum measurements, Monte Carlo Markov chains and bestfit values for the FS and BAO (pre- and postreconstruction) models.  

\section*{Acknowledgement}
We  acknowledge  the DESI collaboration and the GC working group for motivating us to explore  catalog-based blinding for galaxy redshift surveys, and we thank them for ``keeping us real".
We thank Hee-Jong Seo, Lado Samushia and Alexander Mead for useful comments on the draft. 
We thank Gaby Perez and the IT team at ICCUB for their support and dedication in keeping the computer clusters running  despite the difficulties of  the prolonged COVID19 lockdown. We thank an anonymous referee for useful comments that improved the quality of this paper.
Funding for this work was partially provided by the Spanish MINECO under projects AYA2014-58747-P AEI/FEDER, UE,  PGC2018-098866-B-I00,FEDER, UE,   and MDM-2014-0369 of ICCUB (Unidad de Excelencia Mar\'ia de Maeztu).  HGM and SB acknowledge the support from la Caixa Foundation (ID 100010434) which code LCF/BQ/PI18/11630024. 
LV acknowledges support by European Union's Horizon 2020 research and innovation programme ERC (BePreSySe, grant agreement 725327). JLB is supported by  the Allan C. and Dorothy H. Davis Fellowship. 

The massive production of all MultiDark-Patchy mocks for the BOSS Final Data Release has been performed at the BSC Marenostrum supercomputer, the Hydra cluster at the Instituto de Fisica Teorica UAM/CSIC, and NERSC at the Lawrence Berkeley National Laboratory. We acknowledge support from the Spanish MICINNs Consolider-Ingenio 2010 Programme under grant MultiDark CSD2009-00064, MINECO Centro de Excelencia Severo Ochoa Programme under grant SEV- 2012-0249, and grant AYA2014-60641-C2-1-P. The MultiDark-Patchy mocks was an effort led from the IFT UAM-CSIC by F. Prada's group (C.-H. Chuang, S. Rodriguez-Torres and C. Scoccola) in collaboration with C. Zhao (Tsinghua U.), F.-S. Kitaura (AIP), A. Klypin (NMSU), G. Yepes (UAM), and the BOSS galaxy clustering working group. 
Funding for SDSS-III has been provided by the Alfred P. Sloan Foundation, the Participating Institutions, the National Science Foundation, and the U.S. Department of Energy Office of Science. The SDSS-III web site is \href{http://www.sdss3.org/}{http://www.sdss3.org/}.

SDSS-III is managed by the Astrophysical Research Consortium for the Participating Institutions of the SDSS-III Collaboration including the University of Arizona, the Brazilian Participation Group, Brookhaven National Laboratory, Carnegie Mellon University, University of Florida, the French Participation Group, the German Participation Group, Harvard University, the Instituto de Astrofisica de Canarias, the Michigan State/Notre Dame/JINA Participation Group, Johns Hopkins University, Lawrence Berkeley National Laboratory, Max Planck Institute for Astrophysics, Max Planck Institute for Extraterrestrial Physics, New Mexico State University, New York University, Ohio State University, Pennsylvania State University, University of Portsmouth, Princeton University, the Spanish Participation Group, University of Tokyo, University of Utah, Vanderbilt University, University of Virginia, University of Washington, and Yale University.

\appendix

\section{Effect of choice of smoothing scale}
\label{sec:effect-smoothing}
Here we extend the discussion on  the choice of smoothing scale which was touched upon in section \ref{ch:mocks_RSD}.  Figure~\ref{fig:powerspectraRSD}  illustrates  that adopting a too small smoothing scale produces a field dominated by nonlinear velocities, which are hard to model. A too large smoothing scale instead washes out small scale perturbations. Since the $f$ parameter is affected by blinding only at scales above the smoothing scale,  choosing a smoothing scale that is too large compared to the smallest scales included in the analysis would yield an effective scale-dependent $f$ in the blinded catalog. When applying the RSD shift to a catalog different from BOSS, we envision that the choice of smoothing scale should be calibrated on specific survey mocks beforehand. However, we do not expect that using a suboptimal smoothing will have a significant impact on the results. 

\begin{figure}[t]
    \centering
    \includegraphics[width=\textwidth]{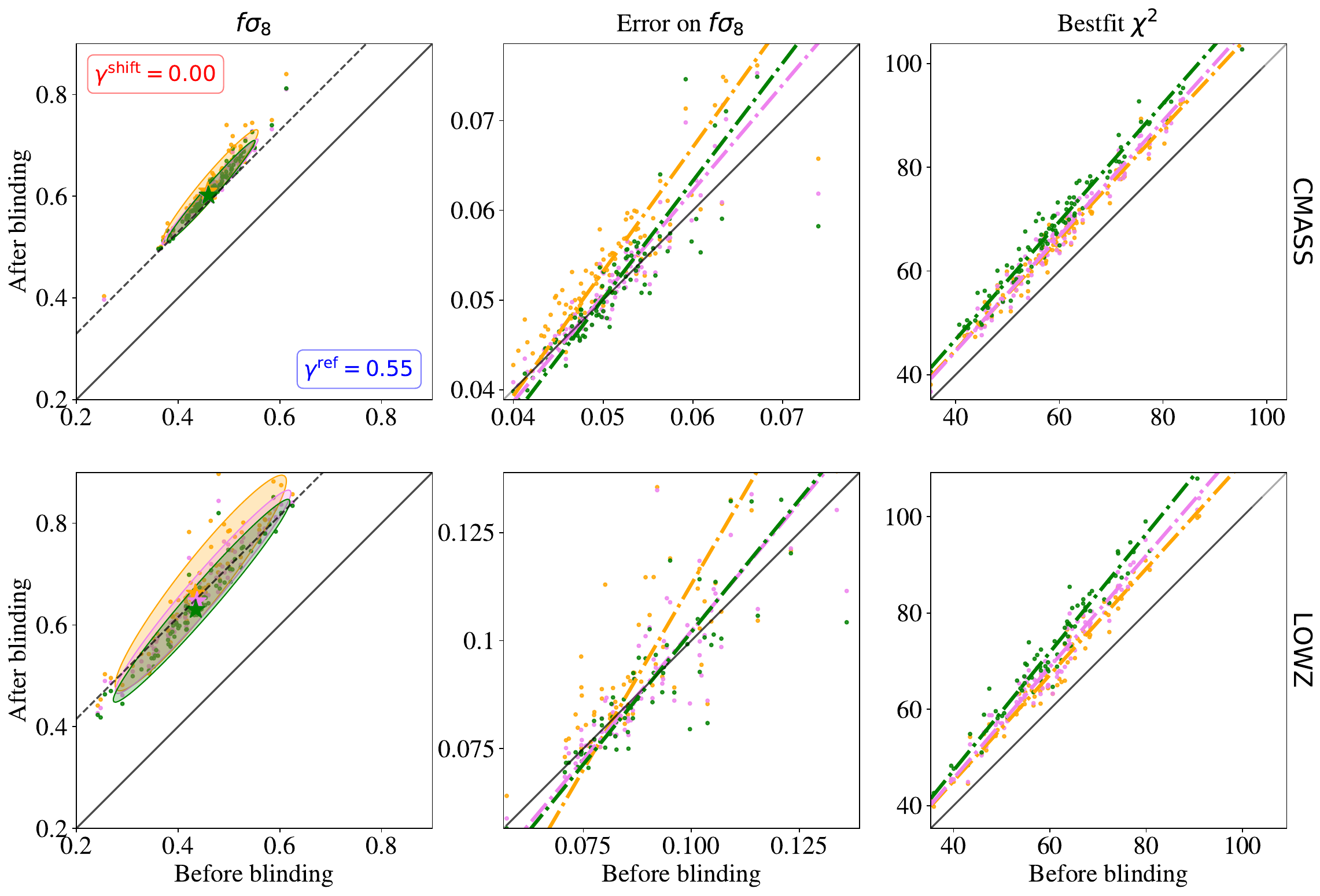}    \includegraphics[width=\textwidth]{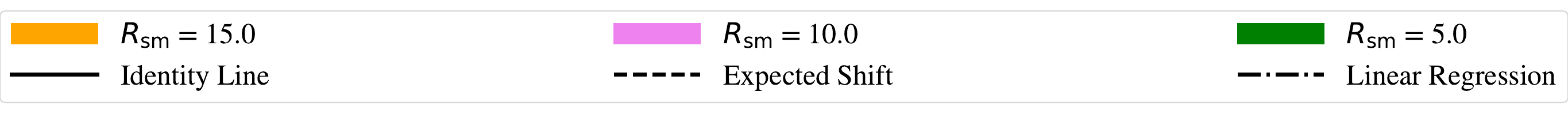}
    \caption{This figure shows FS fit results fo RSD-only blinding on $f \sigma_8$, $\sigma_{f \sigma_8}$ and best fit $\chi^2$ from \Nmocks LOWZ and CMASS realizations. Each panel compares different values of smoothing scale used during the blinding process. The black dashed line in the left panels gives the theoretical expectation for the blinding shift. Colored dashed lines in the middle and right panels are linear regression fits to the mock distributions.}
    \label{fig:comparison_smoothing}
\end{figure}

Figure \ref{fig:comparison_smoothing} shows mean $(f\sigma_8)$ values, errors and best-fit $\chi^2$ before and after blinding for different smoothing scales for LOWZ and CMASS mocks. We can see from the left column, that in all cases the mean values of $(f\sigma_8)$ are nearly identical. Only in the case of $R_\mathrm{sm}=15 \, \mpcoh$ the distribution is slightly broadened. This is also reflected by the error distribution presented in the middle column,  showing a small increase of the errors after blinding. For smaller smoothing scales the error bars do not deviate from the original ones. From the right panels we see that there is a non-significant trend towards higher $\chi^2$ with decreasing smoothing scale. Note that the overall $\chi^2$ distribution is slightly offset from the diagonal. This is related to the fact, that we show here a very extreme blinding shift towards $\gamma=1.0$. We do not observe a systematic discrepancy in $\chi^2$ before and after blinding for more realistic blinding scenarios, as shown in figure \ref{fig:chiscatterBAOdiffzCOMB}.

We conclude  that the  dependence of the  performance of the blinding procedure the specific choice of the smoothing scale is small enough, so that it does not need to be included in the modeling when analysing blinded catalogs. However, we suggest to calibrate the smoothing scale on mocks beforehand in order to minimise  mis-estimation of the error bars.

\section{Impact of deviations from the reference cosmology} \label{ch:deviations_reference}

Here, we study whether the template used for the FS analysis has any impact on parameter constraints when applied to blinded catalogs. Other works (\cite{Carter_BAOtest}, \cite{Bernal_BAObias}, \cite{GilMarininprep}) have shown that the model-independent template fit approach delivers robust BAO and RSD results independent of the choice of template. To validate that these findings also hold for galaxy catalogs, that have been modified using the blinding scheme presented here, we proceed as follows.

In addition to the reference template we generate two linear power spectra with a cosmology far away from the reference one, that we call \textsc{Low-Om} and \textsc{High-Om}. They have either a smaller value with respect to $\Omega_\mathrm{m}\reference$ of $\Omega_\mathrm{m}^\mathrm{low} = 0.286$ or a higher value of $\Omega_\mathrm{m}^\mathrm{high} = 0.35$ and other differences are specified in table \ref{tab:cosmoparams_template}.

\begin{table}[h]
    \centering
    \resizebox{\textwidth}{!}{
    \begin{tabular}{|c||c|c|c|c|c|c|c||c|}
    \hline
        Cosmology & $\Omega_m$ & $\Omega_b h^2$ & $h$ & $\sigma_8$ & $n_s$ & $M_\nu\,[\mathrm{eV}]$ & $r_{\rm d} \,[\mathrm{Mpc}]$ \\ \hline
       \textsc{MD-Patchy} & 0.307115  & 0.02214 & 0.6777 & 0.8288 & 0.9611 & 0 & 147.66 \\
       \textsc{Ref} & 0.31 & 0.022  & 0.676 & 0.8288 & 0.9611 & 0.06 & 147.78 \\
       \textsc{Low-Om} & 0.286  & 0.023 & 0.7 & 0.82 & 0.96 & 0 & 147.15 \\
       \textsc{High-Om} & 0.35  & 0.022 & 0.676 & 0.814 & 0.97 & 0.056 & 143.17 \\
        \hline
    \end{tabular}
    }
    \caption{Values of cosmological parameters and sound horizon at radiation drag for the \textsc{MD-Patchy} mocks cosmology and the cosmologies that were used to generate the different power spectrum templates for this analysis.}
    \label{tab:cosmoparams_template}
\end{table}

Next, we fit the FS model to the mean of 1000 original and blinded Patchy mock catalogs using the 3 different templates. We choose a wavevector range of $0.02 \hompc < k < 0.15 \hompc$. The mocks are blinded using the reference and shifted cosomology of the ``Combined shift" case, specified in table \ref{tab:shiftedparams}. 
 
The performance of the different templates in reproducing the expected shift in the FS model parameters $\left\lbrace \alpha_\parallel, \alpha_\perp, f\sigma_8 \right\rbrace$ due to blinding is shown in table \ref{tab:diff_template_results}. The expectation for the ratio of scaling parameters and difference in $f\sigma_8$ between original and blinded catalogs for LOWZ and CMASS is given by the first row. These values are independent of the choice of template, while the absolute values of the scaling parameters are not, because of their dependence on the fiducial cosmology. The subsequent rows show the results obtained from our MCMC fits to the mean of 1000 mocks using the different templates described by the cosmologies presented in table \ref{tab:cosmoparams_template}. We find a good agreement between theory and expectation for all three choices of template, despite the rather extreme differences among their cosmologies. This shows that galaxy
clustering analyses based on a varying template approach can be applied without a problem to blinded catalogs.
 
\begin{table}[h]
    \centering
    \resizebox{\textwidth}{!}{
    \begin{tabular}{|c||c|c|c|c|c|c|}
    \hline
       \multirow{2}{*}{Case} & \multicolumn{2}{c|}{$\alpha_\parallel^\prime/\alpha_\parallel$} & \multicolumn{2}{c|}{$\alpha_\perp^\prime/\alpha_\perp$} & \multicolumn{2}{c|}{$f\sigma_8 - (f\sigma_8)^\prime$}  \\ \cline{2-7}  
       & $z=0.33$ & $z= 0.60$ & $z=0.33$ & $z=0.60$ & $z=0.33$ & $z=0.60$ \\ \hline
       Expectation & 0.9637 & 0.9502 & 0.9800 & 0.9702 & 0.0809 & 0.0541  \\ \hline
      \textsc{Ref} & 0.960 $\pm$ 0.003 & 0.951 $\pm$ 0.002 & 0.982 $\pm$ 0.001 & 0.973 $\pm$ 0.001 & 0.091 $\pm$ 0.003 & 0.054 $\pm$ 0.002 \\
       \textsc{Low-Om} & 0.965 $\pm$ 0.003 & 0.949 $\pm$ 0.002 & 0.981 $\pm$ 0.002 & 0.972 $\pm$ 0.001 & 0.078 $\pm$ 0.004 & 0.055 $\pm$ 0.003 \\ 
     \textsc{High-Om} & 0.969 $\pm$ 0.003 & 0.956 $\pm$ 0.002 & 0.985 $\pm$ 0.002 & 0.978 $\pm$ 0.001 & 0.083 $\pm$ 0.004 & 0.058 $\pm$ 0.002 \\
        \hline
    \end{tabular}
    }
    \caption{The expected deviations in scaling parameters and $f\sigma_8$ between the original and blinded catalogs compared to the results from the fits on the mean of 1000 LOWZ and CMASS mocks for each template. We show ratios/differences instead of absolute values, because their expected values are independent of the template. All differences between expectation and prediction are below the 1\% level, even for large deviations in cosmological parameters for the \textsc{Low-Om}  and \textsc{High-Om} templates.}
    \label{tab:diff_template_results}
\end{table} 

%\newpage
\bibliography{BlindingPaper}{}

\providecommand{\href}[2]{#2}\begingroup\raggedright\begin{thebibliography}{10}

\bibitem{aghamousa_desi_2016}
A.~Aghamousa, J.~Aguilar, S.~Ahlen, S.~Alam, L.~E. Allen, C.~Allende~Prieto
  et~al., \emph{The {DESI} {Experiment} {Part} {I}: {Science},{Targeting}, and
  {Survey} {Design}}, .

\bibitem{laureijs_euclid_2011}
R.~Laureijs, J.~Amiaux, S.~Arduini, J.-L. Auguères, J.~Brinchmann, R.~Cole
  et~al., \emph{Euclid {Definition} {Study} {Report}}, {\emph{arXiv:1110.3193
  [astro-ph]} (Oct., 2011) }.

\bibitem{collaboration_science-driven_2017}
L.~S. Collaboration, P.~Marshall, T.~Anguita, F.~B. Bianco, E.~C. Bellm,
  N.~Brandt et~al., \emph{Science-{Driven} {Optimization} of the {LSST}
  {Observing} {Strategy}}, {\emph{arXiv e-prints} (Aug., 2017)
  arXiv:1708.04058}.

\bibitem{redbook}
{Square Kilometre Array Cosmology Science Working Group}, D.~J. {Bacon} et~al.,
  \emph{{Cosmology with Phase 1 of the Square Kilometre Array; Red Book 2018:
  Technical specifications and performance forecasts}}, {\emph{arXiv e-prints}
  (Nov., 2018) arXiv:1811.02743}, [\href{https://arxiv.org/abs/1811.02743}{{\tt
  1811.02743}}].

\bibitem{Aghanim:2018eyx}
{\scshape Planck} collaboration, N.~Aghanim et~al., \emph{{Planck 2018 results.
  VI. Cosmological parameters}},  \href{https://arxiv.org/abs/1807.06209}{{\tt
  1807.06209}}.

\bibitem{PhysRevLett.70.1049}
K.~Arisaka, L.~B. Auerbach, S.~Axelrod, J.~Belz, K.~A. Biery, P.~Buchholz
  et~al., \emph{Improved upper limit on the branching ratio
  b(${\mathit{k}}_{\mathit{l}}^{0}$\ensuremath{\rightarrow}${\mathrm{\ensuremath{\mu}}}^{\ifmmode\pm\else\textpm\fi{}}$${\mathit{e}}^{\ensuremath{\mp}}$)},
  \href{http://dx.doi.org/10.1103/PhysRevLett.70.1049}{\emph{Phys. Rev. Lett.}
  {\bf 70} (Feb, 1993) 1049--1052}.

\bibitem{klein_blind_2005}
J.~R. Klein and A.~Roodman, \emph{Blind {Analysis} in {Nuclear} and {Particle}
  {Physics}},
  \href{http://dx.doi.org/10.1146/annurev.nucl.55.090704.151521}{\emph{Annual
  Review of Nuclear and Particle Science} {\bf 55} (2005) 141--163}.

\bibitem{ligo_scientific_collaboration_search_2010}
{LIGO Scientific Collaboration}, {Virgo Collaboration}, J.~Abadie, B.~P.
  Abbott, R.~Abbott, M.~Abernathy et~al., \emph{Search for gravitational waves
  from compact binary coalescence in {LIGO} and {Virgo} data from {S}5 and
  {VSR}1}, \href{http://dx.doi.org/10.1103/PhysRevD.82.102001}{\emph{Physical
  Review D} {\bf 82} (Nov., 2010) 102001}.

\bibitem{conley_measurement_2006}
A.~J. Conley, G.~Goldhaber, L.~Wang, G.~Aldering, R.~Amanullah, E.~D. Commins
  et~al., \emph{Measurement of {Omega}(m), {Omega}(lambda) from a blind
  analysis of {Type} {Ia} supernovae with {CMAGIC}: {Using} color information
  to verify the acceleration of the {Universe}},
  \href{http://dx.doi.org/10.1086/503533}{\emph{Astrophys.J.} {\bf 644} (2006)
  1--20}.

\bibitem{zhang_blinded_2017}
B.~R. Zhang, M.~J. Childress, T.~M. Davis, N.~V. Karpenka, C.~Lidman, B.~P.
  Schmidt et~al., \emph{A blinded determination of {H}0 from low-redshift
  {Type} {Ia} supernovae, calibrated by {Cepheid} variables},
  \href{http://dx.doi.org/10.1093/mnras/stx1600}{\emph{Monthly Notices of the
  Royal Astronomical Society} {\bf 471} (Oct., 2017) 2254--2285}.

\bibitem{des_collaboration_dark_2017}
{DES Collaboration}, T.~M.~C. Abbott, F.~B. Abdalla, A.~Alarcon, J.~Aleksić,
  S.~Allam et~al., \emph{Dark {Energy} {Survey} {Year} 1 {Results}:
  {Cosmological} {Constraints} from {Galaxy} {Clustering} and {Weak}
  {Lensing}}, {\emph{ArXiv e-prints} {\bf 1708} (Aug., 2017) arXiv:1708.01530}.

\bibitem{muir_blinding_2019}
J.~Muir, G.~M. Bernstein, D.~Huterer, F.~Elsner, E.~Krause, A.~Roodman et~al.,
  \emph{Blinding multi-probe cosmological experiments}, {\emph{arXiv:1911.05929
  [astro-ph]} (Nov., 2019) }.

\bibitem{sellentin_blinding_2019}
E.~Sellentin, \emph{A blinding solution for inference from astronomical data},
  {\emph{arXiv:1910.08533 [astro-ph, physics:gr-qc, physics:physics]} (Oct.,
  2019) }.

\bibitem{kuijken_gravitational_2015}
K.~Kuijken, C.~Heymans, H.~Hildebrandt, R.~Nakajima, T.~Erben, J.~T.~A. de~Jong
  et~al., \emph{Gravitational lensing analysis of the {Kilo}-{Degree}
  {Survey}}, \href{http://dx.doi.org/10.1093/mnras/stv2140}{\emph{Monthly
  Notices of the Royal Astronomical Society} {\bf 454} (Dec., 2015)
  3500--3532}.

\bibitem{hildebrandt_kids+viking-450_2019}
H.~Hildebrandt, F.~K\"ohlinger, J.~L. van~den Busch, B.~Joachimi, C.~Heymans,
  A.~Kannawadi et~al., \emph{{KiDS}+{VIKING}-450 and {DES}-{Y}1 combined:
  {Mitigating} baryon feedback uncertainty with {COSEBIs}}, {\emph{arXiv
  e-prints} (Oct., 2019) arXiv:1910.05336}.

\bibitem{Blake:2004tr}
C.~Blake and S.~Bridle, \emph{{Cosmology with photometric redshift surveys}},
  \href{http://dx.doi.org/10.1111/j.1365-2966.2005.09526.x}{\emph{Mon. Not.
  Roy. Astron. Soc.} {\bf 363} (2005) 1329--1348},
  [\href{https://arxiv.org/abs/astro-ph/0411713}{{\tt astro-ph/0411713}}].

\bibitem{2012MNRAS.427.1891A}
J.~{Asorey}, M.~{Crocce}, E.~{Gazta{\~n}aga} and A.~{Lewis}, \emph{{Recovering
  3D clustering information with angular correlations}},
  \href{http://dx.doi.org/10.1111/j.1365-2966.2012.21972.x}{\emph{{Mon. Not.
  Roy. Astron. Soc.}} {\bf 427} (Dec., 2012) 1891--1902},
  [\href{https://arxiv.org/abs/1207.6487}{{\tt 1207.6487}}].

\bibitem{1979Natur.281..358A}
C.~{Alcock} and B.~{Paczynski}, \emph{{An evolution free test for non-zero
  cosmological constant}},
  \href{http://dx.doi.org/10.1038/281358a0}{\emph{Nature} {\bf 281} (Oct.,
  1979) 358}.

\bibitem{kaiser_clustering_1987}
N.~Kaiser, \emph{Clustering in real space and in redshift space},
  \href{http://dx.doi.org/10.1093/mnras/227.1.1}{\emph{Monthly Notices of the
  Royal Astronomical Society} {\bf 227} (July, 1987) 1--21}.

\bibitem{1972MNRAS.156P...1J}
J.~C. {Jackson}, \emph{{A critique of Rees's theory of primordial gravitational
  radiation}},
  \href{http://dx.doi.org/10.1093/mnras/156.1.1P}{\emph{Mon.Not.Roy.Astron.Soc.}
  {\bf 156} (Jan., 1972) 1P}, [\href{https://arxiv.org/abs/0810.3908}{{\tt
  0810.3908}}].

\bibitem{1977ApJ...212L...3S}
W.~L.~W. {Sargent} and E.~L. {Turner}, \emph{{A statistical method for
  determining the cosmological density parameter from the redshifts of a
  complete sample of galaxies.}},
  \href{http://dx.doi.org/10.1086/182362}{\emph{ApJL} {\bf 212} (Feb., 1977)
  L3--L7}.

\bibitem{2012MNRAS.424..564R}
A.~J. {Ross}, W.~J. {Percival}, A.~G. {S{\'a}nchez}, L.~{Samushia}, S.~{Ho},
  E.~{Kazin} et~al., \emph{{The clustering of galaxies in the SDSS-III Baryon
  Oscillation Spectroscopic Survey: analysis of potential systematics}},
  \href{http://dx.doi.org/10.1111/j.1365-2966.2012.21235.x}{\emph{Monthly
  Notices of the Royal Astronomical Society} {\bf 424} (Jul, 2012) 564--590},
  [\href{https://arxiv.org/abs/1203.6499}{{\tt 1203.6499}}].

\bibitem{2014MNRAS.441...24A}
L.~{Anderson}, {\'E}.~{Aubourg}, S.~{Bailey}, F.~{Beutler}, V.~{Bhardwaj},
  M.~{Blanton} et~al., \emph{{The clustering of galaxies in the SDSS-III Baryon
  Oscillation Spectroscopic Survey: baryon acoustic oscillations in the Data
  Releases 10 and 11 Galaxy samples}},
  \href{http://dx.doi.org/10.1093/mnras/stu523}{\emph{Monthly Notices of the
  Royal Astronomical Society} {\bf 441} (Jun, 2014) 24--62},
  [\href{https://arxiv.org/abs/1312.4877}{{\tt 1312.4877}}].

\bibitem{Dawson:2015wdb}
K.~S. Dawson et~al., \emph{{The SDSS-IV extended Baryon Oscillation
  Spectroscopic Survey: Overview and Early Data}},
  \href{http://dx.doi.org/10.3847/0004-6256/151/2/44}{\emph{Astron. J.} {\bf
  151} (2016) 44}, [\href{https://arxiv.org/abs/1508.04473}{{\tt 1508.04473}}].

\bibitem{Lewis:1999bs}
A.~Lewis, A.~Challinor and A.~Lasenby, \emph{{Efficient computation of CMB
  anisotropies in closed FRW models}},
  \href{http://dx.doi.org/10.1086/309179}{\emph{Astrophys. J.} {\bf 538} (2000)
  473--476}, [\href{https://arxiv.org/abs/astro-ph/9911177}{{\tt
  astro-ph/9911177}}].

\bibitem{2011JCAP...07..034B}
D.~{Blas}, J.~{Lesgourgues} and T.~{Tram}, \emph{{The Cosmic Linear Anisotropy
  Solving System (CLASS). Part II: Approximation schemes}},
  \href{http://dx.doi.org/10.1088/1475-7516/2011/07/034}{\emph{Journal of
  Cosmology and Astro-Particle Physics} {\bf 2011} (Jul, 2011) 034},
  [\href{https://arxiv.org/abs/1104.2933}{{\tt 1104.2933}}].

\bibitem{1994ApJ...426...23F}
H.~A. {Feldman}, N.~{Kaiser} and J.~A. {Peacock}, \emph{{Power-Spectrum
  Analysis of Three-dimensional Redshift Surveys}},
  \href{http://dx.doi.org/10.1086/174036}{\emph{Astrophysical Journal} {\bf
  426} (May, 1994) 23}, [\href{https://arxiv.org/abs/astro-ph/9304022}{{\tt
  astro-ph/9304022}}].

\bibitem{PhysRevD.72.043529}
E.~V. Linder, \emph{Cosmic growth history and expansion history},
  \href{http://dx.doi.org/10.1103/PhysRevD.72.043529}{\emph{Phys. Rev. D} {\bf
  72} (Aug, 2005) 043529}.

\bibitem{nusser_prediction_1994}
A.~Nusser and M.~Davis, \emph{On the prediction of velocity fields from
  redshift space galaxy samples},
  \href{http://dx.doi.org/10.1086/187172}{\emph{The Astrophysical Journal} {\bf
  421} (Jan., 1994) L1--L4}.

\bibitem{eisenstein_robustness_2007}
D.~J. Eisenstein, H.-j. Seo and M.~White, \emph{On the {Robustness} of the
  {Acoustic} {Scale} in the {Low}-{Redshift} {Clustering} of {Matter}},
  \href{http://dx.doi.org/10.1086/518755}{\emph{The Astrophysical Journal} {\bf
  664} (Aug., 2007) 660--674}.

\bibitem{padmanabhan_reconstructing_2009}
N.~Padmanabhan, M.~White and J.~D. Cohn, \emph{Reconstructing {Baryon}
  {Oscillations}: {A} {Lagrangian} {Theory} {Perspective}},
  \href{http://dx.doi.org/10.1103/PhysRevD.79.063523}{\emph{Physical Review D}
  {\bf 79} (Mar., 2009) }.

\bibitem{burden_efficient_2014}
A.~Burden, W.~J. Percival, M.~Manera, A.~J. Cuesta, M.~V. Maga\~{n}a and S.~Ho,
  \emph{Efficient {Reconstruction} of {Linear} {Baryon} {Acoustic}
  {Oscillations} in {Galaxy} {Surveys}},
  \href{http://dx.doi.org/10.1093/mnras/stu1965}{\emph{Monthly Notices of the
  Royal Astronomical Society} {\bf 445} (Dec., 2014) 3152--3168}.

\bibitem{Beutler:2016ixs}
{\scshape BOSS} collaboration, F.~Beutler et~al., \emph{{The clustering of
  galaxies in the completed SDSS-III Baryon Oscillation Spectroscopic Survey:
  baryon acoustic oscillations in the Fourier space}},
  \href{http://dx.doi.org/10.1093/mnras/stw2373}{\emph{Mon. Not. Roy. Astron.
  Soc.} {\bf 464} (2017) 3409--3430},
  [\href{https://arxiv.org/abs/1607.03149}{{\tt 1607.03149}}].

\bibitem{Seo:2015eyw}
H.-J. Seo, F.~Beutler, A.~J. Ross and S.~Saito, \emph{{Modeling the
  reconstructed BAO in Fourier space}},
  \href{http://dx.doi.org/10.1093/mnras/stw1138}{\emph{Mon. Not. Roy. Astron.
  Soc.} {\bf 460} (2016) 2453--2471},
  [\href{https://arxiv.org/abs/1511.00663}{{\tt 1511.00663}}].

\bibitem{Gil_Mar_n_2012}
H.~Gil-Mar\'in, C.~Wagner, L.~Verde, C.~Porciani and R.~Jimenez,
  \emph{Perturbation theory approach for the power spectrum: from dark matter
  in real space to massive haloes in redshift space},
  \href{http://dx.doi.org/10.1088/1475-7516/2012/11/029}{\emph{Journal of
  Cosmology and Astroparticle Physics} {\bf 2012} (Nov, 2012) 029–029}.

\bibitem{McDonald_2009}
P.~McDonald and A.~Roy, \emph{Clustering of dark matter tracers: generalizing
  bias for the coming era of precision {LSS}},
  \href{http://dx.doi.org/10.1088/1475-7516/2009/08/020}{\emph{Journal of
  Cosmology and Astroparticle Physics} {\bf 2009} (aug, 2009) 020--020}.

\bibitem{Baldauf_2012}
T.~Baldauf, U.~Seljak, V.~Desjacques and P.~McDonald, \emph{Evidence for
  quadratic tidal tensor bias from the halo bispectrum},
  \href{http://dx.doi.org/10.1103/physrevd.86.083540}{\emph{Physical Review D}
  {\bf 86} (Oct, 2012) }.

\bibitem{Saito:2014qha}
S.~Saito, T.~Baldauf, Z.~Vlah, U.~Seljak, T.~Okumura and P.~McDonald,
  \emph{{Understanding higher-order nonlocal halo bias at large scales by
  combining the power spectrum with the bispectrum}},
  \href{http://dx.doi.org/10.1103/PhysRevD.90.123522}{\emph{Phys. Rev.} {\bf
  D90} (2014) 123522}, [\href{https://arxiv.org/abs/1405.1447}{{\tt
  1405.1447}}].

\bibitem{Gil-Marin:2014sta}
H.~Gil-Mar\'in, J.~Nore\~{n}a, L.~Verde, W.~J. Percival, C.~Wagner, M.~Manera
  et~al., \emph{{The power spectrum and bispectrum of SDSS DR11 BOSS galaxies
  – I. Bias and gravity}},
  \href{http://dx.doi.org/10.1093/mnras/stv961}{\emph{Mon. Not. Roy. Astron.
  Soc.} {\bf 451} (2015) 539--580},
  [\href{https://arxiv.org/abs/1407.5668}{{\tt 1407.5668}}].

\bibitem{Taruya:2010mx}
A.~Taruya, T.~Nishimichi and S.~Saito, \emph{{Baryon Acoustic Oscillations in
  2D: Modeling Redshift-space Power Spectrum from Perturbation Theory}},
  \href{http://dx.doi.org/10.1103/PhysRevD.82.063522}{\emph{Phys. Rev.} {\bf
  D82} (2010) 063522}, [\href{https://arxiv.org/abs/1006.0699}{{\tt
  1006.0699}}].

\bibitem{Percival:2008sh}
W.~J. Percival and M.~White, \emph{{Testing cosmological structure formation
  using redshift-space distortions}},
  \href{http://dx.doi.org/10.1111/j.1365-2966.2008.14211.x}{\emph{Mon. Not.
  Roy. Astron. Soc.} {\bf 393} (2009) 297},
  [\href{https://arxiv.org/abs/0808.0003}{{\tt 0808.0003}}].

\bibitem{Wilson:2015lup}
M.~Wilson, J.~Peacock, A.~Taylor and S.~de~la Torre, \emph{{Rapid modelling of
  the redshift-space power spectrum multipoles for a masked density field}},
  \href{http://dx.doi.org/10.1093/mnras/stw2576}{\emph{Mon. Not. Roy. Astron.
  Soc.} {\bf 464} (2017) 3121--3130},
  [\href{https://arxiv.org/abs/1511.07799}{{\tt 1511.07799}}].

\bibitem{Beutler:2016arn}
{\scshape BOSS} collaboration, F.~Beutler et~al., \emph{{The clustering of
  galaxies in the completed SDSS-III Baryon Oscillation Spectroscopic Survey:
  Anisotropic galaxy clustering in Fourier-space}},
  \href{http://dx.doi.org/10.1093/mnras/stw3298}{\emph{Mon. Not. Roy. Astron.
  Soc.} {\bf 466} (2017) 2242--2260},
  [\href{https://arxiv.org/abs/1607.03150}{{\tt 1607.03150}}].

\bibitem{Carter_BAOtest}
P.~{Carter}, F.~{Beutler}, W.~J. {Percival}, J.~{DeRose}, R.~H. {Wechsler} and
  C.~{Zhao}, \emph{{The impact of the fiducial cosmology assumption on BAO
  distance scale measurements}},
  \href{http://dx.doi.org/10.1093/mnras/staa761}{\emph{MNRAS} {\bf 494} (Mar.,
  2020) 2076--2089}, [\href{https://arxiv.org/abs/1906.03035}{{\tt
  1906.03035}}].

\bibitem{Bernal_BAObias}
J.~L. {Bernal}, T.~L. {Smith}, K.~K. {Boddy} and M.~{Kamionkowski},
  \emph{{Robustness of baryon acoustic oscillations constraints to
  beyond-$\Lambda$CDM cosmologies}}, {\emph{arXiv e-prints} (Apr., 2020)
  arXiv:2004.07263}, [\href{https://arxiv.org/abs/2004.07263}{{\tt
  2004.07263}}].

\bibitem{Briedeninprep}
{Brieden et al.}, \emph{Blinding take 2}, {\emph{In preparation} }.

\bibitem{burden_reconstruction_2015}
A.~Burden, W.~J. Percival and C.~Howlett, \emph{Reconstruction in {Fourier}
  space}, \href{http://dx.doi.org/10.1093/mnras/stv1581}{\emph{Monthly Notices
  of the Royal Astronomical Society} {\bf 453} (Oct., 2015) 456--468}.

\bibitem{kitaura_clustering_2016}
F.-S. Kitaura, S.~Rodr\'iguez-Torres, C.-H. Chuang, C.~Zhao, F.~Prada,
  H.~Gil-Mar\'in et~al., \emph{The clustering of galaxies in the {SDSS}-{III}
  {Baryon} {Oscillation} {Spectroscopic} {Survey}: mock galaxy catalogues for
  the {BOSS} {Final} {Data} {Release}},
  \href{http://dx.doi.org/10.1093/mnras/stv2826}{\emph{Monthly Notices of the
  Royal Astronomical Society} {\bf 456} (Mar., 2016) 4156}.

\bibitem{rodriguez-torres_clustering_2016}
S.~A. Rodr\'iguez-Torres, C.-H. Chuang, F.~Prada, H.~Guo, A.~Klypin,
  P.~Behroozi et~al., \emph{The clustering of galaxies in the {SDSS}-{III}
  {Baryon} {Oscillation} {Spectroscopic} {Survey}: modelling the clustering and
  halo occupation distribution of {BOSS} {CMASS} galaxies in the {Final} {Data}
  {Release}}, \href{http://dx.doi.org/10.1093/mnras/stw1014}{\emph{Monthly
  Notices of the Royal Astronomical Society} {\bf 460} (Aug., 2016) 1173}.

\bibitem{Yamamoto:2005dz}
K.~Yamamoto, M.~Nakamichi, A.~Kamino, B.~A. Bassett and H.~Nishioka, \emph{{A
  Measurement of the quadrupole power spectrum in the clustering of the 2dF QSO
  Survey}}, \href{http://dx.doi.org/10.1093/pasj/58.1.93}{\emph{Publ. Astron.
  Soc. Jap.} {\bf 58} (2006) 93--102},
  [\href{https://arxiv.org/abs/astro-ph/0505115}{{\tt astro-ph/0505115}}].

\bibitem{bianchi_measuring_2015}
D.~Bianchi, H.~Gil-Mar\'in, R.~Ruggeri and W.~J. Percival, \emph{Measuring
  line-of-sight-dependent {Fourier}-space clustering using {FFTs}},
  \href{http://dx.doi.org/10.1093/mnrasl/slv090}{\emph{Monthly Notices of the
  Royal Astronomical Society} {\bf 453} (Oct., 2015) L11}.

\bibitem{gelman1992}
A.~Gelman and D.~B. Rubin, \emph{Inference from iterative simulation using
  multiple sequences},
  \href{http://dx.doi.org/10.1214/ss/1177011136}{\emph{Statist. Sci.} {\bf 7}
  (11, 1992) 457--472}.

\bibitem{Hartlap:2006kj}
J.~Hartlap, P.~Simon and P.~Schneider, \emph{{Why your model parameter
  confidences might be too optimistic: Unbiased estimation of the inverse
  covariance matrix}},
  \href{http://dx.doi.org/10.1051/0004-6361:20066170}{\emph{Astron. Astrophys.}
  (2006) }, [\href{https://arxiv.org/abs/astro-ph/0608064}{{\tt
  astro-ph/0608064}}].

\bibitem{Sherwin:2018wbu}
B.~D. Sherwin and M.~White, \emph{{The Impact of Wrong Assumptions in BAO
  Reconstruction}},
  \href{http://dx.doi.org/10.1088/1475-7516/2019/02/027}{\emph{JCAP} {\bf 02}
  (2019) 027}, [\href{https://arxiv.org/abs/1808.04384}{{\tt 1808.04384}}].

\bibitem{BOSSDR12}
C.~P. {Ahn}, R.~{Alexandroff}, C.~{Allende Prieto}, F.~{Anders}, S.~F.
  {Anderson}, T.~{Anderton} et~al., \emph{{The Tenth Data Release of the Sloan
  Digital Sky Survey: First Spectroscopic Data from the SDSS-III Apache Point
  Observatory Galactic Evolution Experiment}},
  \href{http://dx.doi.org/10.1088/0067-0049/211/2/17}{\emph{ApJS} {\bf 211}
  (Apr., 2014) 17}, [\href{https://arxiv.org/abs/1307.7735}{{\tt 1307.7735}}].

\bibitem{BOSS1}
K.~S. {Dawson}, D.~J. {Schlegel}, C.~P. {Ahn}, S.~F. {Anderson},
  {\'E}.~{Aubourg}, S.~{Bailey} et~al., \emph{{The Baryon Oscillation
  Spectroscopic Survey of SDSS-III}},
  \href{http://dx.doi.org/10.1088/0004-6256/145/1/10}{\emph{AJ} {\bf 145}
  (Jan., 2013) 10}, [\href{https://arxiv.org/abs/1208.0022}{{\tt 1208.0022}}].

\bibitem{EisensteinBOSS2011}
D.~J. {Eisenstein}, D.~H. {Weinberg}, E.~{Agol}, H.~{Aihara}, C.~{Allende
  Prieto}, S.~F. {Anderson} et~al., \emph{{SDSS-III: Massive Spectroscopic
  Surveys of the Distant Universe, the Milky Way, and Extra-Solar Planetary
  Systems}}, \href{http://dx.doi.org/10.1088/0004-6256/142/3/72}{\emph{AJ} {\bf
  142} (Sept., 2011) 72}, [\href{https://arxiv.org/abs/1101.1529}{{\tt
  1101.1529}}].

\bibitem{Bolton12}
A.~S. {Bolton}, D.~J. {Schlegel}, {\'E}.~{Aubourg}, S.~{Bailey}, V.~{Bhardwaj},
  J.~R. {Brownstein} et~al., \emph{{Spectral Classification and Redshift
  Measurement for the SDSS-III Baryon Oscillation Spectroscopic Survey}},
  \href{http://dx.doi.org/10.1088/0004-6256/144/5/144}{\emph{AJ} {\bf 144}
  (Nov., 2012) 144}, [\href{https://arxiv.org/abs/1207.7326}{{\tt 1207.7326}}].

\bibitem{smee13}
S.~A. {Smee}, J.~E. {Gunn}, A.~{Uomoto}, N.~{Roe}, D.~{Schlegel}, C.~M.
  {Rockosi} et~al., \emph{{The Multi-object, Fiber-fed Spectrographs for the
  Sloan Digital Sky Survey and the Baryon Oscillation Spectroscopic Survey}},
  \href{http://dx.doi.org/10.1088/0004-6256/146/2/32}{\emph{AJ} {\bf 146}
  (Aug., 2013) 32}, [\href{https://arxiv.org/abs/1208.2233}{{\tt 1208.2233}}].

\bibitem{alamdr12}
S.~{Alam}, F.~D. {Albareti}, C.~{Allende Prieto}, F.~{Anders}, S.~F.
  {Anderson}, T.~{Anderton} et~al., \emph{{The Eleventh and Twelfth Data
  Releases of the Sloan Digital Sky Survey: Final Data from SDSS-III}},
  \href{http://dx.doi.org/10.1088/0067-0049/219/1/12}{\emph{ApJS} {\bf 219}
  (July, 2015) 12}, [\href{https://arxiv.org/abs/1501.00963}{{\tt
  1501.00963}}].

\bibitem{gil-marin_clustering_2016-1}
H.~Gil-Mar\'in, W.~J. Percival, A.~J. Cuesta, J.~R. Brownstein, C.-H. Chuang,
  S.~Ho et~al., \emph{The clustering of galaxies in the {SDSS}-{III} {Baryon}
  {Oscillation} {Spectroscopic} {Survey}: {BAO} measurement from the
  {LOS}-dependent power spectrum of {DR}12 {BOSS} galaxies},
  \href{http://dx.doi.org/10.1093/mnras/stw1264}{\emph{Monthly Notices of the
  Royal Astronomical Society} {\bf 460} (Aug., 2016) 4210}.

\bibitem{gil-marin_clustering_2016b}
H.~Gil-Marín, W.~J. Percival, J.~R. Brownstein, C.-H. Chuang, J.~N. Grieb,
  S.~Ho et~al., \emph{The clustering of galaxies in the {SDSS}-{III} {Baryon}
  {Oscillation} {Spectroscopic} {Survey}: {RSD} measurement from the
  {LOS}-dependent power spectrum of {DR12} {BOSS} galaxies},
  \href{http://dx.doi.org/10.1093/mnras/stw1096}{\emph{Mon.Not.Roy.Astron.Soc.}
  {\bf 460} (Aug., 2016) 4188--4209}.

\bibitem{DAmico:2019fhj}
G.~D'Amico, J.~Gleyzes, N.~Kokron, D.~Markovic, L.~Senatore, P.~Zhang et~al.,
  \emph{{The Cosmological Analysis of the SDSS/BOSS data from the Effective
  Field Theory of Large-Scale Structure}},
  \href{http://dx.doi.org/10.1088/1475-7516/2020/05/005}{\emph{JCAP} {\bf 05}
  (2020) 005}, [\href{https://arxiv.org/abs/1909.05271}{{\tt 1909.05271}}].

\bibitem{Ivanov:2019pdj}
M.~M. Ivanov, M.~Simonovi\'c and M.~Zaldarriaga, \emph{{Cosmological Parameters
  from the BOSS Galaxy Power Spectrum}},
  \href{http://dx.doi.org/10.1088/1475-7516/2020/05/042}{\emph{JCAP} {\bf 05}
  (2020) 042}, [\href{https://arxiv.org/abs/1909.05277}{{\tt 1909.05277}}].

\bibitem{GilMarininprep}
{Gil-Mar\'in et al.}, \emph{The completed sdss-iv extended baryon oscillation
  spectroscopic survey: measurement of the bao and growth rate of structure of
  the luminous red galaxy sample from the anisotropic power spectrum between
  redshifts 0.6 and 1.0}, {\emph{In preparation} }.

\end{thebibliography}\endgroup
\bibliographystyle{JHEP}

\end{document}